\documentclass[12pt, a4paper]{article} 
\usepackage{graphicx}
\usepackage{amssymb}
\usepackage{amsmath}
\RequirePackage{booktabs}

\usepackage{myverbatim}
\def\listing(#1){\begingroup\tt(#1)\endgroup}

\usepackage[utf8]{inputenc} 
\usepackage{url}
\usepackage[colorlinks]{hyperref}

\newcommand{\optionaltext}[1]{} 
\newcommand{\blankline}{\vspace{5pt}}

\newcommand{\COMMENT}[1]{}

\title{
\textbf{Global Cellular Automata GCA -- A Massively Parallel Computing Model}}

\author{Rolf Hoffmann  \\
	Technische Universität Darmstadt, Germany \\
	}

\date{ } 

\begin{document}

\maketitle

\vspace{-1cm}
\footnotesize
\begin{center}
\today
\end{center}
\normalsize
\blankline

\begin{abstract}
The ``\textit{Global Cellular Automata}'' (GCA) Model  is a generalization of the \textit{Cellular Automata} (CA) Model. 
The GCA  model consists of a collection of cells which
change their states depending on the states of their neighbors, like in the
classical CA  model. In generalization of the CA model, the neighbors are no longer \textit{fixed} and \textit{local}, 
they are \textit{variable} and \textit{global}.
In the \emph{basic} GCA model, a cell is structured into a \emph{data part} and a \emph{pointer part}. 
The pointer part  consists of several pointers that hold addresses to global neighbors. 
The \emph{data rule} defines the new data state, and the \emph{pointer rule} define the new pointer states. 
The cell's state is synchronously or asynchronously updated using the new data and new pointer states.
Thereby the global neighbors can be changed from generation to generation. 
Similar to the CA model, only the own cell's state is modified. 
Thereby write conflicts cannot occur, all cells can work in parallel which makes it a \textit{massively parallel model}. 
The GCA model is related to the CROW (concurrent read owners write) model, a specific PRAM (parallel random access machine) model.
Therefore many of the well-studied PRAM algorithms can be transformed into GCA algorithms.
Moreover, the GCA model allows to describe a large number of data parallel applications in a suitable way. 
The GCA model can easily be implemented in software, 
efficiently interpreted on standard parallel architectures, and 
synthesized/configured into special hardware target architectures.
This article reviews the model, applications, and hardware architectures.

\blankline\noindent
\end{abstract}

\newpage
\tableofcontents
\newpage

\section{Introduction}

Since the beginning of parallel processing a lot of theoretical and practical work has been done in order to  find a \emph{parallel programming model} 
\footnote{Different parallel programming models are reviewed in the survey \cite{Keller2007}.}
(for short \emph{parallel model}) that fulfills the following properties, amongst others

\begin{itemize}
	\item \textit{User-friendly:} Applications are easy to model and to program.

  \item \textit{Platform-independent:} 
  The parallel model can easily programmed, compiled and executed on standard sequential and parallel platforms.

\item \textit{Efficient:} Applications can efficiently be interpreted  on many different parallel target architectures.

	\item \textit{System-design-friendly:} Parallel target architectures supporting the executions of the model
	 (including application-specific processing hardware) are easy to design, to implement, 
	and to program.

\end{itemize}

In the following sections such a parallel model, the \emph{Global Cellular Automata (GCA)} model, is described, and how it can be implemented and used. 
GCA is a model of parallel execution, and at the same time it is a simple and direct programming model. 
A \emph{programming model} 
 is the way how the programmer has to think 
in order to map an algorithm to a certain model which finally is interpreted by a machine. In our case, the
programmer has to keep in mind, that a machine exists which interprets and executes the GCA model.

This model was introduced in \cite{HVW00} (attached, Appendix 2, Sect. \ref{Appendix 2})
and then further investigated, implemented, and applied to different problems.
This article is partly based on the former publications   
\cite{HVW00}--\cite{Jend2016}. 

\blankline
A wide range of applications can easily be modeled as a GCA, 
and efficiently be executed on standard or tailored hardware platforms, 
for instance

\begin{itemize}
	\item 
	Graph algorithms \cite{Ehrt2005}, like Hirschberg's algorithm 
  computing the connected cycles of a graph \cite{Jendrsczok2007Hirschberg,Jendrsczok2008HirschbergFPGA},
  dynamic graphs
  
		\item 
	Vector and matrix operations \cite{Jendrsczok2007APL,Jendrsczok2008HirschbergFPGA,JHE09},
  vector reduction (Sect. \ref{Vector Reduction}), permutations, 
  perfect shuffle operations and algorithms
	
	\item
	Sorting and merging (Sect. \ref{Bitonic Merge}, Sect. \ref{Appendix 2}), 
  sorting with pointers \cite{Hoffmann2003SortingPointer}
  
	\item
	Diffusion with exchange of distant particles \cite{JEH09}
  
  \item
	Fast Fourier Transformation \cite{HVW00} (Sect. \ref{Appendix 2})
	\item
	PRAM (Parallel Random Access Machine (Sect. \ref{Relation to the CROW Model})
  algorithms without concurrent write, converted into GCA algorithms,
  like the
  \textit{Prefix Sum} (Sect. \ref{Prefix Sum, Horn's Algorithm})

	\item
	N-body simulation \cite{JHL09}	   
  
	\item
	Traffic simulation \cite{Schaeck2010Traffic,Schaek2011}           
  
	\item
	Multi-agent simulation \cite{Schaeck2009AgentSim,Schaek2010Multiagent,Schaek2011Multicore,Schak2011Corrected}, 
    logic simulation \cite{WiegandSiemers2004},

\item

Hypercube algorithms\footnote{
Sanjay Ranka and Sartaj Sahni: Hypercube Algorithms.
Eds. Dogramaci, Özay et al. Bilkent University Lecture Series, Springer (1990)}
, combinatorics, communication networks, and neural networks	
	
  \item
	Synchronization related to the \emph{Firing Squad Synchronization Problem}
  \cite{Moore1968}--\cite{WikipediaFSSP2022}, 
	a new application described in Sect \ref{SynchronousFiring}.

\end{itemize}

\noindent
This article is organized as follows:

\begin{enumerate}
\item
(The Global Cellular Automata Model GCA, Sect. \ref{The Global Cellular Automata Model GCA}):
the idea using \textit{pointers} and \textit{pointer rules} in the cells, and the three model variants \textit{basic}, \textit{general} and \textit{plain}

\item
(Relations to Other Models, Sect. \ref{Relations to Other Models}):
the relations to the CROW PRAM model, Parallel Pointer Machines and Boolean Networks

\item
 (GCA Algorithms, Sect. \ref{GCA Algorithms}):
examples for the three GCA variants and a novel application (Synchronous Firing)

\item
 (GCA Hardware Architectures, Sect. \ref{GCA Hardware Architectures}):
\textit{fully parallel}, \textit{sequential}, and \textit{partial parallel} architectures

\item 
(Appendix 0, Sect. \ref{Appendix 0}): 
Pascal program code for the 1D basic and general model
\item 
(Appendix 1, Sect. \ref{Appendix 1}): 
Pascal program code for synchronous firing

\item
(Appendix 2, Sect. \ref{Appendix 2}): 
\textit{first paper} introducing the GCA Model.

\end{enumerate}

\section{The Global Cellular Automata Model GCA}
\label{The Global Cellular Automata Model GCA}

The classical Cellular Automata (CA) model consists of an array of cells arranged in an $n$-dimensional grid. Each cell is connected to its neighbors belonging to a local neighborhood.
For instance, the von-Neumann-Neighborhood 
of a cell under consideration (also called the \textit{\textbf{C}enter} \emph{Cell})
contains its nearest neighbors in the \textit{\textbf{N}orth, \textbf{E}ast, \textbf{S}outh, }and \textit{\textbf{W}est}.
The next state of the center cell is defined by a local rule $f$ residing in each cell: $C \leftarrow f(C, N, E, S, W)$. 
At discrete time $t$ (or ``at time-step $t$''), all cells are applying the same rule synchronously  and thereby a new  generation of cell states (a \textit{configuration}) for the next time $t+1$ is computed. 

As each cell changes only its own state (only self-modification is allowed), no write conflicts can occur.
The model is \textit{inherently parallel}, \textit{powerful} and \textit{simple}. Many applications with local communication can smartly be described as CA, and CAs can easily be simulated in software or realized in parallel hardware.   

The GCA model is a \textit{generalization} of the CA model using a \textit{dynamically computed global neighborhood}.
In order to get a first  impression of the model, the reader may read  the original paper 
\cite{HVW00} first, 
 attached as Appendix 2  
(Sect. \ref{Appendix 2}).

\subsection{The Idea}

The motivation to propose the GCA model was to 
allow a more \emph{flexible communication} between cells
by enhancing the CA model. 

Flexible communications is obtained 
by (i) \emph{selecting neighbors dynamically} through rule computed links
and (ii) by allowing any cell of the whole array to be a direct neighbor, 
a so-called \emph{global neighbor}.
Whereas in principle feature (i) can also be realized  in  classical CA, 
feature (ii) is a major paradigm shift from local data access to global data access.
Thereby parallel algorithms which need instant direct communication can easily be modeled. 

Global access even to the most distant cell is the extreme case of the so-called \emph{long range} or \emph{remote} access. 
Long range access can also be called ``long-range wiring''.
The term ``configurable wiring'' can be used when the wiring can be changed before runtime. 

In our model we allow not only a fixed global wiring before processing but also a \emph{dynamic wiring} / \textit{access} during runtime that can change from generation to generation. It is important to notice that write-conflicts cannot appear, because each cell modifies locally its own state only. Therefore  all new cell states can be computed in parallel, and that is why we attribute the model as ``\textit{massively parallel}''. 
Nevertheless we have to realize that global and dynamic neighborhood are more costly 
than the local and fixed neighborhood of standard CA.

 In order to minimize or limit   the cost of the communication network,
one can (i) implement only the communication links (the access pattern) used by the application,
or (ii) restrict the set of possible neighborhoods (the possible links),  locally or in number. 
In the case (ii), the algorithm for the application has to be adjusted to the available neighborhoods.

\begin{figure}[hbt]
	\centering
		\includegraphics[width=10.0cm]{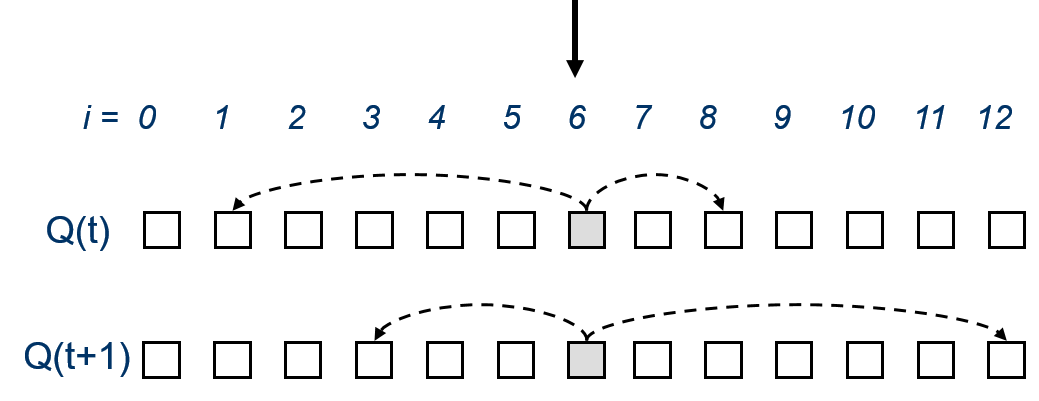}
			\caption{In generation $t$ each cell is connected to $m$ neighbors, and it computes its new neighbors. 
			Then, in  generation $t+1$, each cell is connected to its new neighbors.
			In this example with $m=2$, cell $i=6$ has the neighbors $i=1, 8$ at time-step $t$, and $i=3, 12$ at $t+1$.
			}
	\label{dynlink2d}
\end{figure}

A GCA can informally be described as follows:
A GCA consists of 
an array 
$C=(c_0, c_1,\ldots,c_{n-1})$ of cells $c_i$,
and each cell stores a state $q_i$ which implies an array of
states $Q=(q_0, q_1,\ldots,q_{n-1})$.
The cell's state $q_i=(d_i,P_i)$  consists of a \emph{data part} $d_i$
 and a \emph{pointer part} $P_i = (p_i^1, p_i^2, \ldots, p_i^m)$ 
which contains $m$ \emph{pointers} to neighbors.
The pointers defines the connections (links) to the actual neighbors which are now dynamic.
The local rule does not only update the data part but also the pointer part, and so we use two rules,
the \emph{data rule} and the  \emph{pointer rule}.
Thereby the $m$ neighbors 
can be changed from generation to generation. 
As shown in Fig. \ref{dynlink2d} a cell can change its neighbors between generations. 

All cell states of the array together constitute  a 
\textit{configuration} 
$Q(t)$
at a certain time-step \(t\). 
A GCA is initialized by an initial configuration 
$Q(t=0)$.
The result of the computation is the final configuration  
$Q(t_{final})$.  

\vspace{10pt}
\noindent
Some notions that will be used in the sequel:
\begin{itemize}
\item
\textit{Cell Index: }
The index that identifies a cell.

\item
\textit{Address:}
(\textit{Absolute}) A cell index.
(\textit{Relative}) An offset to the cell's own index.

\item
\textit{Pointer:}
An address pointing to a cell.

\item
\emph{Index Notation:} 
We are mainly using subscripts or superscripts for indexing. 
Alternatively we may use square brackets to denote indexing instead of subscripts (e.g. $q_{pointer} = Q[pointer]$).
We prefer to use square brackets  when dynamic addressing by pointers shall be emphasized.

\end{itemize}

\subsection{The GCA Model Variants}
\label{Section:The Basic GCA Model}

Three model variants are distinguished, 
the \textit{basic model}, 
the \textit{general model} 
and the \textit{plain model}. 
They are closely related and can be transformed into each other to a large extent. 
It depends on the application or the implementation which one will be preferred.
The model variants mainly differ in the way how addresses to the neighbors are stored and computed:

\begin{itemize}
\item \textbf{Basic Model}

Pointers are part of the cell's state which define the global neighbors. 
The are computed at the previous time-step $t-1$ and used at the current time-step $t$.

\item \textbf{General Model}

Pointers are available as in the basic model.
In addition, they can further be modified / specified at the current time-step $t$ before access.

\item \textbf{Plain Model}

The state is not structured into fields, the actual pointers are derived from the current state before access. 
\end{itemize}

The GCA model can easily be programmed.  A compilable PASCAL program is given in Section \ref{Appendix 0} (Appendix 0) that simulates 
the 1D XOR rule with two dynamic neighbors.
The basic model is used in Sect. \ref{Appendix 0-Basic},
and the general model with a common address base is used in Sect. \ref{Appendix 0-General}.

\subsubsection{Basic Model with Stored Pointers} 
\label{Basic Model with Stored Pointers}

\begin{figure}[htb]
	\centering
		\includegraphics[width=11cm]{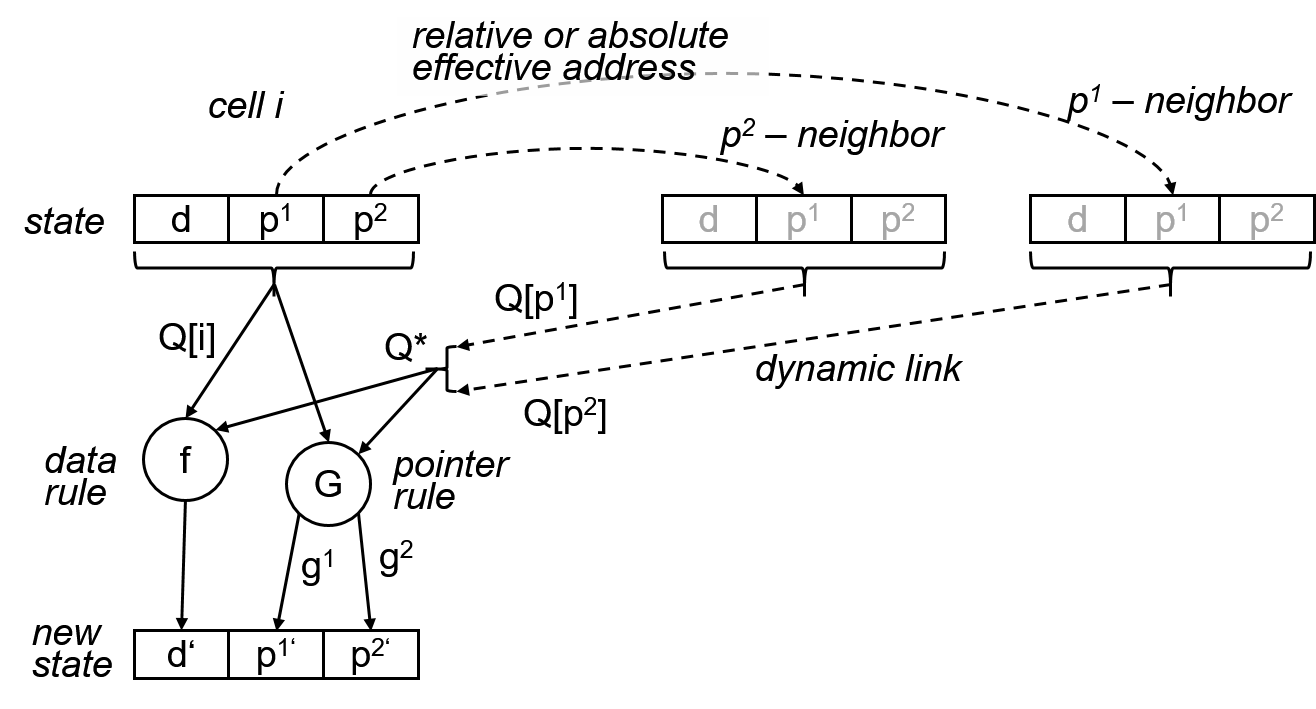} 
			\caption{\textbf{Basic GCA model}, with two pointers.		 
			The cell state is a composition of a data state $d$ 
			and the pointer states $(p^1, p^2)$.
			(Step 1a) Two global cell states 
			are accessed by the pointers and dynamically linked to the cell.
			(Step 1b) The new data state $d'$ and the new pointer states $(p^{1'}, p^{2'})$ are computed by the data rule $f$
			and the pointer rules $G=(g^1,g^2)$.
			(Updating) The new state $(d',p^{1'},p{^2}')$ is copied to the state $(d,p^1,p^2)$.
			Remark: In this figure the cell's  index $i$  of the items was omitted.} 			
	\label{basic-gca}
\end{figure}

The basic model \cite{HVW00,HVWH01} was the first one defined  in order to facilitate the description of cell-based algorithms 
with dynamic long-range interactions. (\cite{HVW00} is attached as Appendix 2, Section \ref{Appendix 2}). 
The cell's state 
consists of two parts, a \textbf{data part} $d$, and a \textbf{pointer part} 
$P$ with $m$ pointers $(p^1,p^2, \ldots, p^m)$. 
The pointers  define directly the global neighbors. 
They are computed in the previous generation $t-1$ to be used in the current generation $t$.
Usually they store \emph{relative} addresses to neighbors, but  \emph{absolute} addresses are allowed, too.

A \emph{basic} GCA is an array
$C=(c_0, c_1,\ldots,c_{n-1})$    
of dynamically interconnected cells $c_i$.
Each cell is composed of storage elements and functions: 

\blankline
$c_i = (q_i, q'_i, f_i, G_i) =((d_i, P_i), (d'_i, P'_i), f_i, G_i)$.

\blankline
For a formal definition we use the elements 

\blankline
$(I, A, D, f, G, q, q', m, u)$ 

as explained in the following:

\begin{itemize}
\item
\emph{I} is a finite \textbf{index set}. 
A unique index (or label, or absolute address) from this set is assigned to each cell.
In the following definitions we want to use only a simple one-dimensional indexing scheme with cell indexes $i\in I=\{0, 1, \ldots, n-1\}$. 
For modeling graph algorithms, we can interpret an index as a label of a node. 
For modeling problems in discrete space, we can map each point in space to a unique index, or we may use a multi-dimensional array and a corresponding indexing scheme.  

\blankline

\item
$m$ is the \textbf{number of pointers} to dynamic neighbors, and $n$ is the number of cells,
where $1 \leq m < n$. 
We call a GCA with $m$ arms/pointers  ``\textit{m-armed GCA"}.

\item
$q_i=(d_i, P_i)\in Q$ is the \textbf{cell's state} and $q'_i=(d'_i,P'_i)$ is its \textbf{new state}.

\item
$Q=D \times A^m$ is the set of cell states.

\item
$d_i\in D$ is the \textbf{data state}, where $D$ is a finite set of data states.

\item
$A$ is the \textbf{address space}.
$p\in A$ is an \emph{address} used  to access a global neighbor. 
It can be \emph{relative} (to the cell's index $i$) or \emph{absolute}. 
Such an address is also called \emph{effective address}. 

$A=I = \{0, \ldots, n-1\}$, is the \textbf{address space for absolute addressing}, or

\blankline
$A=R = \{-n/2, \ldots, (n-1)/2\}$, is the \textbf{address space for relative addressing}, where ``/'' means integer division. That is,

$R=
\begin{cases}
    \{-n/2, \ldots, +(n-2)/2\}     &\textit{if} ~n~ \textit{even}
    \\
    \{-(n-1)/2, \ldots, +(n-1)/2\} &\textit{if} ~n~ \textit{odd}
\end{cases}
$

\item 
$P_i$ is a \textbf{vector of pointers}, the pointer part of the cell's state. 

$P_i=(p_i^1, p_i^2, \ldots, p_i^m)$, where
$p_i^k \in A$.

\item
$f_i$ is the \textbf{data rule}.

$f_i: I \times Q \times  Q^m  \rightarrow D$

It is called \emph{uniform}, if it is index-independent $(\forall{i}: f_i=f)$.

\item
$G_i$ is the \textbf{pointer rule} (also called \textit{neighborhood rule}).

It computes $m$ pointers pointing to the new neighbors at the next time $t+1$
depending on the  cell's state and the  neighbors' states at the current time $t$.

$G_i: I \times Q \times  Q^m  \rightarrow A^m$

It is called \emph{uniform}, if it is index-independent $(\forall{i}: G_i=G)$.

We can split the whole neighborhood rule into a vector of single neighborhood rules each responsible for a single pointer:

$G_i = (g_i^1, \ldots, g_i^m)$ where $g_i^{j=1..m}: I \times Q \times  Q^m  \rightarrow A^m$.

\item
$d_i'=f_i$ is the \textbf{new data state} at time-step $t$ after computation stored temporarily in a memory.

\item
$P_i'=G_i$ is  the \textbf{new vector of pointers} (or the \textit{new neighborhood}) at time-step $t$ after computation  stored temporarily in a memory.

\item
$u \in\{synchronous,asynchronous\}$ is the updating method.

			\underline{\textbf{\emph{u = synchronous}}}

				(\textbf{Phase 1}) Each cells computes its new state $q'_i = (d_i',P_i') $. 

				\begin{itemize}

				\item (\emph{Step 1a})
				The neighbors' states $Q_i^*$ are accessed.
        \footnote{
        In the case that the actual access index is outside its range, it is mapped to it by the \textit{modulo} operation. 
        $Q[i + p]) \mapsto Q[i + p ~mod ~n])$}                

				$Q_i^*=(Q[p^1_i], Q[p^2_i], \ldots Q[p^m_i])$ if  $p^j_i$ is an \textit{absolute} address,

        $Q_i^*=(Q[i + p^1_i], Q[i + p^2_i], \ldots Q[i + p^m_i])$ 
         if  $p^j_i$ is a \textit{relative} address,  

				where $Q$ is the array of cell states:
				
				$Q=(Q[0], Q[1], \ldots, Q[n-1])=(q_0, q_1, \ldots, q_{n-1})$

				\item
				(\emph{Step 1b})
				The new data state and the new neighborhood  are computed by the rules $f_i$ and $G_i$ and stored temporarily. 

				$d'_i\leftarrow f_i(q_i, Q_i^*)$ 

				$P'_i\leftarrow G_i(q_i, Q_i^*)$

				\end{itemize}
				
			
					(\textbf{Phase 2}) 
				For all cells, the new state is copied to the state memory $(q_i \leftarrow q_i')$.
				
The order of computations during Phase 1, and the order of updates during Phase 2 does not matter, but the two phases must be separated.				Parallel computations and parallel updates within each phase are allowed, as it is typically the case for synchronous hardware with clocked registers. 
	
				


       \underline{\textbf{\emph{u = asynchronous}}}

				(\textbf{Only one Phase}) Each cells computes its new state $q'_i = (d_i',P_i') $ which then is copied immediately   to $q_i$.

				\begin{itemize}

				\item (\emph{Step 1a})
				The neighbors' states are accessed, like in the synchronous case. 

				\item
				(\emph{Step 1b})
				The new data state and the new neighborhood are computed by the rules $e$ and $g$ and stored temporarily, like in the synchronous case.  

				\item
				(\emph{Step 1c}) The computed new state  is  immediately stored in the state variable. 
				
				$(d_i,P_i)\leftarrow(d_i',P_i')$.
				
				\end{itemize}

			Every selected cell computes its new state and immediately updates its state. 
			Cells are usually processed in a certain sequential order (including random).
			It is possible to process cells in parallel if there is no data dependency between them.


%
\end{itemize}

\textbf{Relative and Absolute Addressing.}
We have the option to use either \emph{relative} or \emph{absolute} addressing. 
Our understanding is that a pointer $p_i^j$  holds an \emph{effective address} (either relative or absolute), that is ready to access a neighbor.
In the case of absolute addressing, the neighbor's state is $Q[p_i^j]$, and 
in the case of relative addressing, the neighbor's state is $Q[i \oplus  p_i^j]$ where '$\oplus$' means addition $mod ~n$.

This means, that in the case of relative addressing, the cell's index has to be added to the pointer in order to access
the array of states by an absolute address.  
Another way is to use an \emph{index-aware access  network} (or method) that automatically takes into account the cell's position, for instance by an adequate wiring. 
For instance multiplexers can be used where input 0 is connected to the cell $i$ itself, input 1 to the next cell $i+1$, and so on in cyclic order. 
The multiplexer can  then directly be addressed by relative addresses (mapped to positive increments that identify the inputs of the multiplexers).  

Usually \emph{relative addressing} is the first choice, it is more convenient for applications because 
(i) the initial pointer connections are easier to define and often in a uniform way, and
(ii) the initial pointer connections often do not depend on the size $n$ of the array, and
(iii) pointer modifications  are easier to conduct. 

\vspace{10pt}
\textbf{Further Dependencies.}
In some applications, the rules shall further depend on the current time $t$ (counted in every cell, or supplied by a central control),
or on the states $W(i)$ of some additional \emph{fixed local neighbors} as it is standard in classical CA.
Then we can extend the parameter list of the data and pointer rule by $(t, W(i))$, or more general by $(i,t,W(i,t))$.

\vspace{10pt}
\textbf{GCA Implementation Complexity.}
\label{page-complexity}

\begin{itemize}
\item
\textit{Memory Capacity.}
The data part of a cell needs a constant number of bits ~$bit(D)$
where $bit(D)=\delta$ is the number of bits needed to store the data state $D$.
The pointer part needs the capacity 
$m \cdot log_2~n$, it depends on $n$ because the larger the number of cells, 
the larger becomes the address space. 
So the whole memory capacity is

$2n \cdot V(n,m)$~, where $V(n,m)= \delta + m \cdot log_2 ~n$  is the word length of the cell state.

\item
\textit{Data and Pointer Rule.}
The data rule has $m+1$ inputs of word length $V$ and  $\delta$ output bits.

The whole pointer rule  has the same number of inputs bits as the data rule,
but $m \cdot log_2~n$ output bits.
We assume that the internal wiring is included in the rules. 
Then the complexity of the  rules is in

$O(nV \times V)=O((n+1)\cdot V(n,m))$.

\item
\textit{Communication Network.}
\begin{itemize}

\item
\textit{Interconnections.}
The number of links between cells is $n \cdot m(n-1)$
because each cell can have $m(n-1)$ neighbors.
The average link length is $n/4 \times \textit{space-unit}$
for a ring layout structure. 
Each link is $V(n,m)$ bit wide. 
Then we get for the overall effort 
(considering wire length and bit width capacity) 
$O(mn^3 \times V(n,m))$.

\item
\textit{Switches.}
In addition, $mn$  switches or multiplexers are necessary for selecting the neighbors.
Each multiplexer has $n$ inputs and one output with a word length of $V$ bits.
For each  bit of $V$, a simple one-bit multiplexer with a complexity of $O(V)$
is needed. So a word multiplexer has the complexity $O(nV)$.
The complexity for all $nm$ multiplexers is then 
$O(mn^2  \times V(n,m))$.

\end{itemize}

In order to keep the effort for the communication network low, the number $m$ of pointers/arms should be small, especially equal to one, and the 
really used neighbors by the algorithm should be analyzed in order to identify unused links.
The effort for the communication network can be reduced by implementing only 
the required access pattern for a certain set of applications,
or one could restrict the set of possible neighborhoods (links to neighbors) in advance per design and then 
use only the available links for programming the algorithm.
\footnote{For instance, only hypercube connections could be supplied. Then hypercube
algorithms can  directly be implemented, and other algorithms have to be transformed/programmed into 
a ``pseudo" hypercube algorithm, if possible.
}
In principle, any network  with an affordable complexity can be used that allows to read information from remote locations,
not necessarily in one time-step. 
-- The problem of GCA wiring was partially addressed in \cite{WiegandSiemers2004,Siemers2012}.

\end{itemize}

\subsubsection{General Model with Address Modification}
%
\begin{figure}[htb]
	\centering
		\includegraphics[width=11cm]{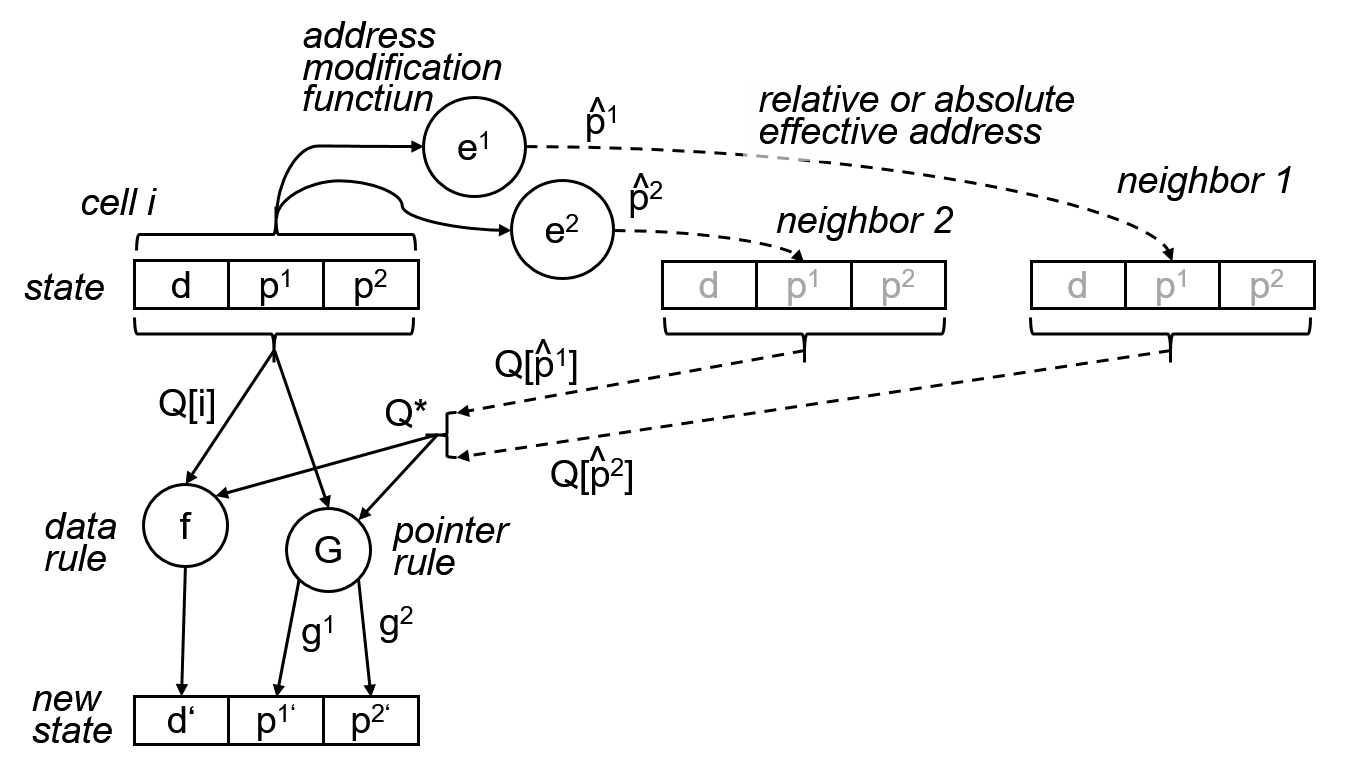} 
			\caption{
			\textbf{General GCA model}, with address modification. Example with two effective addresses.
			Address modification functions $e^1,e^2$ are added to the basic model that allow to modify the addresses before access, at the beginning of the current time-step.
			}
	\label{modi-gca}
\end{figure}
\noindent 
Now we add to the basic model (Sect. \ref{Basic Model with Stored Pointers}) an \emph{address modification function} and call this model \emph{general model}.
In the  \emph{basic model},  the pointers 
store effective addresses
that are directly used to access the neighbors, 
and they
are computed and fixed in the preceding generation $t-1$.
In the\emph{ general model}, the former 
stored pointer values $p^{k=1..m}$ get a different meaning, they represent now
\emph{address bases} that will undergo  additional modifications into  real \emph{effective addresses} $\hat{p}^{k=1..m}$.
The effective addresses  $\hat{p}^k$  are  computed at the beginning of each time-step $t$ 
by an extra \emph{address modification function} $e^k$ for each address $k=1\ldots m$:

\blankline
$\hat{p}_i^k = e^k(p_i^1, p_i^2, \ldots, p_i^m, d_i)$.

\blankline
Further parameters may be taken into account, like the cell index $i$, the current time $t$, or the current state of additional locally fixed neighbors $W(i,t)$. Then we yield the more general formula

\blankline
$\hat{p}_i^k(t) = e^k(p_i^1(t), p_i^2(t), \ldots, p_i^m(t), d_i(t), i, t, W(i,t))$.

\blankline
Usually, only a subset of all possible arguments will be used, for instance

\blankline
$\hat{p}_i^k(t) = e^k(p_i^k(t), d_i(t), i, t, W(i,t))$,   not depending on $p_i^{i\neq k}$

\blankline
$\hat{p}_i^k(t) = e^k(p_i^k(t),  d_i(t))$, not depending on $p_i^{i\neq k},i,t, W$

\blankline
$\hat{p}_i^k(t) = e^k(p_i^k(t),  d_i(t), W(i,t))$, not depending on $p_i^{i\neq k},i,t$

\blankline
$\hat{p}_i^k(t) = e^k(p_i^k(t), i)$, not depending on $p_i^{i\neq k}, d_i, t, W$, index-dependent

\blankline 
$\hat{p}_i^k(t) = e^k(p_i^k(t),  t)$, not depending on $p_i^{i\neq k}, d_i, i, W$, time-dependent.

\blankline 
Compared to the basic model, the \emph{general model} has the advantage that  a GCA algorithm  can immediately (in the same time-step, without a one-step delay) specify its global neighbors, for instance depending on the states of local neighbors. 
To summarize, an effective address is (i) partly computed in the preceding generation (in particular as address base in the same way as pointers are computed in the basic model), and then (ii) further specified by an address modification function in the current generation. 

\vspace{10pt}
\textbf{Examples.} We assume relative addressing and one pointer only (single-arm GCA).
The used operator $\oplus$ denotes an addition $mod ~n$ where the result is mapped into the defined relative address space,
$a \oplus b = (a+b) ~mod ~n - \lfloor n/2 \rfloor$. Examples for address modifications:

\begin{itemize}
\item
The effective address depends on the current data state.

$~\textbf{if}~ d_i=0 \textbf{~then~} \hat{p}_i= p_i \textbf{~else~} \hat{p}_i= p_i \oplus 1$

\item
The effective address depends on the current time.

$~\textbf{if}~ odd(t) \textbf{~then~} \hat{p}_i= p_i \oplus (+1) \textbf{~else~} \hat{p}_i= p_i \oplus (-1)$
 
\item
The effective address depends on the current data state of the left and right neighbor, which are additional fixed neighbors as we have in classical CA.

$~\textbf{if}~ (d_{i-1}=0) \textbf{~and~} (d_{i+1}=0) \textbf{~then~} \hat{p}_i= p_i \oplus 1 \textbf{~else~} \hat{p}_i= p_i$

\item
The effective address depends on the current pointer states of the left and right neighbor, which are fixed neighbors.

$ ~\hat{p}_i= p_{i-1} \oplus p_{i+1}$

\end{itemize}

\vspace{10pt}
\textbf{Variant of the General Model with a Common  Address Base.} 
Instead of using $m$  separate address bases, it is possible to combine them into one common $p_i$ only.
Then $p_i$ can be termed  ``\textit{common address base}'' or \emph{neighborhood address information}.
All $m$ effective addresses are then derived from this common address base:
$\hat{p}_i^k = e^k(p_i, d_i)$ for $k= 1 \ldots m$. This variant can save storage capacity  if only a few special neighborhoods are 
used by the algorithm.

\subsubsection{Plain Model}

\begin{figure}[htb]
	\centering
		\includegraphics[width=11cm]{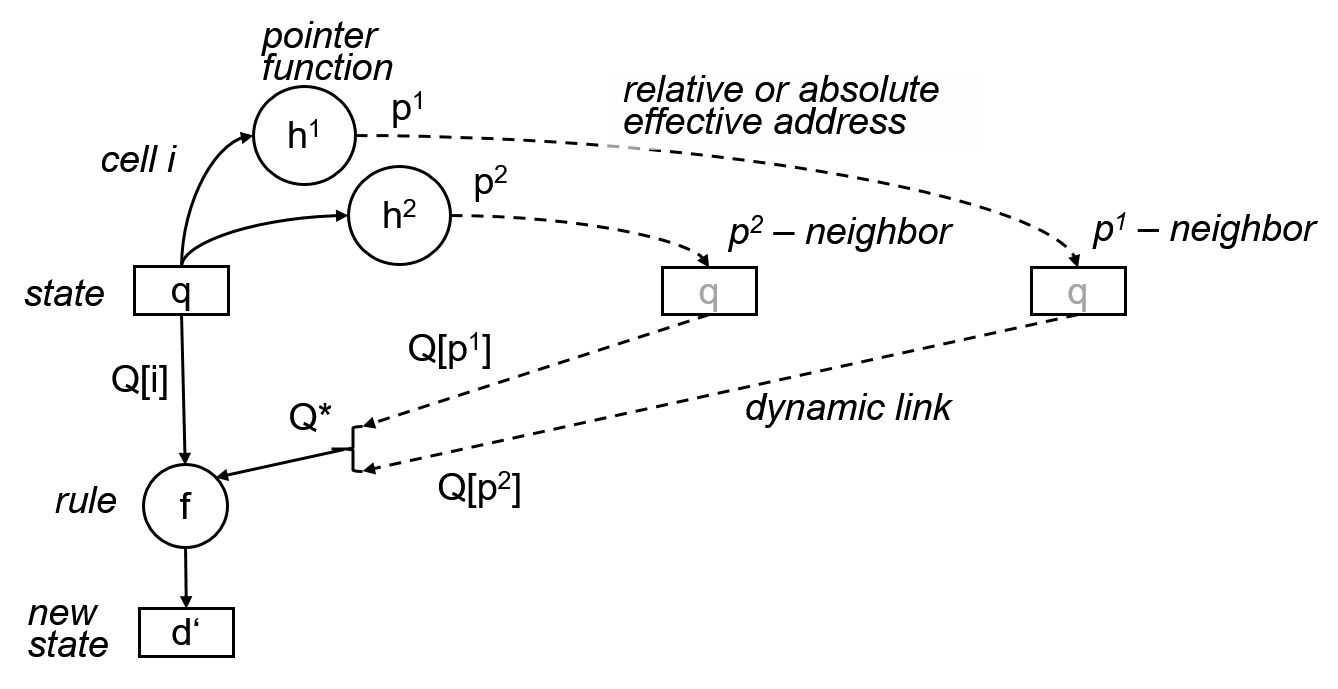} 
			\caption{\textbf{Plain GCA model.} 
			Example with two effective addresses.
			They are computed by the pointer functions $h^1,h^2$ at the beginning of the current time-step before accessing the neighbors.
			}
	\label{plain-gca}
\end{figure}

\noindent
In the \emph{plain} GCA model, the 
pointers are encoded in the cell's state and therefore must be decoded before neighbors can be accessed.
The cell's state is not structured into separate parts (data, pointer) as in the basic and the general model.
(The plain model was also called \emph{condensed} GCA model in a former publication  \cite{Hoffmann2010ISCIA}.)

A \emph{plain} GCA is an array
$C=(c_0, c_1,\ldots,c_{n-1})$    
of dynamically interconnected cells $c_i$.
Each cell $i$ is composed of storage elements and functions: 

$c_i = (q_i, q'_i, f_i, H_i)$.

\blankline
For a formal definition we use the elements $(I, Q, q_i, q'_i, m, A,  P_i,  H_i, f_i, u)$ as explained in the following:

\begin{itemize}
\item
\emph{I} is a finite \textbf{index set} which supplies to each cell a unique index $i$ (label, absolute address) .

$i\in I=\{0, 1, \ldots, n-1\}$. 

\blankline

\item
$Q$ is a finite set of states. They are \textit{not} separated into data and pointer states.

\item
$q_i \in Q$ is the \textbf{cell's state} and $q'_i \in Q$ is its \textbf{new state}.
Storage elements (memories, registers) are provided that can store the cell's state and its new state. 

\item
$m$ is the \textbf{number of pointers} to dynamic neighbors, and $n$ is the number of cells,
where $1 \leq m < n$.

\item
$A$ is the \textbf{address space}.
$p\in A$ is an addr\emph{}ess used to to access a global neighbor. 
It can be relative (to the cell's index $i$) or absolute. 

$A=I = \{0, \ldots, n-1\}$ is the \textbf{address space for absolute addressing}, or

\blankline
$A=R = \{-n/2, \ldots, (n-1)/2\}$ is the \textbf{address space for relative addressing}, where ``/'' means integer division. That is,

$R = \{-n/2, \ldots, +(n-2)/2\}$ if $n$ even, or 

$R=\{-(n-1)/2, \ldots, +(n-1)/2\}$ if $n$ odd.

\item
$P_i$ is a \textbf{vector of pointers}, $P_i=(p_i^1, p_i^2, \ldots, p_i^m)$, where
$p_i^k \in A$.

The pointers are defined by the pointer function 

$P_i=H_i$
~ ($\forall{k=1..m}:p_i^k=h_i^k$), explained next.

\item

$H_i=(h^1_i,h^2_i, \ldots,h^m_i)$ is the \textbf{pointer function}  
(also called \emph{neighborhood selection function, addressing function}).
It computes $m$ pointers (relative or absolute effective addresses)
pointing to the current neighbors depending on the cell's state $q$ at the current time $t$ before access.

$H_i: I \times Q \rightarrow A^m$

It is called \emph{uniform}, if it is index-independent $(\forall{i}: H_i=H)$.

We can split the whole pointer function into a vector of single pointer functions, 
each responsible for a single pointer separately:

$H_i = (h_i^1, \ldots, h_i^m)$ where $h_i^{k=1..m}: I \times Q  \rightarrow A$.

\item
$f_i$ is the \textbf{cell rule}, taking the states of its global neighbors $Q^*_i \in Q^m$  into account.

$f_i: I \times Q \times Q^m  \rightarrow Q$ 

It is called \emph{uniform}, if it is index-independent $(\forall{i}: f_i=f)$.

\item
$u \in\{synchronous,asynchronous\}$ is the updating method.

			\underline{\textbf{\emph{u = synchronous}}}
			
				(\textbf{Phase 1}) Each cells computes its new state $q'_i = f_i$. 

				\begin{itemize}

				\item (\emph{Step 1a})
				The neighbors' states are accessed.
                \footnote{
        In the case that the actual access index is outside its range, it is mapped to it by the \textit{modulo} operation.  $Q[i + p]) \mapsto Q[i + p ~mod ~n])$}

				$Q_i^*=(Q[h^1_i], Q[h^2_i], \ldots Q[h^m_i])$  if $h_i^j$ is an absolute address,

				$Q_i^*=(Q[i + h^1_i], Q[i + h^2_i], \ldots Q[i + h^m_i])$ if $h_i^j$ is a relative address,%
				
				where $Q$ is the vector of cell states:
				
				$Q=(Q[0], Q[1], \ldots, Q[n-1])=(q_0, q_1, \ldots, q_{n-1})$

				\item
				(\emph{Step 1b})
				The new state is computed by the cell rule $f_i$  and stored temporarily. 

				$q'_i\leftarrow f_i(q_i, Q_i^*)$

				\end{itemize}
				
			
					(\textbf{Phase 2}) 
				For all cells the new state is copied to the state memory $(q_i \leftarrow q_i')$.
				
The order of computations during Phase 1 and the order of updates during Phase 2 does not matter, but the phases must be separated.				Parallel computations and parallel updates within each phase are allowed, as it is typically the case in synchronous hardware with clocked registers. 
	
				


       \underline{\textbf{\emph{u = asynchronous}}}

				(\textbf{Only one Phase}) Each cell computes its new state $c'_i$ which is then immediately copied to $c_i$.

				\begin{itemize}

				\item (\emph{Step 1a})
				The neighbors' states are accessed, like in the synchronous case. 

				\item
				(\emph{Step 1b})
				The new cell state is computed by the rules $f_i$  and stored temporarily, like in the synchronous case.  

				\item
				(\emph{Step 1c}) The computed new state is immediately copied to the state variable. 
				
				$q_i\leftarrow q_i'$.
				
				\end{itemize}

			Every selected cell computes its new state and  updates immediately its state. 
			Cells are usually processed in a certain sequential order (including random).
			It may be possible to process some cells states in parallel if there is no data dependence between them. 
			

%
\end{itemize}

\noindent

In some applications the rules and functions may further depend on the current time $t$ (counted in each cell or in a central control),
or on the states $W(i)$ of some additional fixed local neighbors.
Then we can extend the parameter list of the cell rules by $(t, W(i))$.
 
A typical application modeled by GCA needs only one or two pointers, and the set of really addressed cells during the run of a GCA algorithm (Sect. \ref{GCA Algorithms}) -- the \emph{access pattern} -- is often quite limited.
This means that the neighborhood address space needed by a specific algorithm is only a subset of the full address space.
Then the cost to store the address information and for the communication network can be kept low.
Therefore whole GCA can be designed / minimized / configured with regard to a specific application or a class of applications. 
 
\begin{figure}[hbt]
	\centering
		\includegraphics[width=7.0cm]{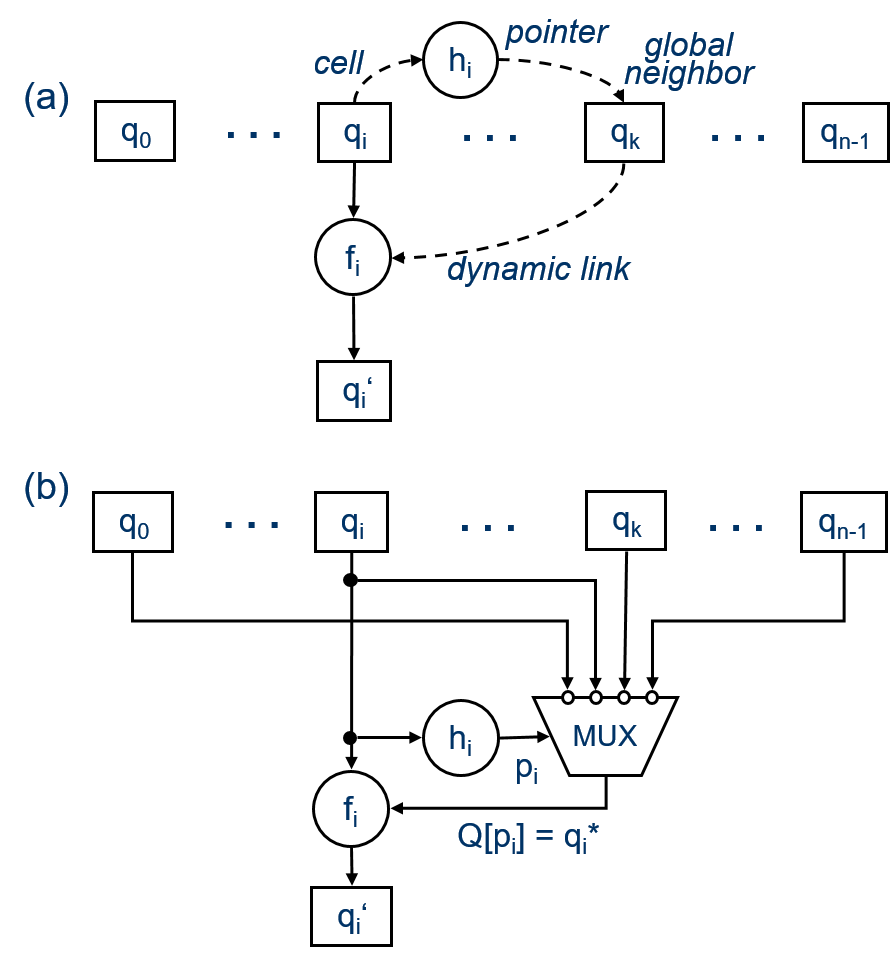}
			\caption
			{\textbf{Plain GCA model}, single-arm.
			(a) Each cell $i$ can select any other cell as its actual neighbor.
			(b) A possible implementation in hardware, absolute addressing. 
			All cell states are inputs to a multiplexer. 
			The actual cell is selected by the pointer $p_i=h_i(q_i)$.
			The rule $f_i(q_i, Q[p_i])$ computes the new state $q'_i$.		
			}
	\label{single-arm-plain}
\end{figure}

 Is a GCA an array of automata as CA are?
Yes, because we can use a CA with a global neighborhood  (fixed connections to every cell) and embedd a GCA.
We can also construct a digital synchronous circuit as for example shown in Fig. \ref{single-arm-plain}.

\vspace{10pt}
\textbf{Single-arm.}
For many applications it is sufficient to use one neighbor only. Then we have 

\blankline
$q_i':= f_i(q_i,q_i^*)$ where

\hspace{5mm}
$q_i^* = Q[p_i]$ for absolute addressing, and 

\hspace{5mm}
$q_i^* = Q[i \oplus p_i]$ for relative addressing,

\hspace{5mm}
where $p_i=h_i(q_i)$   with the declaration  $h_i = h_i^1$ and $p_i=p_i^1$.

\blankline
\noindent 
The principal structure of such a \textit{single-arm} GCA is shown in Fig. \ref{single-arm-plain}.
All cell states are inputs to a multiplexer. 
The actual neighbor is selected by the pointer $p_i = h_i(q_i)$.
Then  the rule $f_i(q_i, Q[p_i])$ computes the new state.

\section{Relations to Other Models}
\label{Relations to Other Models}
\subsection{Relation to the CROW Model}
\label{Relation to the CROW Model}

The GCA model is related to the CROW (concurrent read, owner write) model 
\cite{DYMOND86,DYMOND87,Noam1991,OsterlohKeller2009}, 
a variant of the PRAM (parallel random access machine) models. 

The PRAM is a set of random access
machines (RAM), called processors, that execute the instructions of a program in synchronous lock-step mode and communicate via a global shared memory. Each PRAM instruction
takes one time unit regardless whether it performs a local or a global (remote) operation.
Depending on the access of global variables, variants of the models are distinguished,
CRCW (concurrent read, concurrent write), CREW (concurrent read, exclusive write),
EREW (exclusive read, exclusive write), and CROW. 

The CROW model consists of a \emph{common global memory} and $P$ processors, and each memory location may only be written by its assigned owner processor. In contrast, the GCA model consists of $P$ cells, each with its \emph{local state memory} (data and pointer part) and its local rule (together acting as a small processing unit updating the data and pointer state). Thus the GCA model is (i) ``\emph{cell based}", meaning that the state and processing unit are \emph{distributed} and encapsulated, similar to objects as in the object oriented paradigm, and (ii)
the cells are structured into (data fields, pointer fields, data and pointer rules (for the basic and general model)) according to the application. 
A processing unit of a GCA can be seen as special configured finite state automaton, having just the processing features which are needed for the application. On the other hand, the CROW model is ``\textit{processor based}'', it uses \emph{universal} processors with a standard instruction set independent of the application. Furthermore, in the GCA   the data and pointer state are computed in parallel through the defined rules in one time-step, whereas in the PRAM model several instructions (and time-steps) of a program have to be executed to realize the same effect. 

There is a lot of literature about PRAM models, algorithms and their computational properties, like 
\cite{
JaJa1992,
Lange1993,
Keller2001,
Keller2007}.
The models EROW (exclusive read, owner write) 
\cite{Goyal2005} 
and OROW (owner read, owner write) 
\cite{Rossmanith1991,Gomm1991} 
may also be of interest in this context. 

In this paper we will not investigate the computational properties such as complexity classes for time and space of the GCA model. 
Nevertheless we can see a close relationship to the CROW model, because 
we can (i) distribute the global memory cells with ``owner's write'' property to distinct GCA cells,
and (ii) we can translate a CROW algorithm with several instructions to a GCA algorithm with a few data and pointer rules. 
When we want to compare these models in more depth we have to specify whether we allow an unbounded number of processors and global memory vs. the number of GCA cells and their local memory size.

\subsection{Relation to Parallel Pointer Machines}

The term ``Parallel Pointer Machines'' is ambiguous and stands for different models using processors and memory cells linked by pointers. 
Among them are the KUM (Kolmogorov-Uspenskii machine 1953, 1958) and the SMM (Storage Modification Machine, Schönhage 1970, 1980).
While the KUM operates on an undirected graph with bounded degree, the SMM operates on  a directed graph of bounded out-degree but possibly unbounded in-degree.
Another model similar to SMM is the \emph{Linking Automaton}
(Knuth, The Art of Computer Programming, Vol. 1: Fundamental Algorithms, 1968, 1973).
More details about parallel pointer machines are given in  
\cite{Tromp1985}--
\cite{Petersen2012}.

%
%
These models were mainly defined in the context of graph manipulation. 
The HMM model \cite{LamRuzzo1987}
uses a global memory with exclusive write similar to the CROW model with $n$ processors and with dynamic links between them. 
Our GCA model differs in the way how the pointers are stored, interpreted and manipulated. It comes along in three variants, it is cell-based without a common memory, and it is an easy understandable extension of  the classical  CA.

\subsection{Relation to Random Boolean Networks}

Random Boolean Networks (RBN) were originally proposed by Kauffmann in 1969
\cite{Kauffmann1969,Kauffmann1993}
as a model of genetic regulatory networks.
A RBN consists of $N$ nodes storing a binary state $s\in\{0,1\}$, 
where each node $i\in\{0 \ldots N-1\}=I$ receives $K$ states $s_{i_j}$ (at time $t$) from the connected nodes $i_{j\in\{1 \ldots K\}}$
and computes its next state (valid at time $t+1$) by a 
 boolean function $f_i$: 

\blankline
$\forall i: s_i(t+1)= f_i(s_{i_1}(t), s_{i_2}(t), \ldots s_{i_k}(t))$~~.

\blankline
Considered as a directed graph, each node is a \textit{computing node} 
that receives $K$ inputs via the arcs from  the connected \textit{source nodes}.
In other words, the \textit{fan-in} (\textit{in-degree}) of a node is $K$, equal to the number of arrows
pointing to that node, the head ends adjacent with that node.
Arcs can be seen as \textit{data-flow connections} from source nodes to 
computing nodes.
There can be defined some special nodes dedicated for data input and output.
The network graph  can also be called ``wiring diagram''.
In terms of CA, a node is a \textit{cell} that can have read-connections to any other cell.
In RBN, the connections and functions are fixed during the dynamics, but randomly chosen. 
If the connections and functions are designed / configured for a special application, then
the network is called \textit{Boolean Network} (BN).
So a RBN is a randomly configured BN. RBN are often considered as large sets of different 
configured instances which then are used for statistical analysis.
Normally the fan-in  $K$ is much smaller than $N$, but in the extreme case a node can be affected by all others. 
Usually the number $K$ is constant for all nodes, but it can be node dependent (non-uniform), too.

The  GCA model described in the following sections is a more general model that includes BN. 
In the GCA model, nodes are called \textit{cells}  
and source nodes  $s_{i_j}$ are called \textit{neighbors}.
A cell can point to any global neighbor, and the pointers can be changed dynamically by pointer rules.
Pointers in a GCA graph represent the actual read-access to a neighbor, whereas in a BN graph the pointers are inverted and 
represent the data-flow.

The  GCA model provides dynamically computed links, 
whereas in BN the links are fixed/static.
The rules of GCA tend to be cell/space/index independent, 
whereas in BN the boolean functions tend to be node/index dependent.
Another minor difference is that in the GCA model the own state $s_i$ is always available as parameter in the
next state function,
meaning that in GCA self-feedback is always available, 
whereas in BN self-feedback it intentional by a defined wire (self-loop in the graph).

More information about RBN and BN can be found e.g. in 
\cite{Derrida1986}--\cite{Schwab2020}.

\COMMENT{
\bibitem{Derrida1986}
Derrida, B., Pomeau, Y.:
Random Networks of Automata: A Simple Annealed Approximation.
Europhys. Lett. 1(2) (1986), 45-49

\bibitem{Luque1999}
Luque, B., and Ferrera, A. :
Measuring mutual information in random Boolean networks. 
arXiv preprint adap-org/9909004 (1999).

\bibitem{Shmulevich2002}
Shmulevich, I., Dougherty, E. R., and Zhang, W. :
 From Boolean to probabilistic Boolean networks as models of genetic regulatory networks. 
Proceedings of the IEEE, 90(11) (2002), 1778-1792.

\bibitem{Gershenson2004}
Gershenson, C. : 
Introduction to random Boolean networks. 
arXiv preprint nlin/0408006 (2004).

\bibitem{Bornholdt2008}
Bornholdt, S.:
 Boolean network models of cellular regulation: prospects and limitations. 
Journal of the Royal Society Interface, 5(suppl 1) (2008), 85-94.

\bibitem{Wang2012}
Wang, R. S., Saadatpour, A., and Albert, R. :
Boolean modeling in systems biology: an overview of methodology and applications.
 Physical biology 9.5 (2012): 055001.

\bibitem{Schwab2020}
Schwab, Julian D., et al. : 
Concepts in Boolean network modeling: What do they all mean?.
Computational and structural biotechnology journal 18 (2020): 571-582.

\bibitem{Serra2007}
Serra, R., Villani, M., Damiani, C., Graudenzi, A., Colacci, A., and Kauffman, S. A. : 
Interacting random boolean networks. In Proceedings of ECCS07: European Conference on Complex Systems (pp. 1-15) (2007, October).
}

\section{GCA Algorithms}
\label{GCA Algorithms}
Several GCA algorithms were already described in 
\cite{HVW00} (Reprint, Appendix 2, Sect. \ref{Appendix 2}), and in \cite{HVWH01}--\cite{Jend2016}.

Examples for GCA Algorithms are presented in the following Sections:

\blankline
\ref{Distribution of the Maximum} (Distribution of the Maximum),

\ref{Vector Reduction} (Vector Reduction),

\ref{Prefix Sum, Horn's Algorithm} (Prefix Sum, Horn's Algorithm),

\ref{Bitonic Merge} (Bitonic Merge),

\ref{2D XOR with Dynamic Neighbors} (2D XOR with Dynamic Neighbors),

\ref{Space Dependent XOR Algorithms} (Space Dependent XOR Algorithms),

\ref{1D XOR Rule with Dynamic Neighbors} (1D XOR Rule with Dynamic Neighbors),

\ref{Plain Model Example} (Plain Model Example). 

\blankline
New GCA algorithms about synchronization are presented in the Sections

\blankline
\ref{Synchronous Firing Using a Wave} (Synchronous Firing Using a Wave),

\ref{Synchronous Firing with Spaces} (Synchronous Firing with Spaces),

\ref{Synchronous Firing with Pointer Jumping} (Synchronous Firing with Pointer Jumping).

\subsection{What is a GCA Algorithm?}

We will use the notion ``\emph{GCA algorithm}",  
meaning a specific GCA that computes a sequence of configurations (global states) that is not constant allover. 
As in CA, we start with an initial configuration and expect a dynamic evolution of different configurations.
We distinguish \emph{decentralized} algorithms from 
\emph{controlled} algorithms. We call a decentralized algorithm also \emph{uncontrolled},
\emph{autonomous}, \textit{standalone}, or \textit{(fully) local}.
If not further specified, we mean with a GCA algorithm a decentralized GCA algorithm. 

\vspace{10pt}
\noindent What is a 
\emph{decentralized GCA algorithm}?

\begin{itemize}
\item
\emph{Decentralized GCA algorithm:}
There is no central control which influences the cells behavior. The cells decide themselves about their next state.
The only influence is the central clock that synchronizes parallel computing and updating
when we are using synchronous mode and not asynchronous mode. 
Starting with an initial configuration at time $t=0$, a new generation at $t+1$ is repeatedly computed from the current generation at $t$.
We may require or observe that the global state converges to an attractor (a final configuration or an orbit of configurations), 
or that it changes randomly. 
\end{itemize}

\vspace{10pt}
\textbf{Controlled GCA algorithms.} 
We may enhance our model for more general applications by adding a \emph{central controller} that can be a finite state automaton.
We distinguish three types. 
The properties of these model types is a subject of further research.

\begin{itemize}
\item
\emph{With simple control.}
There is a central control that sends some basic common control signals to the cells.
Typical signals are \textsc{Start}, \textsc{Stop}, \textsc{Reset}, 
a global \textsc{Parameter}, the actual time $t$ given by a central \textsc{Time-Counter},
or a time-dependent \textsc{Control Code}.

\item
\emph{With simple loop control.}
In addition, the control unit is able to maintain simple control structures like loops. 
There can be several loop counters and the number of loops may depend on parameters or on the size $n$ of the cell array.
The control unit may send different instruction codes depending on the control state.
These codes are interpreted by the cells in order to activate different rules.
Not allowed is the feedback of conditions from the cells back to the control unit. 

\item
\emph{With feedback.} 
In addition to the case before, the cells may send conditions back to the control.
Thereby central conditional operations (\emph{if}) and conditional loops (\emph{while, repeat}) can be realized.
A condition can be translated into different instruction codes 
or used to terminate a loop.
More complex control units may be defined if necessary, 
programmable, or supporting the management of subroutines or recursion. 

\end{itemize}

\subsection{Basic Model Examples} 
\label{Basic Model Examples} 

\subsubsection{Distribution of the Maximum}
\label{Distribution of the Maximum}
\begin{figure}[hbt]
	\centering
		\includegraphics[width=0.9\textwidth]{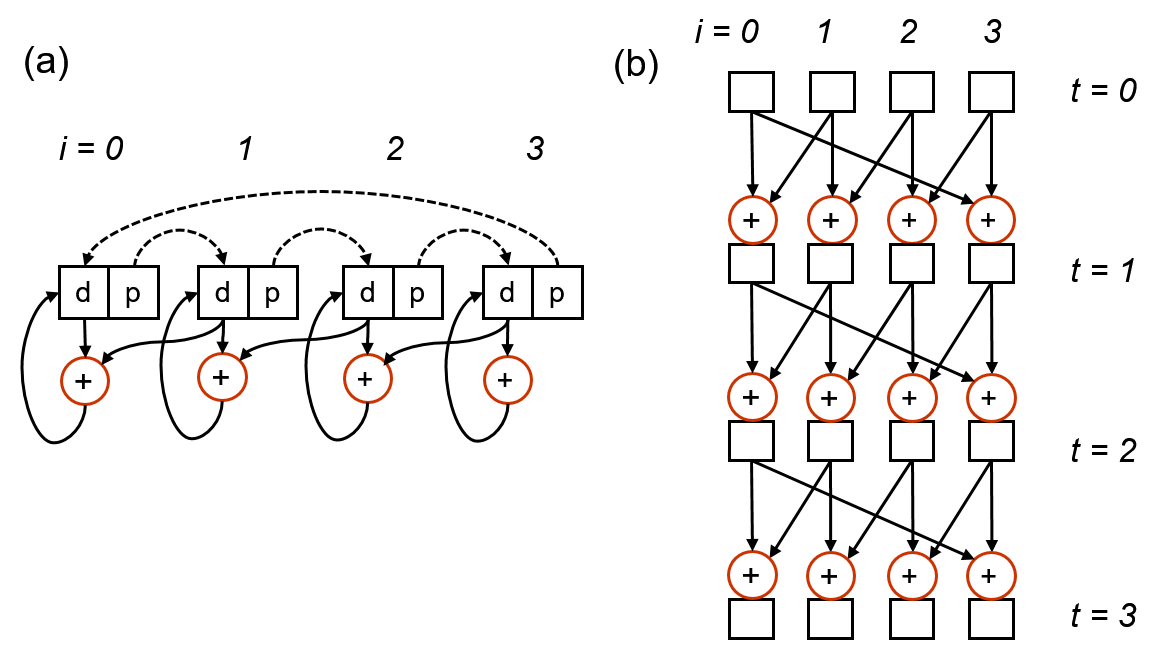}
			\caption
			{\textbf{Maximum}.
			(a)
      Each cell computes the maximum (operator ``+'') of all data elements.
      The pointer to the neighbor is  constant ($p=1$),
          meaning that here always the right neighbor is taken into account. 
      (b)
         The data flow. 
        The algorithm takes $n-1$ parallel steps.
			}
	\label{Maximum}
\end{figure}

All cells shall change their data state into the maximum value of all cells.
The GCA algorithm is rather trivial. The cell's state is
$q=(d,p)$,  where $d$ is an integer and $p$ is a relative pointer. 
Initially $p=1$ for all cells, each cells points to its right neighbor. 
The neighbor's data is $d^*=D[abs(p)]$, where ${abs()}$ maps a relative address to the (absolute) index range 
$\{0, \cdots ,n-1\}$.
If it is clear from the context, then ${abs()}$ may be omitted, and we can simply write  $d^*=D[p]$,
or in ``dot-notation'' : $p.d = d^*=D[p]$.

The data rule is $d'=max(d,d^*)$, and the pointer rule may be a constant $p'=1$. 
The algorithm  takes on the value from the right if it is greater. 
The implementation corresponds to a cyclic left shift register, if the data rule were $d'=d^*$.
The algorithm takes $n-1$ steps. In a conventional way we can write the rules as follows

\blankline
$d_i(t+1) =  max(d_i(t),d_{i+p_i}(t)) = max(d_i(t),d_{i+1}(t))$

$p_i(t+1)=p_i(t)=1$.\\

We can notice that is algorithm can also be described by a classical CA because a fixed local neighborhood is used.
Indeed, the GCA model includes the CA model.
But we  leave the CA model and come to the GCA model when we make use of the global neighborhood (up to $p=n$)
and use the dynamic neighborhood feature.  
Therefore we 
yield a \textit{real} GCA algorithm when we 
use a ``real'' GCA pointer rule 
$p'=f(p,d,n, ...)$, ~ for example\\
 
\blankline
$p'=p+1 ~\emph{mod}~ n$ 

$p'=2p ~\emph{mod} ~n$

$p'= n/2$

$p'= \emph{random}$.\\

We will not investigate these alternatives here further,
and whether they perform better or worse for distributing the maximal value. 
The following real GCA algorithm can also be used to compute the maximum, and it needs only $log_2 ~n$ steps.

\subsubsection{Vector Reduction}
\label{Vector Reduction}

\begin{figure}[hbt]
	\centering
		\includegraphics[width=0.99\textwidth]{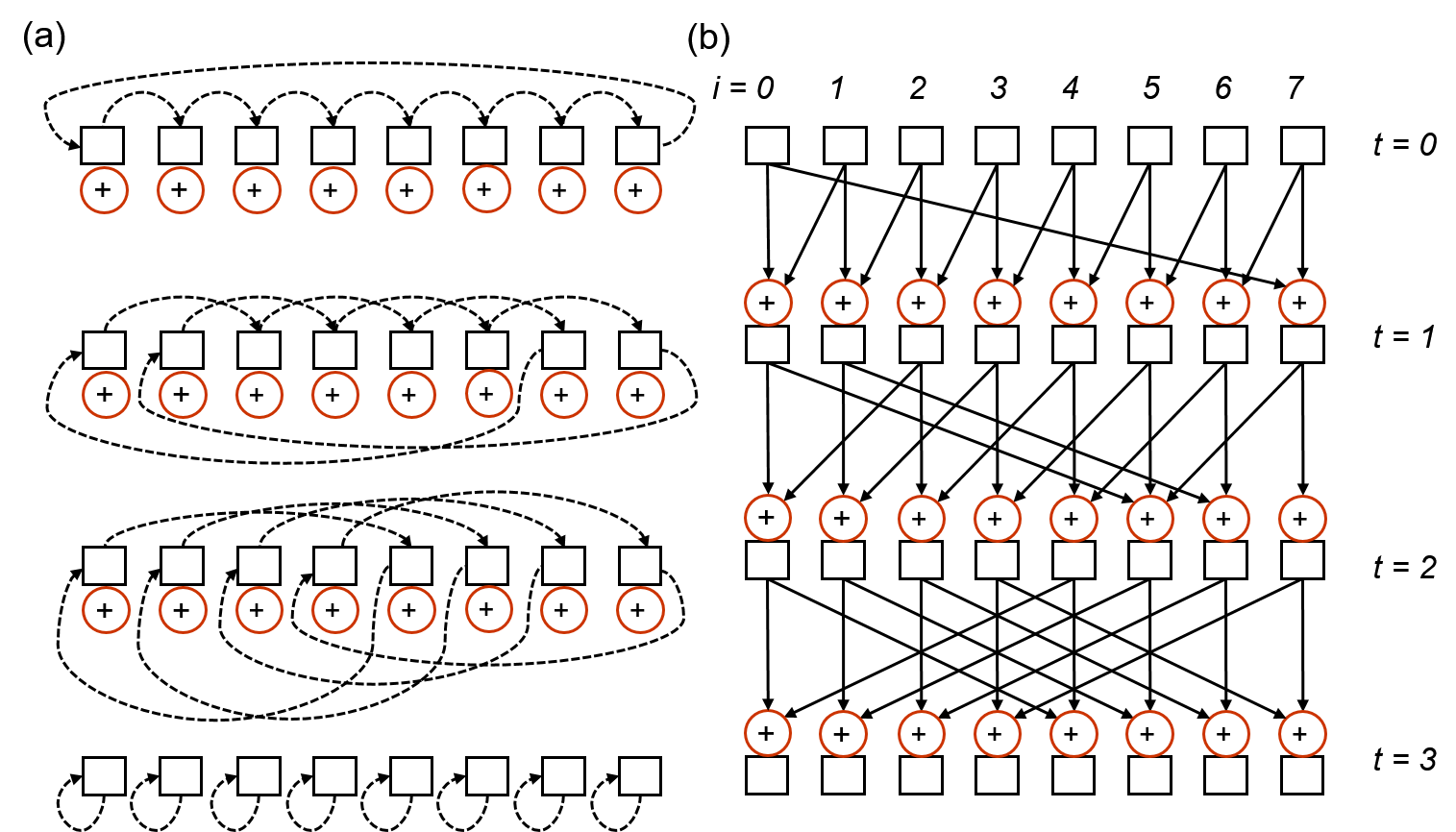}
			\caption
			{\textbf{Vector Reduction}.
			The algorithm computes the sum of all elements.
     Each cell computes the sum in a tree-like fashion.     
      In the first time-step $(t=0) \rightarrow (t=1)$ 
      each cells adds the data value of its right neighbor (with relative pointer value +1).
      In the following generations the distance to the neighbor is doubled ($p=2, 4, \ldots$).
      (a) The  cells with their pointers, dynamically changing. 
      (b) The data flow (inverse to the pointers). 
			}
	\label{Reduction}
\end{figure}

Given a vector $D=(d_0, d_1, \ldots d_{n-1})$.
The reduction function $reduce()$ is\\

$reduce(D)= d_0+d_1+ \ldots +d_{n-1}$ \\

\noindent where '+' denotes any dyadic reduction operator, like \emph{max, min, and, or, average}. 

In order to show the principle, we consider the simplified case where the number of cells is a power of two, $n=2^k$.
Then the reduction can be described as a data parallel algorithm\\

\textbf{for} $t = 1$ \textbf{to} $k$ \textbf{do}

~~~\textbf{parallel for all} $i$

~~~~~~~$d'_i=d_i+d_{i+2^{k-1} ~mod ~n}$  

~~~\textbf{end parallel}   

\textbf{end for}\\

The data elements are accumulated in a tree like fashion and after $k=log_2~n$  steps every cell contains the sum. 
The algorithm can be modified if the number of cells is not a power of two, or if the result shall appear only in one 
distinct cell.

We can easily transform the data parallel algorithm into a GCA algorithm:\\

\begin{tabular}{ll}
$q=(d,p)$        & cell state, $p$ is a relative pointer, initially set to +1\\

$d^*=D[abs(p)]$  & neighbor's data state\\

$d'=d+ (p \neq 0)\cdot d^*$      & data rule, if $(p \neq 0)$ then add\\

$p'=2p  ~mod ~n$      & pointer rule, $p= 1,2,4 \ldots , n/2,0$~~.\\
\end{tabular}\\

\blankline
The problem of controlling the algorithm (Initialize, Start, Stop/Halt) can be implemented differently. 
We assume always an initial configuration at time $t=0$ to be given, and we don't care how it is established. 
Then we assume that a hidden or visible  central time counter $t:=t+1$ is automatically incremented generation by generation.
In some time-dependent algorithms the central time counter can be used, or a separate counter is supplied in
every cell in order to keep the algorithm decentralized. 
The final configuration is reached when the pointer's value changes to 0 by the modulo operation. 
Then $p'=p=0$ holds.
The algorithm may be further active, but the cell's state is not changing any more. 
The algorithm can halt automatically in a decentralized way when
all cells  decide to change into an inactive state when $p=0$.

\subsubsection{Prefix Sum, Horn's Algorithm}
\label{Prefix Sum, Horn's Algorithm}

\begin{figure}[hbt]
	\centering
		\includegraphics[width=0.65\textwidth]{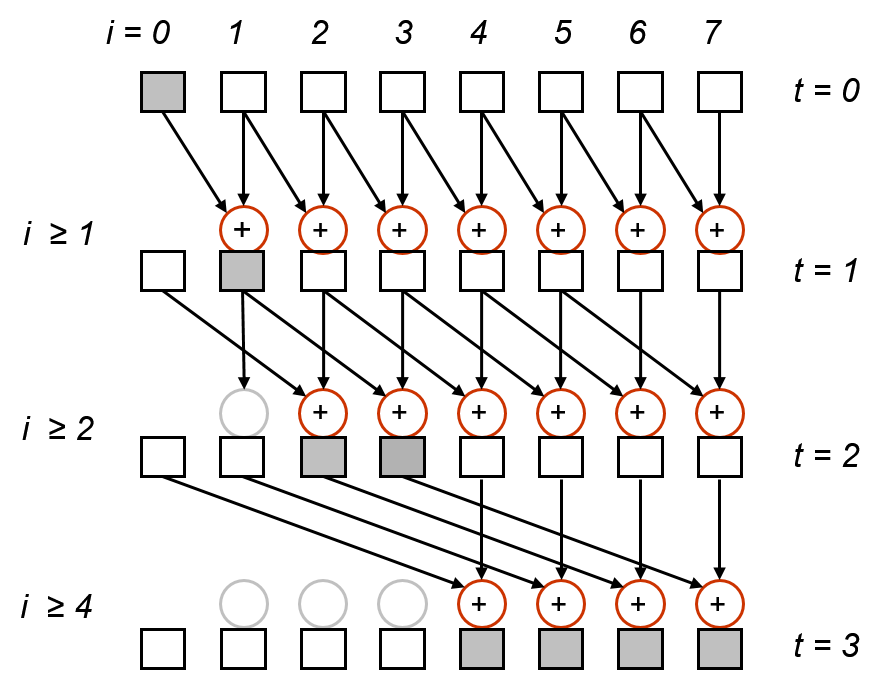}
			\caption
			{\textbf{Horn's Algorithm}.
			The algorithm computes the prefix sum.
      In the first time-step $(t=0) \rightarrow (t=1)$ 
      each cells $i \geq 1$ adds the data value of its left neighbor (relative pointer value -1).
      In the following generations, the distance to the dynamic neighbor is $-2, -4, \ldots$,
      and the number of active adding cells is decreased by 1 until $n/2$.
      The figure shows the data flow. The shaded data elements mark already computed results. 
			}
	\label{Horn}
\end{figure}

Given a vector $D=(d_0, d_1, \ldots d_{n-1})$.
The prefix sum is the vector $(s_i)$ where\\

$s_0=d_0$

$s_1=s_0+d_1 = d_0+d_1$

$s_2=s_1+d_2 = d_0+d_1+d_2$ 

$\ldots$

$s_{n-1}=s_{n-2}+d_{n-1}~~.$ \\

The prefix sum can be computed in different ways. 
\emph{Horn's} algorithm is a CREW data parallel algorithm for $n=2^k$ elements:\\

\textbf{for} $t = 1$ \textbf{to} $k$ \textbf{do}

~~~\textbf{parallel for} $i= 1$ \textbf{to} $n-1$

~~~~~~~\textbf{if}  $i \geq 2^{t-1}$  \textbf{then}    $d'_i=  d_i + d_{i-2^{t-1}}$  

~~~\textbf{endparallel}   

\textbf{endfor}~~.\\

The number of additions (active processors/cells) decreases step by step, it is $(n-1, n-2, n-4, \ldots n/2)$.
The data parallel algorithm can be transformed into the following GCA algorithm straight forward. \\

\begin{tabular}{ll}
$q=(d,p)$        & cell state, $p$ is a relative pointer, initially -1\\

$d^*=D[abs(p)]$  & neighbor's data state\\

$d'= d+ (i \geq -p)\cdot d^*$      & data rule, if $(i \geq -p)$ then add\\
   
$p'=2p  ~mod ~n$      & pointer rule, $p= -1,-2,-4 \ldots , -n/2,0$\\
\end{tabular}\\

An advantage of this algorithm is that the number of simultaneous read accesses (fan-out) is not more than two.
There exists another algorithm where the number of active cells and the maximal fan-out are equal to $n/2$.

\subsection{General Model Examples} 
\label{General Model Example}

\subsubsection{Bitonic Merge} 
\label{Bitonic Merge} 

\begin{figure}[hbt]
(a)\hspace{6.3cm} (b)
\vspace{-1cm}
  \begin{center}
    \includegraphics[width=0.48\textwidth]{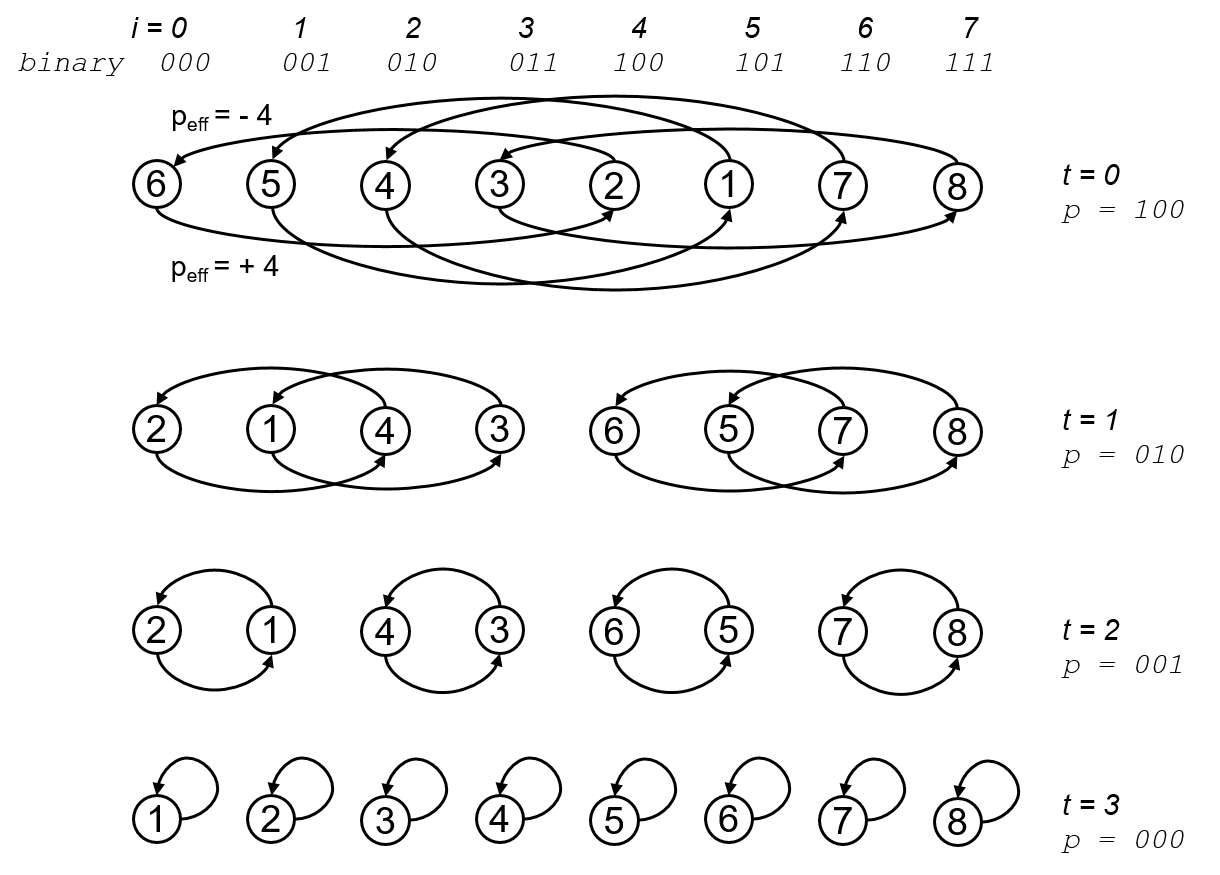}
~~~
    \includegraphics[width=0.38\textwidth]{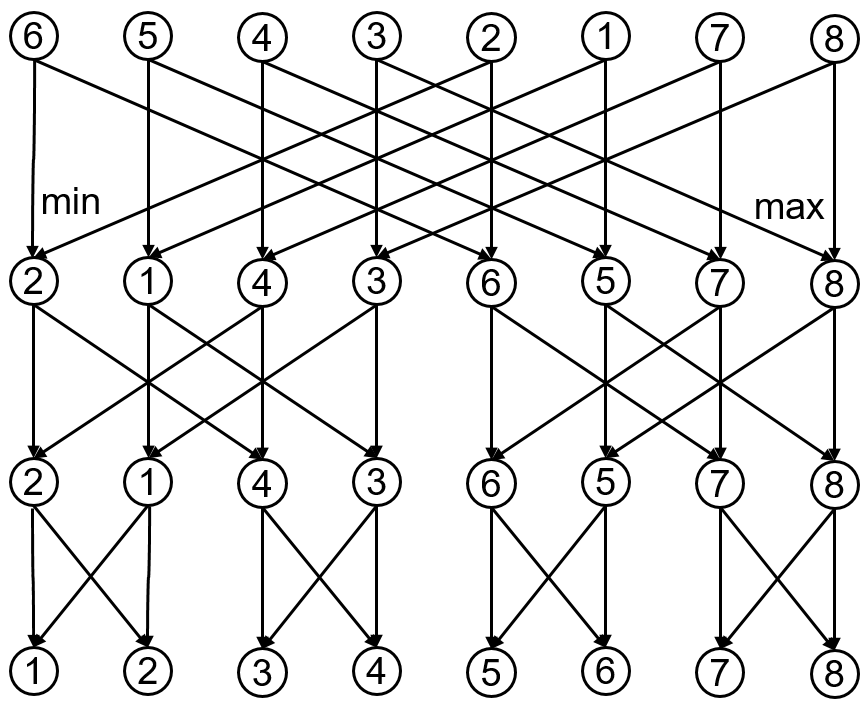}
  \end{center}
  \caption{(a) Initial at $t=0$ a bitonic sequence of length $n=8$ is given. Cells  $0,1,2,3$ access cells $4,5,6,7$ and vice versa.
  The initial pointer base is 4 (binary 100), and it is used to mask the cell's index in order to select either
  $\hat{p}=p_{\textit{eff}}=+4$ or $p_{\textit{eff}}=-4$. Iteratively the pointer base is shifted to the right (division by 2) yielding
  $p_{\textit{eff}}=\pm 2,1,0$. If the right neighbor's value is smaller, it is copied. 
   If the left neighbor's value is greater, it is copied. 
  (b) The data flow. Cells with right neighbors compute the minimum, 
  cells with left neighbors compute the maximum. The graph also shows which cells are accessed during the run, the
  \emph{access pattern} (the inverted arrows, the time-evolution of the pointers). 
  }
  \label{fig:bito}
\end{figure}

The bitonic merge algorithm sorts a bitonic sequence. A sequence of 
numbers is called \emph{bitonic}, if the first part of the sequence is 
ascending and the second part is descending, or if the sequence is 
cyclically shifted.
Consider a sequence of length $n = 2^k$. In the first step, cells with 
distance $2^{k-1}$ are compared, Fig.~\ref{fig:bito}.
Their data values are exchanged if necessary to get the minimum to the
left and the maximum to the right. In each of the following steps
the distance between the cells to be compared is halve of the distance 
of the preceding step. Also with each step the number of sub-sequences 
is doubled. There is no communication between different sub-sequences.
The number of parallel steps is $k = log_2~n$.

The cell' state is a record $q=(d,i,p)$, where $d\in$ \emph{DataSet}, 
$i\in I$ is the cell's identifier, 
and $p\in {0,1,2,...,2^{k-1}}$ is the pointer base, initially set to $2^{k-1}$.\\

\vspace{1.4cm}
The following abbreviations are used in the description of the GCA rules:\\

-- the data and the pointer base: $~d= d_i, ~p= p_i$, 

-- the global neighbor's data state: $~d^*=d^*_i=D[abs(\hat{p_i})]$, 
where $\hat{p_i}$ is the

~~~effective relative address computed from the relative address base.\\

The address modification rule computing the effective address is

\[
  \hat{p} =\left\{\begin{array}{lll} 
	   
	        +p   & \textbf{if} ~~  (i~  \textbf{and}~ p)=0 \hspace{2mm} \\  

         -p    & \textbf{if} ~~  (i~  \textbf{and}~ p)=1 \hspace{2mm}       

			\end{array}
	      \right. .
\]

The data rule is 
\[
  d' =  
  \left\{
  \begin{array}{lll} 
	    
	d^*       &\textbf{if} ~~~~~~(i ~\textbf{and}~  p=0) ~\textbf{and}~   (d^*<d)                \\
			     &~~~~\textbf{or} ~(i ~\textbf{and}~  p=1) ~\textbf{and}~   (d<d^*)     \\

  d       & \textbf{otherwise}                                               \\
      
	\end{array}	      
  \right. .
\]

The pointer rule is $p'=p/2$~.\\
\\
The algorithm can also be described in the  cellular automata language CDL, as follows.

\footnotesize
\begin{myverbatim}
cellular automaton bitonic_merge; 
const dimension = 1; 
      distance  = infinity;  {global access to any cell}
 
type celltype=record            
      d: integer; {initialized by a bitonic sequence to be merged}
      i: integer; {own position initialized by 0..(2^k)-1} 
      {p = pointer base to neighbor, mask initialized by 2^(k-1)}
      p: integer; {2^(k-1), 2^(k-2) ... 1}
     end; 
 
var peff : celladdress; {eff. relative address of global neighbor}
    dneighbor, d: integer;  {neighbor's and own data}
 
#define cell *[0]  {the cell's own state at rel. address 0}

rule begin 
 if ((cell.i and cell.p) = 0 ) then 
  begin 
    {cell id is smaller than bit mask / base pointer}
    {use the neighbor to the right with distance given by base}
    peff := [cell.p]; {use base address without change}
    dneighbor := *peff.i; d := cell.i;  {data access}
    {if neighbor's data is smaller / not in order} 
    if (d > dneighbor) then cell.d := dneighbor; 
  end 
 else 
  begin 
    {cell id is greater than bit mask / base pointer}
    {use the neighbor to the left with distance given by -base}
    peff := [-cell.p];  {address modification}
    dneighbor := *peff.i; d := cell.i; {data access}
    {if neighbor's data is greater / not in order} 
    if (dneighbor > d) then cell.d := dneighbor; 
  end; 

{access-pattern 2^(k-1),...,4,2,1, where n=2^k}
p := p / 2;  
end;
\end{myverbatim}
\normalsize

The general algorithm can be transformed into a \emph{basic} GCA algorithm.
Then the address calculation has to be performed already in the previous generation $t-1$.
Initially the pointers of the left half are $+n/2$, and $-n/2$ for the right half of cells. 
The pointer rule then needs to compute the requested access pattern for the next time-step
using in principle the method used in the former address modification rule. 

Then there arises a principle difference between the general and basic GCA algorithm for this application.
In the general algorithm, the address base is the same for every cell (but time-dependent) and could be supplied 
by a central unit. 
In the basic GCA algorithm, the effective address has to be stored and computed in each cell because it depends on time and index.

\subsubsection{2D XOR with Dynamic Neighbors}
\label{2D XOR with Dynamic Neighbors}

\textbf{CA XOR Rule.} Firstly, for comparison, we want to describe the classic CA 2D XOR rule computing the \textit{mod} 2 sum of their 
four orthogonal neighbors.
Given is a 2D array of cells\\

$D = \textit{array} ~[0~..~n-1, 0~..~n-1] \textit{~of~binary}$, where \textit{binary} $=\{0,1\}$~~.\\

The data state of cell $(x,y)$ is $D[x,y]=d_{(x,y)}$.
The data state of  a neighbor with the relative address $p=(px,py)$ is $d_{(x,y)+(px,py)} = d_{(x+px,y+py)} $.
The nearest NESW neighbors' relative addresses are\\

$p^{North}=(0,-1), ~p^{East}=(1,0), ~p^{South}=(0,1), ~p^{West}=(-1,0)$.\\

The data rule is (written in different notations)\\

$d'_{(x,y)} = d_{(x,y)+p^{North} }  + d_{(x,y)+p^{East}} + d_{(x,y)+p^{South}} + d_{(x,y)+p^{West}} ~~mod~2  $\\

$d' = p^{North}.d  + p^{East}.d + p^{South}.d + p^{West}.d~~mod~2  $\\

$d' = d^{North}  + d^{East} + d^{South} + d^{West}~~mod~2  $~~.\\

\textbf{GCA Rule with dynamic neighbors.}
Now we want to use dynamic neighbors which can change their distance to the center cell.\\

\begin{itemize}
\item
cell state

$q=(d,p)$   

\noindent where $d\in D=\{0,1\}$ is the data part, and  $p$ is the \textit{common address base} (a distance, a relative pointer), initially set to 1.
   
\item 
effective relative addresses to neighbors
\footnote{
\textit{Remark}. The pointer $p$ is used four times in a simple symmetric way, meaning that we use  the \textit{general GCA model} with the \textit{common address base} $p$.
If we would prefer to use the \textit{basic} model, we had to use the cell state
$q=(d,~p^{North}, ~p^{East}, ~p^{South}, ~p^{West})$, 
and we would need four pointer rules, just simple variations of each other.} 

$p^{North}=(0,-p), ~p^{East}=(p,0), ~p^{South}=(0,p), ~p^{West}=(-p,0)$.

\item
neighbors' data states

$d^{North}=p^{North}.d, ~d^{East}=p^{East}.d, ~d^{South}=p^{South}.d, ~d^{West}=p^{West}.d$

\item
data rule

$d'=    d^{North} + d^{East}~ + d^{South}+ d^{West} ~mod~2$    
  
\item pointer rule 1, emulating the classical CA rule

$p'=p=1$     

\item 
pointer rule 2, $p= (1,2,3 \ldots ,n-1)^*$ 



$p'=\begin{cases}
(p+1) \textit{mod } n       &  \textit{ if } (p+1) \textit{mod } n >0\\
1                           &  \textit{ if } (p+1) \textit{mod } n =0                                     \\
\end{cases}$

\item pointer rule 3, 4, 5, 6: $\Delta=2,3,4,5;   ~p=(1,1+\Delta, 1+2\Delta,\ldots )^*$

$p'=\begin{cases}
(p+\Delta) \textit{mod } n       &  \textit{ if } (p+\Delta) \textit{mod } n >0\\
1                                &  \textit{ if } (p+\Delta) \textit{mod } n =0                                     \\
\end{cases}$

\item pointer rule 7, $p= 1,2,4, \ldots 0$

$p'=2p  ~mod ~n$    

\item pointer rule 8, $p= 1,3,9, \ldots 0$

$p'=3p  ~mod ~n$   

\end{itemize}

\begin{figure}[htbp]
	\centering
  		    \includegraphics[width=0.0715\textwidth]{Figures/timescale} 
		    \includegraphics[width=0.1\textwidth]{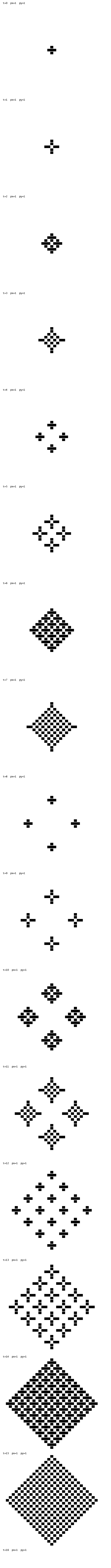}  
		    \includegraphics[width=0.1\textwidth]{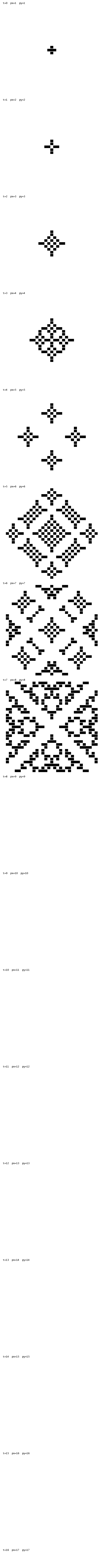} 
    		\includegraphics[width=0.1\textwidth]{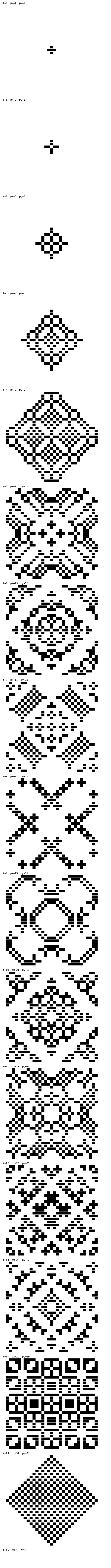} 
        \includegraphics[width=0.1\textwidth]{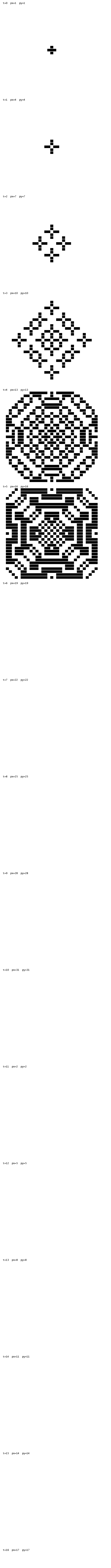} 
        \includegraphics[width=0.1\textwidth]{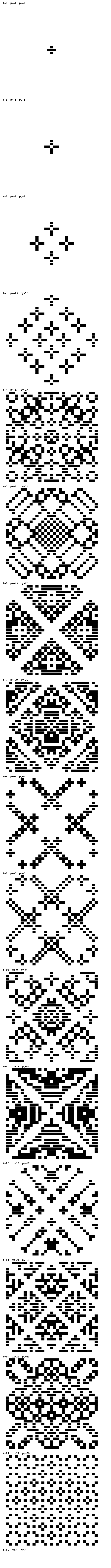} 
        \includegraphics[width=0.1\textwidth]{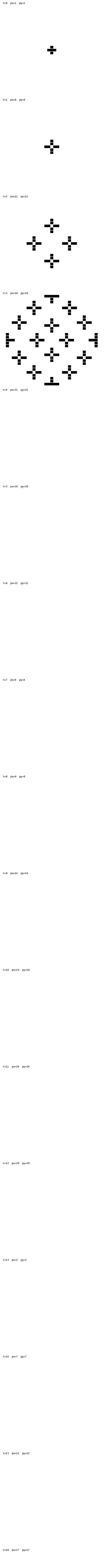}
        \includegraphics[width=0.1\textwidth]{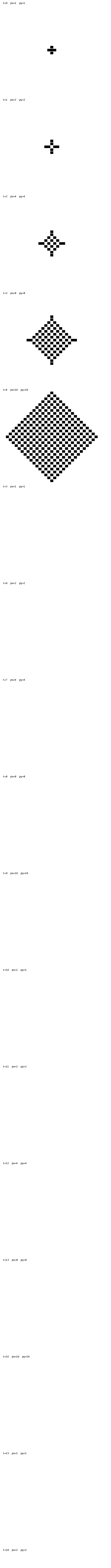}         
        \includegraphics[width=0.1\textwidth]{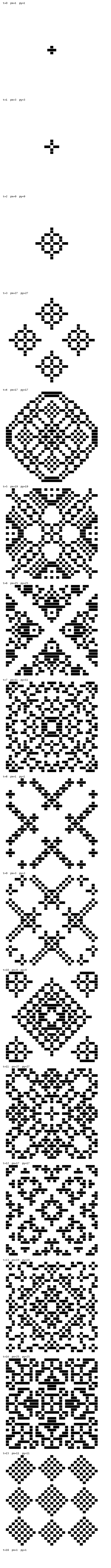}   
                                
          $ ~~~~~~~p'=1  ~~~~p+1  ~~~~p+2 ~~~~p+3  ~~~~~p+4  ~~~p+5 ~~~~~~2p  ~~~~~~~~~3p~~~~$                   
                        
			\caption{
      \footnotesize
      The evolution of the XOR rule with dynamic neighbors.
      ($p'=1$, rule 1) The classical XOR rule with local NESW neighbors. 
      ($p+1$, rule 2) The pointer to the neighbors is incremented by one.
      ($p+\Delta$, rule 3, 4, 5, 6) The pointer is incremented by $\Delta=2,3,4,5$.
      $(2p,$ rule $ 7)(3p,$ rule $8)$ The pointer is multiplied by 2, 3, respectively.
      \normalsize
			}
	\label{p+2}
\end{figure}

\begin{figure}[htbp]
	\centering
  		(a)\includegraphics[width=0.28\textwidth]{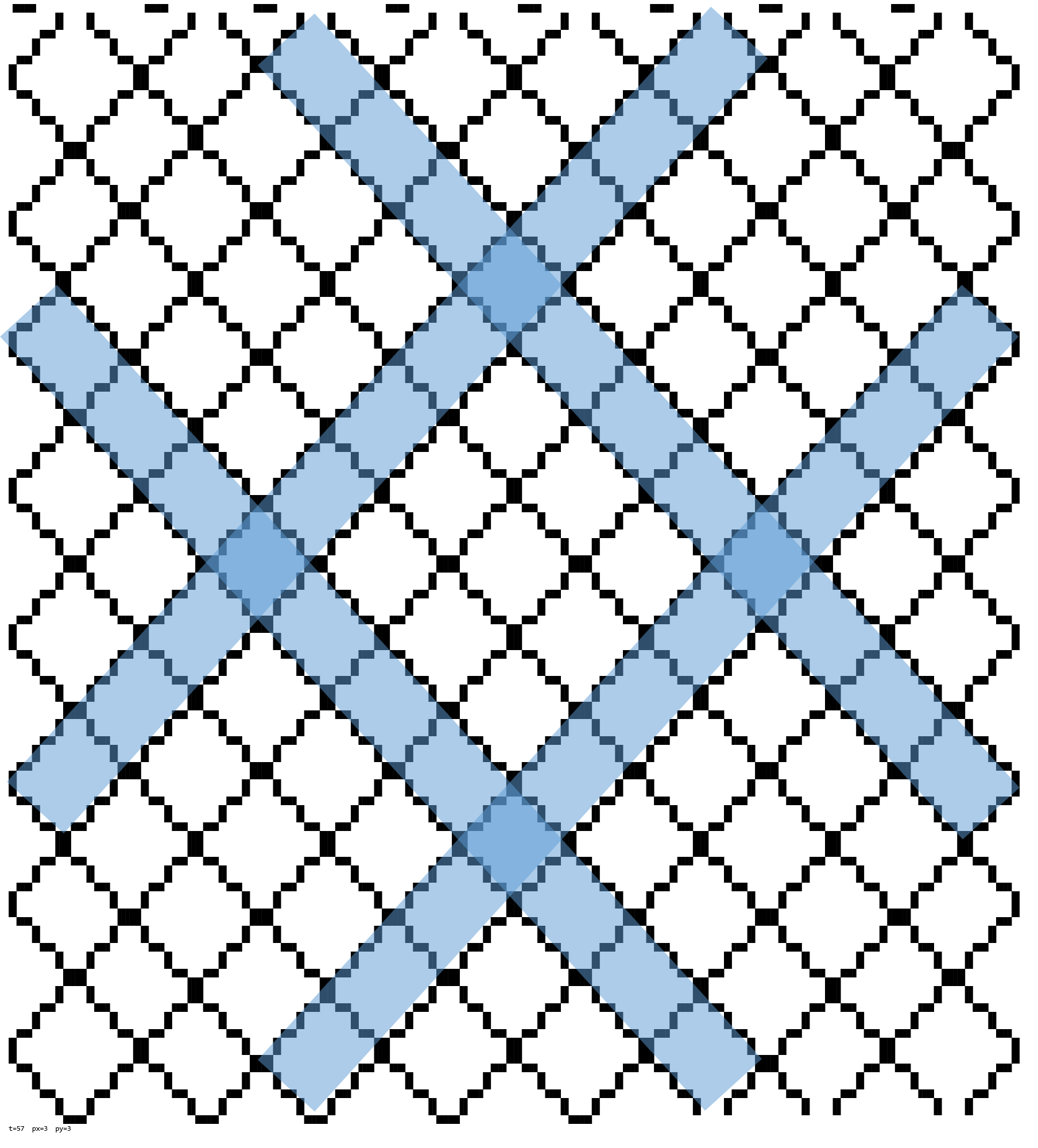}                 
  		(b)\includegraphics[width=0.28\textwidth]{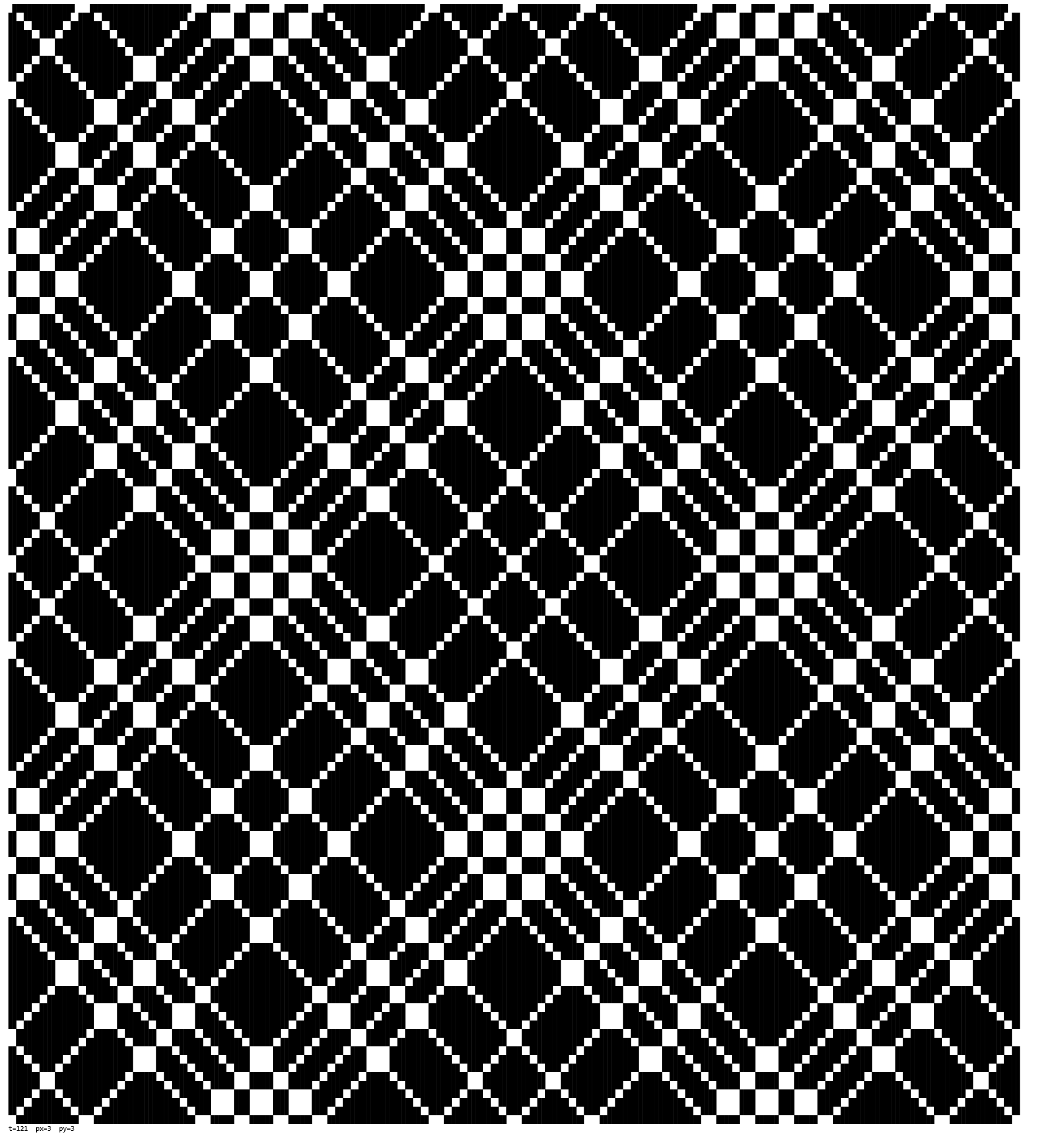}  
  		(c)\includegraphics[width=0.28\textwidth]{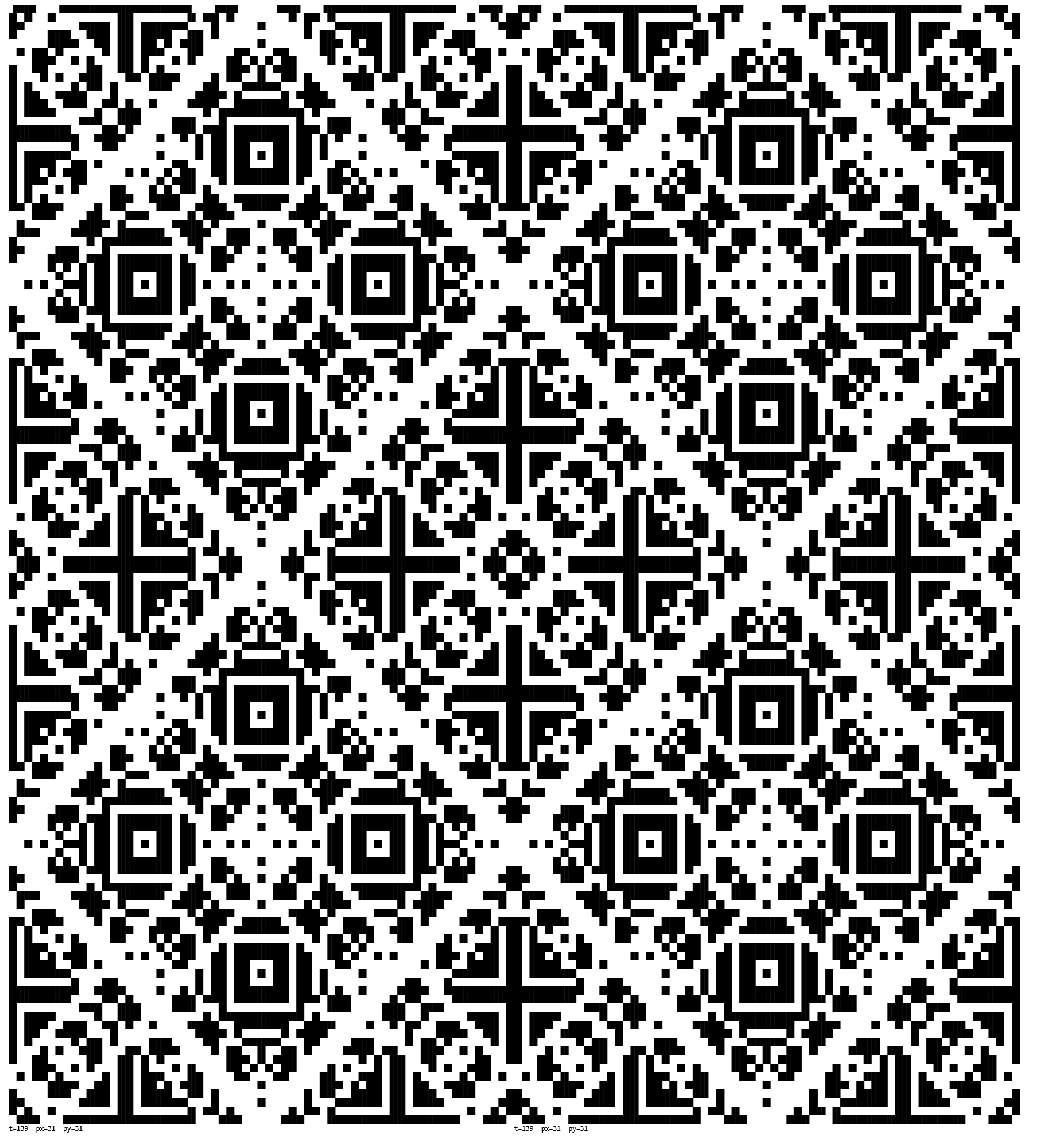}                              
			\caption{
      Some special patterns evolved by XOR rules with four distant orthogonal neighbors.  
      $n=65$. The initial configuration is a cross like in Fig. \ref{p+2}.
      The patterns are of size $130 \times 130$, by doubling the $65\times 65$ pattern in x- and y- direction
      in order to exhibit better the inherent structures. 
(a) Pointer rule $p'=p=3$, at $t=57$. 
(b) Pointer rule $p'=3p$ mod $n$ [if $3p$ mod $n < n$] +1 ~[if $3p$ mod $n =0$], 
at $t=121$; and (c) at t=139.}
	\label{mesh}
\end{figure}

Depending on the actual pointer rule, the evolution of configurations (patterns) differs.
For $n=32$, as depicted in Fig. \ref{p+2}, the  evolution starts initially with a cross
(5 cells with value 1) in the middle. 
For all pointer rules, the evolution converges to a blank (all zero) configuration at a time-step $t\leq16$.
Equal or relative similar pattern can be observed for the different pointer rules,
for example look at the following patterns, for
\blankline

$(t=3,~p'=1) \equiv (t=2,~p'=p+1)$

$(t=7,~p'=1) \equiv (t=3,~p'=2p)$

$(t=8,~p'=p+2) \equiv (t=8,~p'=p+4) \equiv (t=8,~p'=3p) $

$(t=15,~p'=1) \equiv (t=15,~p'=p+2) \equiv (t=4,~p'=2p)$~.\\

We can conclude from these examples that dynamic neighbors (given by the pointer rules)
can produce more complex patterns.
By ``complex pattern'' we mean here a pattern that is more difficult to understand 
(needs more attention for interpretation) because it contains more
different subpatterns compared to the simple CA XOR rule. 
For example, 
the pattern $(t=5,~p'=p= 1)$ contains 1 sub-patterns (a cross),
whereas pattern $(t=5,~p'=p+1)$ contains 4 sub-patterns (plus their rotations).

Three selected patterns are shown in
Fig. \ref{mesh}.
The data rule is the XOR rule with four orthogonal neighbors, as before.
The size of the pattern is   $65 \times 65$.    The initial configuration is a cross like in Fig. \ref{p+2}.
      The patterns shown are of size $130 \times 130$, by doubling the $65 \times 65$ pattern in x- and y- direction
      in order to exhibit better the inherent structures. 
      The used pointer rules are (a) $p'=p=3$, and 
(b, c) $p'=3p \textit{~mod} ~n  ~[\textit{if}  ~3p \textit{~mod} ~n < n]$, or 
   $p'=1                    ~[\textit{if}  ~3p \textit{~mod} ~n = 0$].

\subsubsection{Time-Dependent XOR Algorithms} 
\label{Time Dependent XOR Algorithms}

\begin{figure}[htbp]
	\centering
  		    \includegraphics[width=0.0715\textwidth]{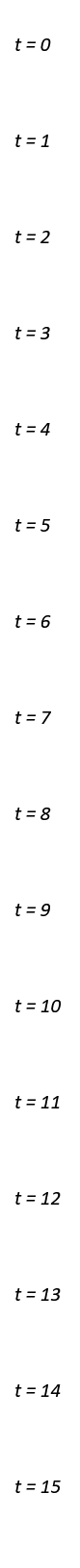} 
		    \includegraphics[width=0.1\textwidth]{Figures/p=1}  
 		    \includegraphics[width=0.1\textwidth]{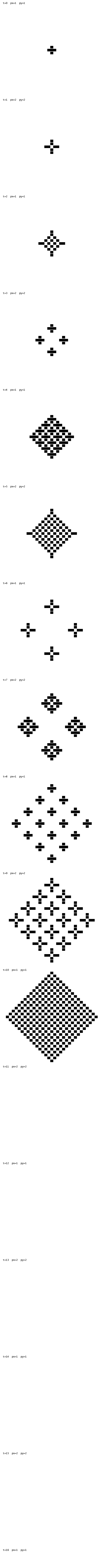} 
    		\includegraphics[width=0.1\textwidth]{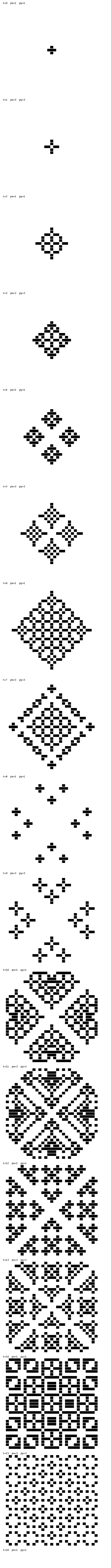} 
        \includegraphics[width=0.1\textwidth]{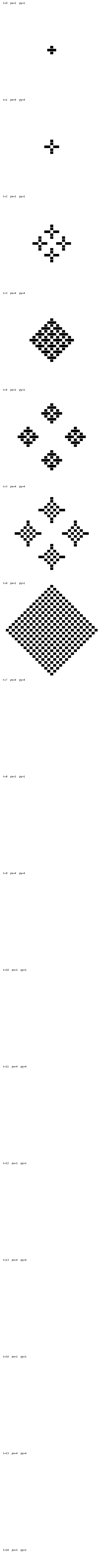} 
        \includegraphics[width=0.1\textwidth]{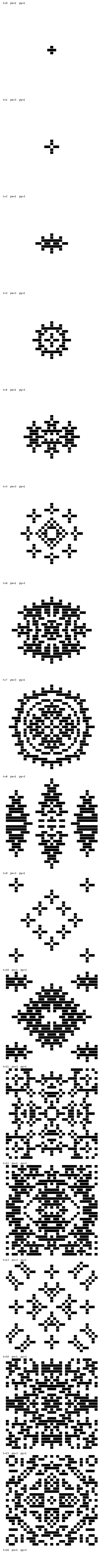} 
        \includegraphics[width=0.1\textwidth]{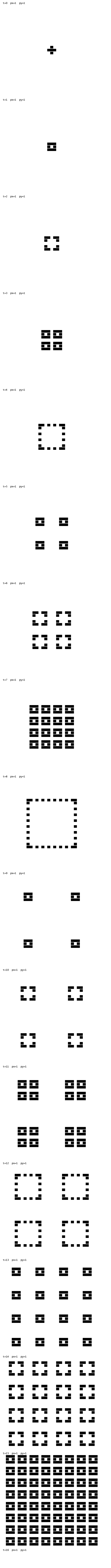}
        \includegraphics[width=0.1\textwidth]{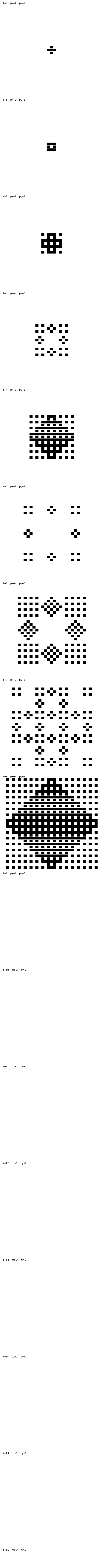}    
        \includegraphics[width=0.1\textwidth]{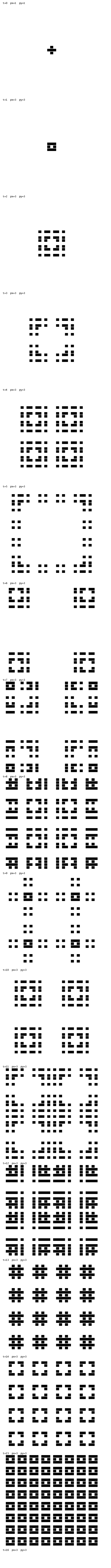}   
                     
          $ ~~~~rule ~A     ~~~~~~B ~~~~~~~~~C ~~~~~~~~D ~~~~~~~~~E ~~~~~~~~~F ~~~~~~~~~G ~~~~~~~~~H$                   
                        
			\caption{
      \footnotesize
      The evolution of the XOR rule with dynamic neighbors, time and space dependent. 
      (A) The classical XOR rule with local NESW neighbors for comparison.
      (B, C, D, E) The pointer alternates in time.
      (B) $p'=(1,2)^*$.
      (C) $p'=(1,3)^*$.
      (D) $p'=(1,4)^*$.
      (E) $px',py'=((1,3),(3,1))^*$. The distance to neighbors is different in x- and y-direction.
      (F, G, H) The pointer is space dependent, different neighbors defined by pointers are used where checkerboard is black or white.
      Either the orthogonal neighbors or the diagonal neighbors are used. 
      (F, G, H) pointers to neighbors are $px=py=1,2,3$.
      \normalsize
			}
	\label{TimeDependentGCA}
\end{figure}

We want to give an example where the pointer rule depends on the time $t$.
Either a central or a local clock can be used. 
In the case of a local clock, the cell's state needs to be extended.
We use the XOR rule of the preceding section. 

\begin{itemize}
\item
cell state. $p$ is the address base, a relative pointer, initially set to 1.

$q=(d,p)$

\item 
effective relative addresses to neighbors

$p^{North}=(0,-p), ~p^{East}=(p,0), ~p^{South}=(0,p), ~p^{West}=(-p,0)$.

\item
data rule

$d'=    d^{North} + d^{East}~ + d^{South}+ d^{West} ~mod~2$    
  
\item pointer rule A, emulating the classical CA rule, for comparison

$p'=p=1$     

\item 
pointer rule B:  $p=1+ t ~mod ~2$, ~$p= (1,2, ~1,2, \ldots)$

pointer rule C:  $p=1+ 2(t ~mod ~2)$, ~$p= (1,3, ~1,3, \ldots)$

pointer rule D:  $p=1+ 3(t ~mod ~2)$, ~$p= (1,4, ~1,4, \ldots)$

\item pointer rule E: $(px,py)=((1,3),(3,1))^*=((1,3),(3,1), (1,3),(3,1), \ldots)$

where

$p^{North}=(0,-py), ~p^{East}=(px,0), ~p^{South}=(0,py), ~p^{West}=(-px,0)$.

$px=1+ 2(t ~mod ~2)$, ~$py=1+ 2((t+1) ~mod ~2)$.

\end{itemize}

The evolution of these time dependent XOR rules are shown in Fig.	\ref{TimeDependentGCA} (B, C, D, E).
Rule E exhibits  more irregular patterns because the distance to the neighbors is different in $x$- and $y$-direction,
and alternating.

\subsubsection{Space-Dependent XOR Algorithms} 
\label{Space Dependent XOR Algorithms}

We want to give an example where the pointer rule depends on the space
given by the two-dimensional cell index $(x,y)$.
We use the same XOR rule and definitions as in the preceding section.

\begin{itemize}

\item pointer rules F, G, H

A checkerboard is considered, where white 0-cells are defined by the condition $[(x+y) ~mod ~2 = 0]$,
and black 1-cells  by the condition \\ $[(x+y) ~mod ~2 = 1]$.

The pointer rules for \textit{white} cells defines their orthogonal neighbors:

$p^{North}=(0,-py), ~p^{East}=(px,0), ~p^{South}=(0,py), ~p^{West}=(-px,0)$.

The pointer rules for \textit{black} cells defines their diagonal neighbors:

$p^{North}=(px,-py), ~p^{East}=(px,py), ~p^{South}=(-px,py), \\~p^{West}=(-px,-py)$.

with $px=py=p=1,2,3$ for rule F, G, H.

Note that for \textit{black} cells,
$p^{North}$ addresses \textit{NorthEast}, 
$p^{East}$ addresses \textit{SouthEast},
$p^{South}$ addresses \textit{SouthWest}, and
$p^{West}$ addresses \textit{NorthWest}.

\end{itemize}

The space-dependent rules F, G, H 
(Fig.	\ref{TimeDependentGCA})
show different patterns and  sub-patterns compared to the time dependent rules B -- E.
These examples show that different and more complex patterns can be generated if the neighbors are  changed
in time or space by an appropriate pointer rule.

\subsubsection{1D XOR Rule with Dynamic Neighbors} 
\label{1D XOR Rule with Dynamic Neighbors} 

Two compilable PASCAL program are given in Section \ref{Appendix 0} (Appendix 0) that simulate 
the 1D XOR rule with two dynamic neighbors.
The basic model is used in Sect. \ref{Appendix 0-Basic},
and the general model with a common address base is used in Sect. \ref{Appendix 0-General}.

\subsection{Plain Model Example} 
\label{Plain Model Example} 

\begin{figure}[htbp]
	\centering
        \includegraphics[width=0.15\textwidth]{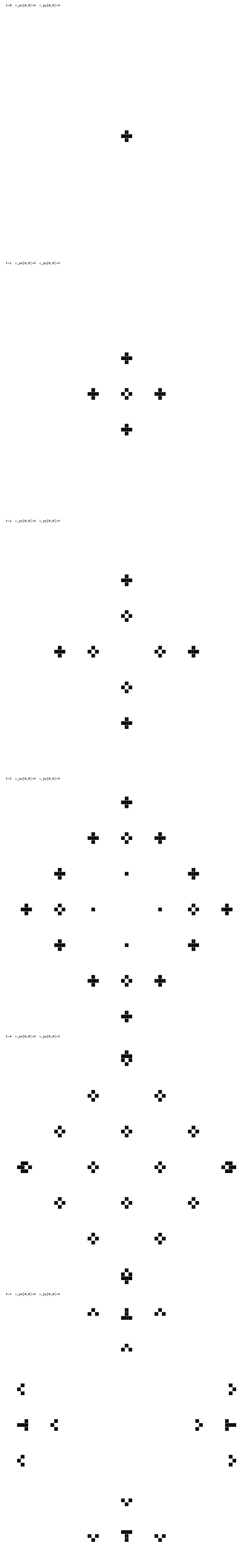}
  		  \includegraphics[width=0.15\textwidth]{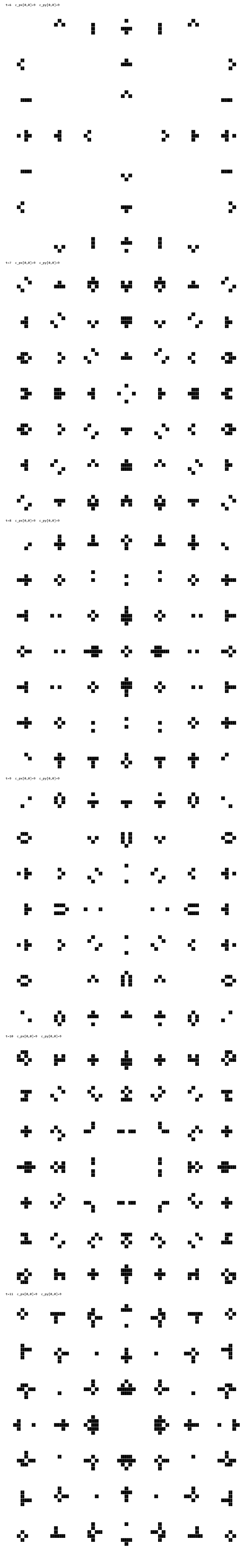} 
  		  \includegraphics[width=0.15\textwidth]{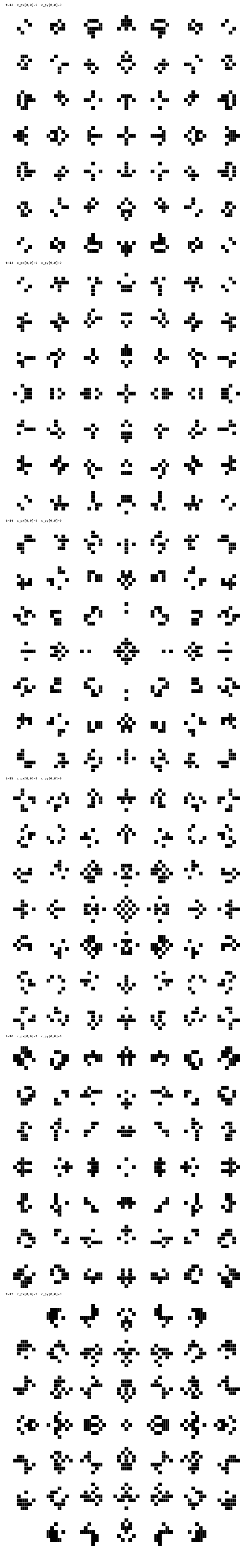} 
  		  \includegraphics[width=0.15\textwidth]{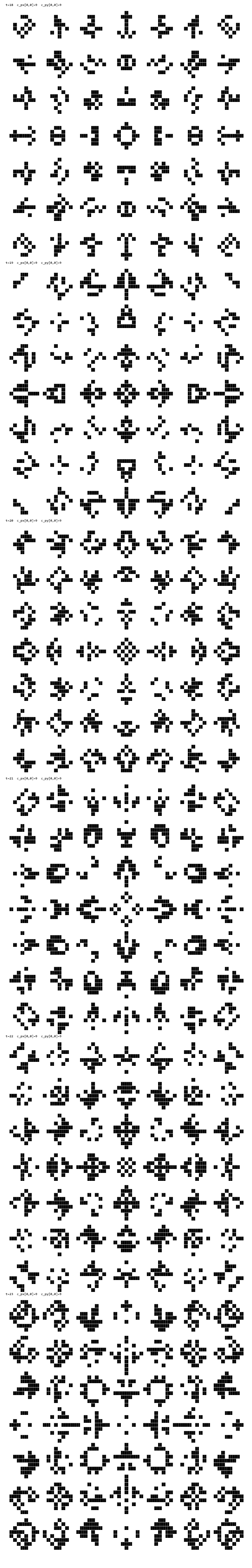}                 
  		  \includegraphics[width=0.15\textwidth]{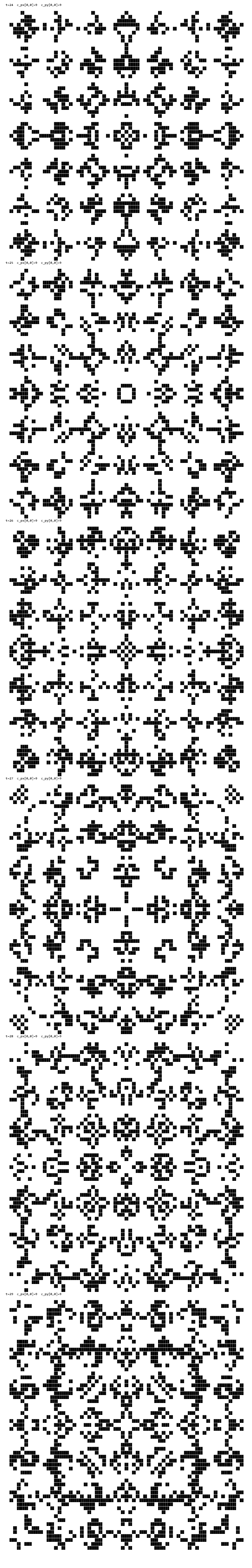}
  		  \includegraphics[width=0.15\textwidth]{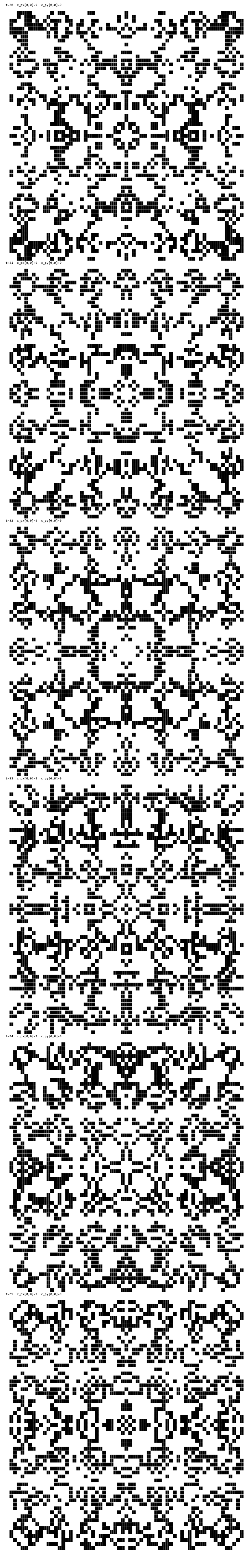}                
                                
          $t=0-5     ~~~~~6-11 ~~~~~~~12-17 ~~~~~~18-23 ~~~~~~24-29 ~~~~~~30-35$                   
                        
			\caption{
Plain GCA Model, XOR rule with data dependent pointers. 
The pointer value is $p=9$ if the cell's state is 0 (white), and is $p=1$  if it  is 1 (black).
For $t=7-24$ we observe small  $49$ sub-patterns placed regularly at $7 \times 7$ distinct positions.
The sub-patterns are changing and slowly increasing
until they merge. 
			}
	\label{datadependent}
\end{figure}

\begin{figure}[htbp]
	\centering
        \includegraphics[width=0.15\textwidth]{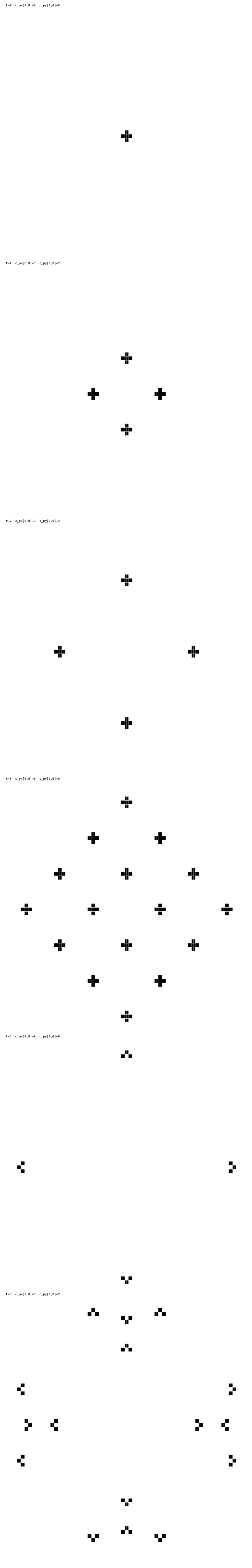}
  		  \includegraphics[width=0.15\textwidth]{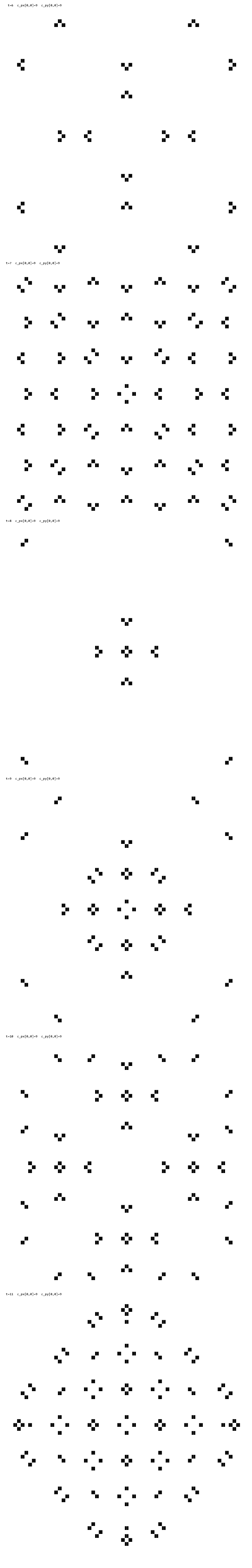} 
  		  \includegraphics[width=0.15\textwidth]{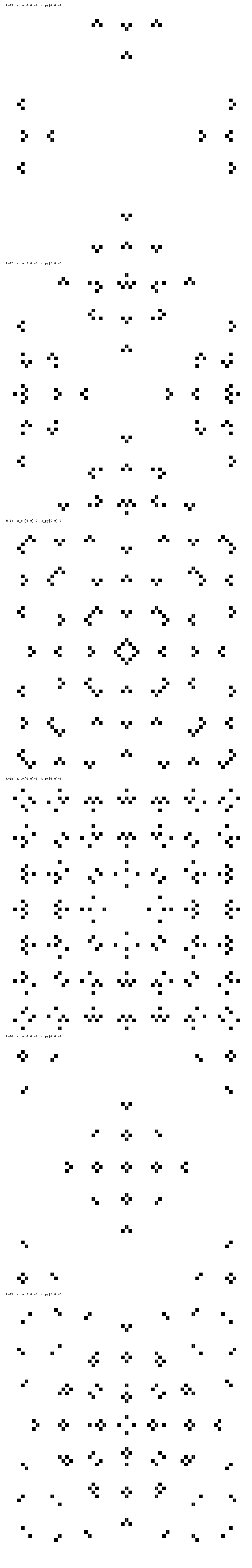} 
  		  \includegraphics[width=0.15\textwidth]{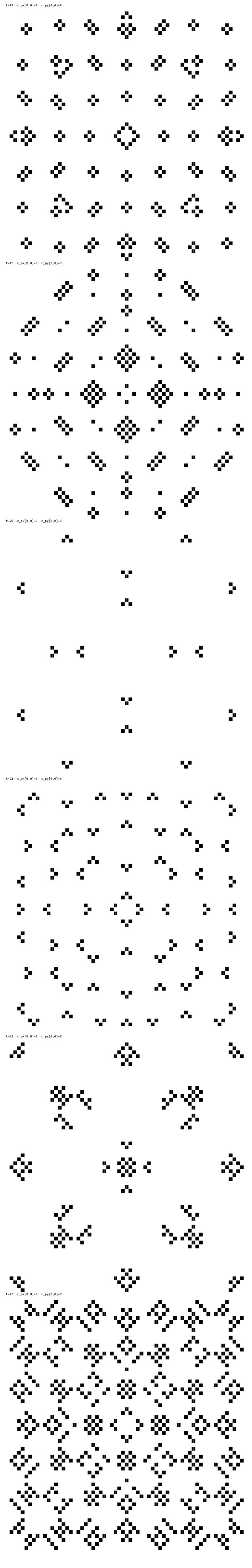}                 
  		  \includegraphics[width=0.15\textwidth]{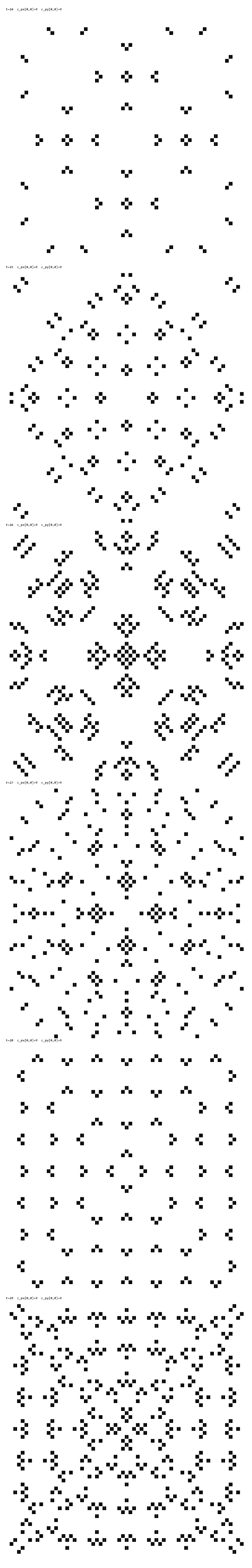}
  		  \includegraphics[width=0.15\textwidth]{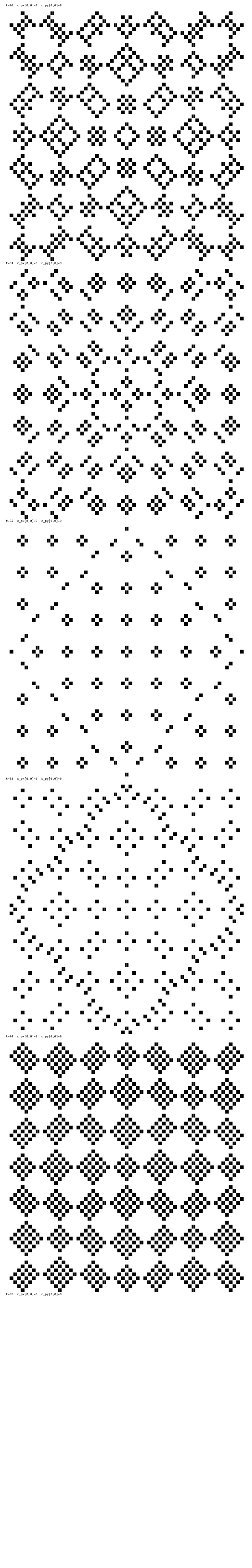}                
                                
          $t=0-5     ~~~~~6-11 ~~~~~~~12-17 ~~~~~~18-23 ~~~~~~24-29 ~~~~~~30-35$                   
                        
			\caption{
Plain GCA Model, XOR rule with data dependent pointers. 
The pointer value is $p=9$ if the cell's state is 0 (white), and is $p=3$  if the cell's state is 1 (black).
For $t\geq35$ all cells stay black. 
			}
	\label{datadependent2}
\end{figure}

In the plain GCA model, the cell's state $q$ is not structured into a data and pointer part. 
The pointer(s) are computed from the state. In our example, we use again the XOR rule with remote
NESW neighbors, and the cell's state is binary. The distance to the neighbors is 
directly related to the cell's state, here it is defined as

%

\[
  p = (1-q)A + q B = \left\{\begin{array}{ll} 
	   
	       A  & \textbf{if} ~~  q=0   \\
         B  & \textbf{if} ~~  q=1,  ~~\textit{where} ~1 \leq A,B \leq n/2. 

			\end{array}
	      \right. .
\]	

The effective relative addresses to the distant NESW neighbors are\\

$p^{North}=(0,-p), ~p^{East}=(p,0), ~p^{South}=(0,p), ~p^{West}=(-p,0)$.\\

Fig. \ref{datadependent} and Fig. \ref{datadependent2}
show the evolution of this rule with data dependent pointers. 
Fig. \ref{datadependent}:
The pointer value is $p=9=A$ if the cell's state is 0 (white), and is $p=1=B$  if the cell's state is 1 (black).
For $t=7-24$ we observe small  $49$ sub-patterns placed regularly at $7 \times 7$ distinct positions.
The sub-patterns are changing and slowly increasing until they merge. 
The density of black cells is roughly increasing during the evolution, but the pattern does not converge into 
a full black configuration.  

Fig. \ref{datadependent2}:
The pointer value is $p=9=A$ if the cell's state is 0 (white), and is $p=3=B$  if the cell's state is 1 (black).
For $t \geq 35$ all cells remain white. 
Note that the interesting pattern for $t=34$ is not a true checkerboard, the white areas are squares of two different sizes, or rectangles.  

\newpage
\subsection{A New Application: Synchronous Firing} 
\label{SynchronousFiring}

Our problem is similar to the  \emph{Firing Squad Synchronization Problem} (FSSP) that is a well studied classical Cellular Automata Problem
\cite{Moore1968, Mazoyer1987, Umeo2009, WikipediaFSSP2022}.
Initially at time $t=0$ all cells in a line are ``quiescent", the whole system is quiescent.
Then at $t=1$, a dedicated cell (the \emph{general}) becomes active by a special external or internal event. 
The goal is to design a set of states and a local CA rule such that, no matter how long the line of cells is, 
there exists a time $t_{fire}$ such that every cell changes into the firing state at that time simultaneously.

Here we are modifying the problem because we aim at GCA modeling, allowing pointer manipulation and global access. 
In order to avoid confusion, we call our problem ``\emph{Synchronous Firing}" (SF).
Applying the GCA model, the problem becomes easier to solve, although not necessarily simple. 
We expect a shorter synchronization time.

The cell's state is $q=(d,p)$, where $d$ is the data state and $p$ the pointer.
We can easily find a trivial solution. 
The cells ($i=0, 1, \ldots, n-1$) are arranged in a ring, all of them are quiescent \emph{soldiers} (state S) at time $t=0$.
All cells contain a pointer pointing to cell $i=0$.
At $t=1$ a \emph{general} (state G) is installed at position $i=0$.
Now all cells read the state of their global neighbor which is G for all of them. 
Then, at $t=2$ all cells change into the \emph{Firing state} (F).
Although trivial, this solution is somehow realistic. The soldiers observe the general, and when he gives a signal, all of them fire at the next time-step, for instance after one second.

This solution of the problem is not general enough because the soldiers must know the position of the general in advance. 
We aim at more general/non-trivial solutions.

\subsubsection{Synchronous Firing Using a Wave}
\label{Synchronous Firing Using a Wave}

As before, all cells are arranged in a ring and initially they are in the quiescent state S (Soldier).
Then, by an external force, \emph{any one of the soldiers changes its state into G} (General).
Now we want to find a solution where all cells fire simultaneously, independently of the general's position.
Furthermore we want to allow only one pointer per cell (one-armed GCA) and the \emph{initial values of the relative pointers to be the same}.

In the solution we use the data states S, G, and F. 
Initially all pointers are set to the value -1, meaning that every cell points to its left neighbor in the ring.

\begin{figure}[htb]
	\centering
		\includegraphics[width=8cm]{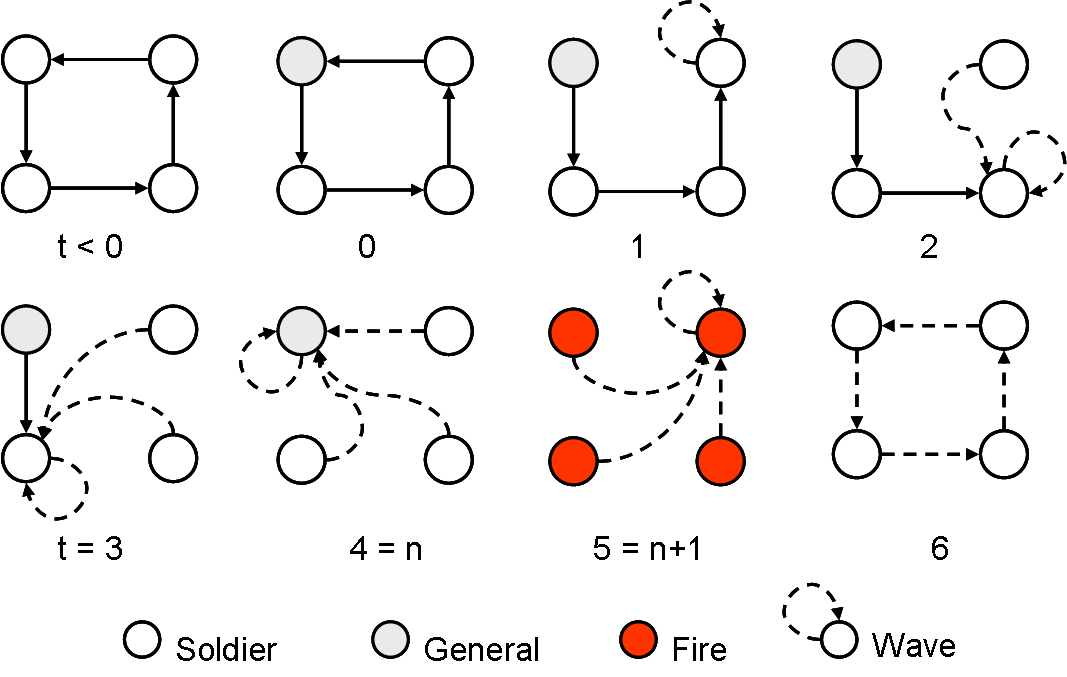} 
			\caption{Synchronous Firing using pointers. 
			At $t<0$ the system is quiescent. 
			Then, at  $t=0$, one of the soldiers becomes a general and will produce a self-loop at $t=1$. 
			From  $t=1$ to  $t=5$ a wave propagates clockwise.
			When it reaches the general, the cells know that they have to fire at the next time-step. 
      Solid arrows depict pointers, dotted arrows depict pointers that were modified.}
	\label{firingsquad}
\end{figure}

\begin{figure}[h!]
	\centering
\textsc{GCA-ALGORITHM 1}

Synchronous Firing using a Wave


\begin{tabular}{l| c | llr}
\toprule
\addlinespace[2pt]

\cmidrule{2-2} 

x.1
& ~$t<0~$                    
& $~~p_i=-1$                          
& $~~d_i=S$    
&initial

\\
\cmidrule{2-2} 

x.2
& ~$t=0~$                    
&                                     
& $~~d_k = G$   
&$\exists k\in I$

\\[0pt]
\cmidrule{2-2} 

A.1
& ~$t>0~$                    
& $~~p_i \leftarrow g(p_i,d_i,p^*_i,d^*_i)  $   
& $~~d_i \leftarrow f(p_i,d_i,d^*_i)$
& $\forall{i}$

\\
\cmidrule{2-2} 

y.1
& ~$t=n+1~$                    
& $~~p_i = k-i$                    
& $~~d_i = F$
& $\forall{i}$

\\
\cmidrule{2-2} 

y.2
& ~$t=n+2~$                    
& $~~p_i = -1$                    
& $~~d_i = S$
& $\forall{i}$

\\
\cmidrule{2-2} 
\bottomrule
\end{tabular}		
			\caption{
			At $t<0$ the system is quiescent. 
			Then at  $t=0$ a general is introduced. 
			From  $t=1$ to  $t=n+1$ a wave propagates clockwise.
			When it reaches the general each cell knows that it has to fire at the next time-step $t=n+1$.}
	\label{algorithm-syncfiring-wave}
\end{figure}

The GCA algorithm consists of a pointer rule and a data rule. 
The following abbreviations are used: 

$p= p_i, ~d= d_i, ~p^*=p^*_i=P_{rel}[abs(p_i)], ~d^*=d^*_i=D[abs(p_i)]$. 

\blankline
The pointer rule:\footnote{$~a\oplus b = a + b ~\textit{mod} ~n$}
\[
  p' = g= \left\{\begin{array}{lll} 
	      p \oplus 1  & \textbf{if} ~(d=S,G) ~\textbf{and}~ ((d^*=G) ~\textbf{or}~ (p^*\neq -1)) \hspace{2mm}  &(1a) \\

       -1           &\textbf{if} ~(d=F)  \hspace{2mm}                                                        &(1b) \\
			\end{array}
	      \right. .
\]

The data rule: 
\[
  d' = f= \left\{\begin{array}{lll} 
	    
	 F  \hspace{3mm}  &\textbf{if} ~(d^*=G) ~\textbf{and}~  ((p\neq -1) \textbf{~or~}  (p^*=0)) \hspace{2mm} &(2a) \\
			
   S                & \textbf{if} ~(d=F)  \hspace{2mm}                                                    &(2b) \\
      
			\end{array}
	      \right. .
\]

The algorithm works as follows, as shown for $n=4$ in Fig. \ref{firingsquad}:

\begin{itemize}
\item $t < 0$:
Initially the configuration is quiescent. \\
$\forall i \in I: p_i=-1, d_i=S$.

\item $t=0$:
A general is assigned.\\
$\exists! i\in I: d_i=G$

\item $t=1$:
A wave is starting. 
The first soldier in the ring whose left neighbor is the general forms a self-loop which marks (the head of) the wave.\\
($p=p+1 \textit{~if~} (d=S) \textit{~and~} (d^*=G)$, Rule 1a)

\item $t=2 ,\ldots, n+1$:
The wave moves clockwise. 
Cells that recognize the wave follow it.
The cell's pointer is incremented if the neighbor’s pointer $p^*$ does not point to the left anymore.\\
($p=p+1 \textit{~if~}  (p^*\neq -1)$, Rule 1a)

\item $t=n$:
The wave has reached the general and all cells point to it ($d^*=G$). 
This situation signals that all cells shall fire. (Rule 2a). 
Then the General and the Soldiers (except one) fire if 
their pointers are not equal to -1 (the initial condition).
The Soldier to the right of the General is prevented to fire by the condition $p\ne -1$
because the condition  $p = -1$   is  true at the beginning and in the pre-firing state.
Therefore the excluded Soldier needs to be included by an additional condition $p^*=0$
that detects the self-loop of the General.

\item $t_{fire}=n+1$:
All cells are in the firing state. The whole system can be reset into the quiescent state (Rule 2b), 
or another algorithm could be started, for instance repeating the same algorithm with a general at another position.

\end{itemize}

\noindent We can describe this algorithm in a special tabular notation as shown in Fig. 
\ref{algorithm-syncfiring-wave}.
The first column shows a numbering scheme.
Preconditions and inputs before starting the algorithm are marked by ``\emph{x.i}''.
The algorithmic actions are marked by ``\emph{A.i}''.
Predicates and outputs are marked by ``\emph{y.i}'', they are no actions. 
They show intermediate or final results of algorithmic actions and serve also for a better understanding of the algorithm.
They are not necessary to describe the algorithm, they are optional and may also be true at another time.
In the second column a temporal precondition is given. 
We assume that the time proceeds stepwise but we do not give an implementation for that. 
There may be a time counter in every cell, or there may be a central time-counter that can be accessed by any cell.
The third column specifies the change of the pointer according to the pointer rule $g$.
The fourth column specifies the change of the data according to the data rule $f$.
The fifth column is reserved for comments or additional assertions. 

The classical CA solution of Mazoyer  
\cite{Mazoyer1987}
with local neighborhood needs $t_{fire}=2n-1$. 
So the GCA solution is only nearly twice as fast. 
The purpose was not find the fastest GCA algorithm but to show how a GCA algorithm can be described and works in principle.

\subsubsection{Synchronous Firing with Spaces}
\label{Synchronous Firing with Spaces}

Our next solution is based on the former algorithm using a wave as described in Sect. \ref{Synchronous Firing Using a Wave}.
Now the number of cells shall be larger than the number of active cells (General, Soldiers),
empty (inactive) cells (spaces) can be placed at arbitrary positions between them. 
So an active ring of cells is embedded into a larger ring of cells. 
Our algorithm will have the following features:

\begin{itemize}
\item
Any number of inactive cells can be placed between active cells.
\item
The ordering scheme used for connecting the active cells by pointers needs not to follow the indexing scheme. 
\item
Several rings of active cells can be embedded in the space and processed in parallel. 
\end{itemize}     

The algorithm uses two pointers per cell, $p^1$ and  $p^2$.
Initially active cells are connected in one or more rings (circular double linked lists).
Pointer $p^2$ remains constant, thereby a loop exist always in one direction. 
Pointer $p^1$ is variable and is used to mark the wave.
Inactive (constant) cells are marked by self-loops, their pointers are set to zero ($p^1=0$ and  $p^2=0$).
(Another way to code inactive cells were to use an extra data state.)

We associate the index range with a horizontal line of cells, where
cell index 0 corresponds to the leftmost position and  index $n-1$ to the rightmost position.
In our later example and for explanation we connect initially a cell to its left neighbor by $p^1$ and to its right neighbor by $p^2$.
(The connection scheme can be arbitrarily as long as the cells are connected in a ring.)

\blankline
The pointer rule for $p^2$ is  $p^2{'}= p^2$ (no change after initialization).

\blankline
The pointer rule for $p^1$ is

\[
  p^1{'} = g= \left\{\begin{array}{lll} 

			  p^1   & \textbf{if~not~}\emph{Active} &(3a) \\

              &~~ \textbf{otherwise} &\\

				0     &~~ \textbf{if~}(p^1.d=G) ~\textbf{and}~(p^1 \neq 0) ~\textbf{and}~ (p^1.p^1 \neq 0)  &(3b) \\

			  p^1 \oplus p^1.p^2\hspace{1mm}  &~~ \textbf{if~}((p^1 = 0) ~\textbf{or}~ (p^1.p^1 =0)) &(3c) \\

			\end{array}
	      \right.
\]

The data rule is 
\[
  d' = f= \left\{\begin{array}{lll} 
	      d  \hspace{2mm}  & \textbf{if~not~}\emph{Active} &(4a) \\
				
				              &~~ \textbf{otherwise} &\\

	      F            &\hspace{3mm}\textbf{if} ~(p^1.d=G) ~\textbf{and}~  ((p^1\neq -p1.p2) \textbf{~or~}  (p^1.p^1=0)) \hspace{2mm}  &(4b) \\

			\end{array}
	      \right. 
\]

The algorithm works as follows.

\begin{itemize}
\item $t < 0$: 
\emph{Initialization}.
All data states are set to $d_i=S$.
Inactive cells are represented by ($p^1=0$ and  $p^2=0$).
Rings consisting of active cells to be synchronized are formed. 
A cell may belong to one ring only, i.e. rings are mutually exclusive. 
Neighboring cells $c_j$,  $c_i$, and $c_k$ of a ring are connected by pointers.
Cell $c_i$ points to the ``left'' cell $c_j$ by $p^1$ and to the ``right'' cell $c_k$ by $p^2$.
The conditions 
$c_i.p^1=-c_j.p^2$ and $c_i.p^2=-c_k.p^1$ are true.

\item $t=0$:
A General is assigned in each ring by setting  $d_{i(k)}=G$,  
where $i(k)$ is the index of the General in the ring $k$.

\item $t=1$:
A wave is starting in each ring. 
The soldier in each ring whose $p^1$ neighbor is the General forms a self-loop ($p^{1}=0$) which marks the wave (Rule 3b).

\item $t>1$: (Rule 3c).
The wave move along in the direction of $p^2$. 
The pointer $p^1$ is set to $p^2$ (the next position of the wave) when the cell itself is the head of the wave 
(self-loop $p^1 = 0$) because then $p^1 \oplus p^1.p^2 = p^2$.
The pointer $p^1$ follows the wave through $p^1 \oplus p^1.p^2$ when the $p^1$ neighbor is the head of the wave
(self-loop $p^1.p^1 =0$).

\item $t(k)=L(k)$:
The wave has reached the General of a ring $k$, where $t=L(k)$ is the length of the ring $k$.
This situation signals that all cells shall fire (Rule 4a). 
All cells of the ring $k$ point to the General ($p^1.d=G$), this is the precondition to fire. 
The Soldiers (except one) fire only if 
their pointers are not equal to the initial condition
$p^1\neq -p^1.p^2$, which is an indirect self-loop of length 2.
But a self-loop of length 2 is true for the Soldier \texttt{S} next to the General \texttt{G} 
via  $p^1$ at the beginning and in the pre firing state.
($\texttt{G} \rightarrow {p^2} \rightarrow \texttt{S} ~/~ \texttt{G} \leftarrow {p^1} \leftarrow \texttt{S}$).
So by adding the condition $p^1.p^1=0$ (\texttt{S} points via $p^1$ to \texttt{G} showing a self-loop), \texttt{S} will also fire.  
The General is allowed to fire when the self-loop of length 2 ($p^1=-p^1.p^2$) has changed into a self-loop ($p^1=0$), and then the condition $p^1\neq -p^1.p^2$ holds.

\item $t_{fire}(k)=L(k)+1$:
All cells of ring $k$ are in the firing state. 

\end{itemize}

\textbf{Example.}
The number of cells is $n=9$, index $i\in\{0, 1, \ldots, 8\}$.
Two rings with bidirectional links to their neighbors are embedded in the array.
Ring \textbf{A} is the connection of the cells $(2, 4, 6)$.
Ring \textbf{B} is the connection of the cells $(1,3,5,7)$.
Cells 0 and 9 are passive cells that can be seen as the borders of the array.
The $p^1$ pointers (relative values) of \textbf{A} are $(-5,-2,-2)$.
The value -5 is the (cyclic) distance from cell 2 to 6.
The $p^2$ pointers of \textbf{A} are $(2, 2, 5)$.
The value 5 is the (cyclic) distance from cell 6 to 2.
--
The $p^1$ pointers of \textbf{B} are $(-3,-2,-2,-2)$.
The $p^2$ pointers of \textbf{B} are $(2, 2, 2, 3)$.

\begin{figure}[htbp]
	\centering
		\includegraphics[width=12cm]{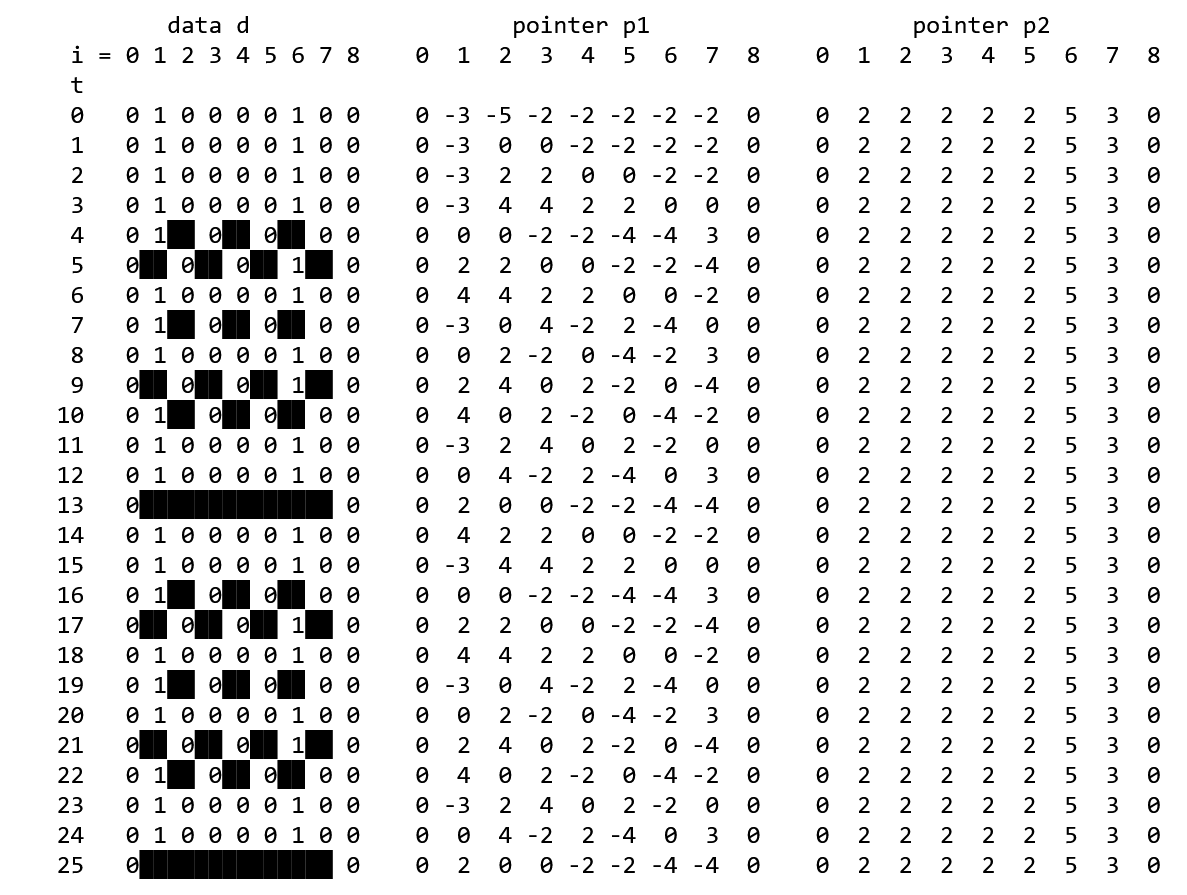} 
			\caption{Synchronous Firing of two rings embedded in an 1D array.
			The cyclic connected cells  $(2, 4, 6)$ form ring \textbf{A}, and the cells  $(1,3,5,7)$ form ring \textbf{B}.
			\textbf{A}t $t=0$ we can observe the connections by the pointers and one General for ring \textbf{B} at $i=1$, and another at $i=6$ for ring \textbf{A}. 
			Then two waves are starting, one in ring \textbf{A} and one in ring \textbf{B}.
			Ring \textbf{A} fires at $t=4,7,10,13, \ldots$ (firing states are represented by black squares),
			and ring \textbf{B} fires at $t=5,9,13,17, \ldots$.
			All cells except the border cells fire at  $t=13,25, \ldots$.	
			}
	\label{syncp1p2}
\end{figure}

\subsubsection{Synchronous Firing with Pointer Jumping}
\label{Synchronous Firing with Pointer Jumping}

\textbf{Solution 1.}
The question is whether the synchronization time can be reduced by using the \emph{pointer jumping} (or \emph{pointer doubling}) technique.
This technique is well-known from PRAM (parallel random access machine) algorithms. 
It means for the GCA model that an \emph{indirect} neighbor, a neighbor of a neighbor, becomes a \emph{direct} neighbor. 
This can be accomplished by pointer substitution ($p\leftarrow p^*$)
 in the case of absolute pointers, 
or pointer addition ($p\leftarrow p+p^*$) 
in the case of relative pointers (or pointer vectors where the cells are identified by their
coordinates in the $n$-dimensional space),  
or simply by pointer doubling ($p\leftarrow 2p$)
 in the case of relative pointers when the cells are ordered by a consecutive 1D array index. 
For instance, this technique allows us to find the maximum of data items stored in a line of cells in logarithmic time.

A first algorithm is given in Fig. \ref{algorithm-syncfiring-pointerjumping1}.
Initially at $t=0$ we assume that there is one general among all remaining soldiers. 
Then the following rules are applied.
Pointer Rule:

\[
p' = g(p,p^*,n)= 
\left.
\begin{array}{ll} 

p \oplus p^* = (p+p^*) \textit{~mod} ~n       & ~~~~(5) \\

\end{array}
\right. 
\]

Alternatively the rule 
$g(p,n) = 2p \textit{~mod~} n$ could be used because  $p=p^*$ holds here.
Data Rule:

\[
  d' = f(p,d,d^*)= \left\{
	     \begin{array}{lll} 
			
				d^*  
				&\hspace{6mm} \textbf{if} ~(p \neq 0) ~\textbf{and}~  (d < d^*)  
				&\hspace{2mm}  \hfill (6a) \\	     
			
	      2  
				&\hspace{6mm} \textbf{if} ~(p=0)~\textbf{and}~  (d=1) 
				&\hspace{2mm}  \hfill (6b) \\
				
				d
				&\hspace{6mm} \textbf{otherwise} 
				&\hspace{2mm}  \hfill (6c) \\
   
			\end{array}
	      \right. .
\]

The algorithm in tabular form is shown in Fig. \ref{algorithm-syncfiring-pointerjumping1}.
The time evolution of the pointers and the data are  shown in the following for $n=8$:

\begin{figure}[htbp]
	\centering
\textsc{GCA-ALGORITHM 2}

Synchronous Firing with Pointer Jumping
\begin{tabular}{l| c | llr}
\toprule
\addlinespace[2pt]

\cmidrule{2-2} 

x.1
& ~$t<0~$                    
& $~~p_i=1$                          
& $~~d_i=S=0$    
& $\forall i \in I$

\\
\cmidrule{2-2} 

x.2
& ~$t=0~$                    
& $~~p_i=1$                                    
& $~~d_k = G=1$   
&$\exists ! k\in I$

\\[0pt]
\cmidrule{2-2} 

A.1
& ~$t>0~$                    
& $~~p_i \leftarrow g(p_i,d_i,p^*_i,d^*_i)  $   
& $~~d_i \leftarrow f(p_i,d_i,d^*_i)$
& $\forall{i}$

\\
\cmidrule{2-2} 

y.1
& ~$t=log_2 ~n~$                    
& $~~p_i = 0$                    
& $~~d_i = G=1$
& $\forall{i}$

\\
\cmidrule{2-2} 

y.2
& ~$t \geq 1+ log_2 ~n~$                    
& $~~p_i = 0$                    
& $~~d_i = F=2$
& $\forall{i}$

\\
\cmidrule{2-2} 
\bottomrule
\end{tabular}	

			\caption{
			The system starts working at $t=0$ when one of the soldiers is assigned to be a general.
			The information $G=1$ is exponentially distributed among the neighbors by pointer jumping.
			At $t=log_2 ~n$ all the pointers become 0, and the data is  \emph{G} everywhere which is the signal to fire. 
      At $t=log_2 ~n + 1$ alls cells change into the firing state $d=2$.
			}
	\label{algorithm-syncfiring-pointerjumping1}
\end{figure}

\small 
\begin{verbatim}
     Pointer           Data
 i=  0 1 2 3 4 5 6 7   0 1 2 3 4 5 6 7
 t
 0  >1 1 1 1 1 1 1 1  >1 0 0 0 0 0 0 0
 1   2 2 2 2 2 2 2 2   1 0 0 0 0 0 0 1
 2   4 4 4 4 4 4 4 4   1 0 0 0 0 1 1 1
 3  >0 0 0 0 0 0 0 0  >1 1 1 1 1 1 1 1
 4   0 0 0 0 0 0 0 0  >2 2 2 2 2 2 2 2
\end{verbatim}
\normalsize

The algorithm works as follows, according to Fig. \ref{algorithm-syncfiring-pointerjumping1}:

\begin{itemize}
\item $t < 0$:
Each cell points to its right neighbor in the ring. 
Every cell is in state $S$.

\item $t=0$:
A general is assigned at any position. 

\item $t>0$:
The pointer and data rule are applied.
The pointer value is doubled at each step until $0=2^n ~mod~ n$ is reached $(1, 2, ... 2^{n-1}, 0)$.
The data value 1 propagates exponentially to all cells until the system will be ready to fire. 

\item $t=log_2 ~n$:
This situation ($\forall i: (p_i=0) ~and~ (d_i=1)$) signals that all cells are ready to fire.

\item $t_{fire}=1+log_2 ~n$:
All cells change into the firing state. 

\end{itemize}

There are two shortcomings of this solution.
(1) The number $n$ must be a power of 2.
(2) When the General is assigned, the pointers must have the value +1.
So it is not possible to introduce the general at a later time when the pointers were already changed by the rule. 
Therefore we look for a more general solution without these restrictions.

\vspace{10pt}
\textbf{Solution 2.}
The following solution works for any $n$, and the General can be introduced at any time at any position. 
Pointer Rule:

\[
  p' = g(p,n)= \left\{
	    \begin{array}{lll} 
			
		   1
			 & \textbf{if~}p=0
			 & ~~~~(7a) \\

			 0  
			 & \textbf{if~} p<0 
			 & ~~~~(7b) \\	
							
	     2p ~mod ~n 
			 &\textbf{otherwise}
			 & ~~~~(7c) \\	
			\end{array}
	      \right. .
\]	

\blankline
This rule ensures that the pointers run in a cycle with values that are powers of 2.
The cyclic sequence is $(1,2,4, \ldots, N/2,0)$ where $N$ is the next power of 2 boundary for $n$ : $2^{k-1} < n \leq N=2^k$.
Rule (7a) implicates that the sequence is repeated when 0 is reached. 
Rule (7c) doubles the pointer by default.
Rule (7b) is used if $n$ is not a power of two. Then, in the last step of the cycle, zero cannot be the result of pointer doubling. The result of doubling modulo $n$ would be less then $p$ which is the criterion to force the pointer to take on the value 0, and so to mark the end of the cycle.

\blankline
Data Rule: 
\[
  d' = f(p,d,d^*)= \left\{
	     \begin{array}{lll} 
			
				d^*  
				&\hspace{6mm} \textbf{if} ~(p \neq 0) ~\textbf{and}~  (d < d^*)  
				&\hspace{2mm}  \hfill (8a) \\	     
			
	      2  
				&\hspace{6mm} \textbf{if} ~(p=0)~\textbf{and}~  (d=1) 
				&\hspace{2mm}  \hfill (8b) \\
				
	      3  
				&\hspace{6mm} \textbf{if} ~(p=0)~\textbf{and}~  (d=2) 
				&\hspace{2mm}  \hfill (8c) \\				
				
				d
				&\hspace{6mm} \textbf{otherwise} 
				&\hspace{2mm}  \hfill (8d) \\
   
			\end{array}
	      \right. .
\]

The data states are:
$0=S$ (Soldier),
$1=G$ (General), 
$2=A$ (Attention),
$3=F$ (Fire).
Rule (8a) is used to propagate exponentially the states 1 and 2.
Rule (8b) changes the state into 2 when the last value (0) of the cyclic pointer sequence is detected.
Firing Rule (8c) is applied when all states are 2 at the end of the cycle.
Otherwise the state remains unchanged (8d). 

Note that the pointers are running in a cycle, the system waits (busy waiting) for the General to be introduced.
This system state can be interpreted as a ``quiescent state'' that is in fact an orbit.
After the General was introduced the algorithm starts working until the system fires.

Compared to the algorithm before, we need now around two cycles instead of one 
but the algorithm is much more general. 

The maximal firing time is $t_{fire}^{max}= 2+2log_2~n$ if the general is introduced when the pointers are in the state 00...0.
The minimal firing time is $t_{fire}^{min}= 2+log_2~n$ if the general is introduced when the pointers are in the state 11...1.

The time evolution of the pointers and the data are shown in the following for $n=9$:

\scriptsize 	 	
\begin{verbatim}
 (a)  Pointer             Data               (b) Pointer             Data
 i=   0 1 2 3 4 5 6 7 8   0 1 2 3 4 5 6 7 8      0 1 2 3 4 5 6 7 8   0 1 2 3 4 5 6 7 8
 t 
-1    0 0 0 0 0 0 0 0 0   0 0 0 0 0 0 0 0 0     -1-1-1-1-1-1-1-1-1   0 0 0 0 0 0 0 0 0       
 0    1 1 1 1 1 1 1 1 1  >0 0 0 0 1 0 0 0 0     >0 0 0 0 0 0 0 0 0  >0 0 0 0 1 0 0 0 0 
 1    2 2 2 2 2 2 2 2 2   0 0 0 1 1 0 0 0 0      1 1 1 1 1 1 1 1 1  >0 0 0 0 2 0 0 0 0
 2    4 4 4 4 4 4 4 4 4   0 1 1 1 1 0 0 0 0      2 2 2 2 2 2 2 2 2   0 0 0 2 2 0 0 0 0
 3   -1-1-1-1-1-1-1-1-1   1 1 1 1 1 0 1 1 1      4 4 4 4 4 4 4 4 4   0 2 2 2 2 0 0 0 0
 4   >0 0 0 0 0 0 0 0 0  >1 1 1 1 1 1 1 1 1     -1-1-1-1-1-1-1-1-1   2 2 2 2 2 0 2 2 2
 5    1 1 1 1 1 1 1 1 1  >2 2 2 2 2 2 2 2 2     >0 0 0 0 0 0 0 0 0  >2 2 2 2 2 2 2 2 2
 6    2 2 2 2 2 2 2 2 2   2 2 2 2 2 2 2 2 2      1 1 1 1 1 1 1 1 1  >3 3 3 3 3 3 3 3 3
 7    4 4 4 4 4 4 4 4 4   2 2 2 2 2 2 2 2 2      
-1   -1-1-1-1-1-1-1-1-1   2 2 2 2 2 2 2 2 2      
 9   >0 0 0 0 0 0 0 0 0  >2 2 2 2 2 2 2 2 2      
10    1 1 1 1 1 1 1 1 1  >3 3 3 3 3 3 3 3 3   
\end{verbatim}
\normalsize

On the left (a) a case with  $t_{sync}^{max}$ is shown, and on the right (b) a case with   $t_{sync}^{min}$.
All pointers are equal and they are running permanently in the cycle: $(1,2,4,-1,0)^*$.

\section{GCA Hardware Architectures}
\label{GCA Hardware Architectures}

We have to be aware that an architecture \textit{ARCH} may consists of three parts
\textit{ARCH = (FIX, CONF, PROGR)}
where 
\textit{CONF} and \textit{PROG} are optional.
FIX is the \textit{fixed} hardware by construction/production,
\textit{CONF} is the \textit{configurable} part (typically the logic and wiring 
as in a FPGA (field programmable logical array)), and 
\textit{PROG} means programmable, usually by a loadable program into a memory before runtime. 

There are four possible general types of architectures

\vspace{8pt}
\begin{tabular}{lll}
Architecture & Parts &  Description   \\
Type         &       &     \\
\hline

1& \textit{FIX }        & special processor\\
\hline

2& \textit{FIX, CONF} & configurable processor\\

\hline
3& \textit{FIX, PROG} & programmable processor\\

\hline
4&  \textit{FIX, CONF, PROG} & config. \& progr. processor\\

\end{tabular}
\vspace{8pt}

After configuration and programming the architecture turns into a special (configured \& programmed) processor.
In general a ``processor'' can be complex and built by interconnected sub processors, like
a multicore or multiprocessor system with a network.

\begin{figure}[tb]
	\centering
		\includegraphics[width=10cm]{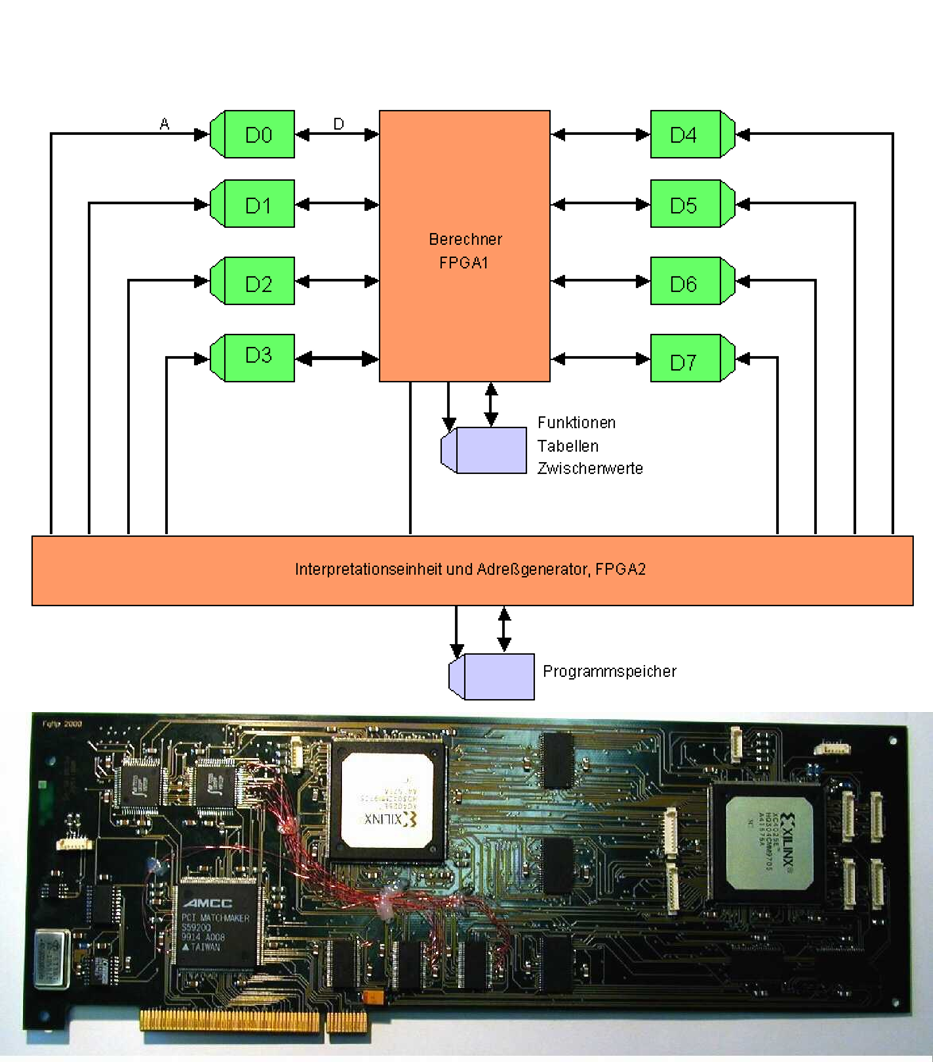}
			\caption{CEPRA-S supporting CA and GCA models. 
8 data memories, program memory, temporary memory,
computational unit (FPGA1), interpretation and address generator (FPGA2), PCI-Interface.}
	\label{CEPRA-S}
\end{figure}

\begin{figure}[tb]
	\centering
		\includegraphics[width=10cm]{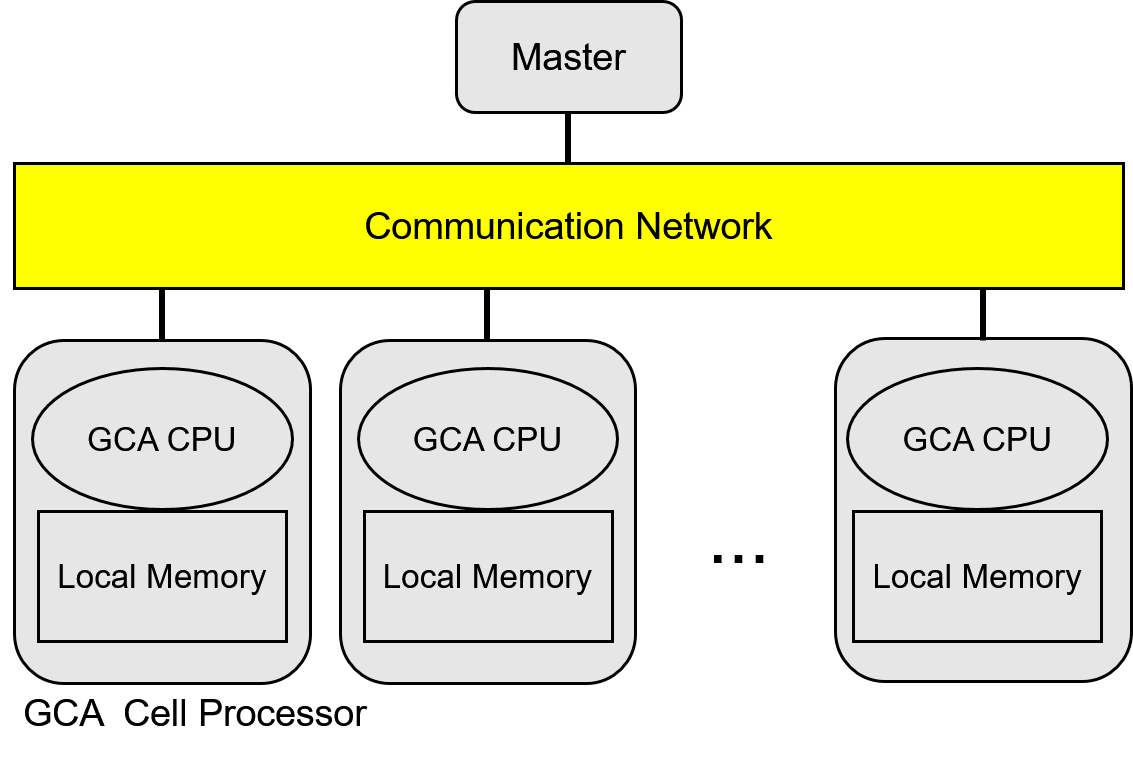}
			\caption{
      Multiprocessor Architecture with cell processors that may offer GCA support 
      (address modification, accessing global neighbors, optimized network, special GCA instructions).       
      }
	\label{GCA-CELL-Processor}
\end{figure}

A variety of architectures can be used or designed to support the GCA model. 
In our research group (Fachgebiet Rechnerarchitektur, FB20 Informatik, Technische Universität Darmstadt)
we developed special hardware support using FPGAs, firstly for the CA model 
(CEPRA (Cellular Processing Architecture) series,
 CEPRA-3D 1997, CEPRA-1D 1996, CEPRA-1X 1996, CEPRA-8D 1995,
CEPRA-8L 1994,
CEPRA-S 2001),
and then for the GCA model (2002--2016) 
\cite{Hoffmann2002}--\cite{Jend2016}.
The CEPRA-S (Fig. \ref{CEPRA-S}) was designed not only for CA but also for GCA.

There are mainly three fundamental GCA architectures:

\begin{itemize}
\item
\textbf{Fully Parallel Architecture}.
A specific GCA algorithm is directly mapped
into the hardware using registers, operators and hardwired links which may
also be switched if necessary. The advantage of such an implementation is a
very high performance 
\cite{Heenes2005Neu,Jendrsczok2007Hirschberg,Jendrsczok2008HirschbergFPGA} 
(Sect. \ref{Fully Parallel Architecture}), 
but the problem size is limited by the hardware
resources, and the flexibility to apply different rules is low.

\item
\textbf{Data Parallel Architecture with Memory Banks and Pipelining (DPA).} 
This partial parallel architecture 
\cite{
HHV03,
Hoffmann2003WMPP,
HHH04,
Heenes2005Neu,
Jendrsczok2007APL,
JEH09,
JHE09,
JHL09}
offers a high performance, is scalable and it can process a large
number of cells. The flexibility to cope with different 
and complex applications is restricted.

\item
\textbf{Multiprocessor Architecture.}
This architecture (Fig. \ref{GCA-CELL-Processor})
is not as powerful as the
above mentioned, but it has the advantage that it can be tailored to any
GCA problem by programming. It also allows integrating standard or other
computational models. Standard processors can be used, or special ones supporting GCA features, 
see \cite{
Heenes2005Neu,
Heenes2005Pars,
HHJ06,
Heenes2007,
SHH09a,
SHH09b,
Schaeck2009AgentSim,
Schaek2010Multiagent,
Schaeck2010Traffic,
Schaek2011,
Schaek2011Multicore,
Schak2011Corrected,
Milde2011,
Jend2016}. 

Standard multiprocessor platforms, like standard multicores or GPUs, can also execute efficiently the GCA model. 
In \cite{Milde2011} a speedup of 13 for bitonic merging was reached on an NVIDIA GFX 470 compared to an Intel Q9550@3GHz with 4 threads, and 150 for a diffusion algorithm. 
\end{itemize}

\subsection{Fully Parallel Architecture}
\label{Fully Parallel Architecture}

\begin{figure}[t]
	\centering
		\includegraphics[width=0.8\textwidth]{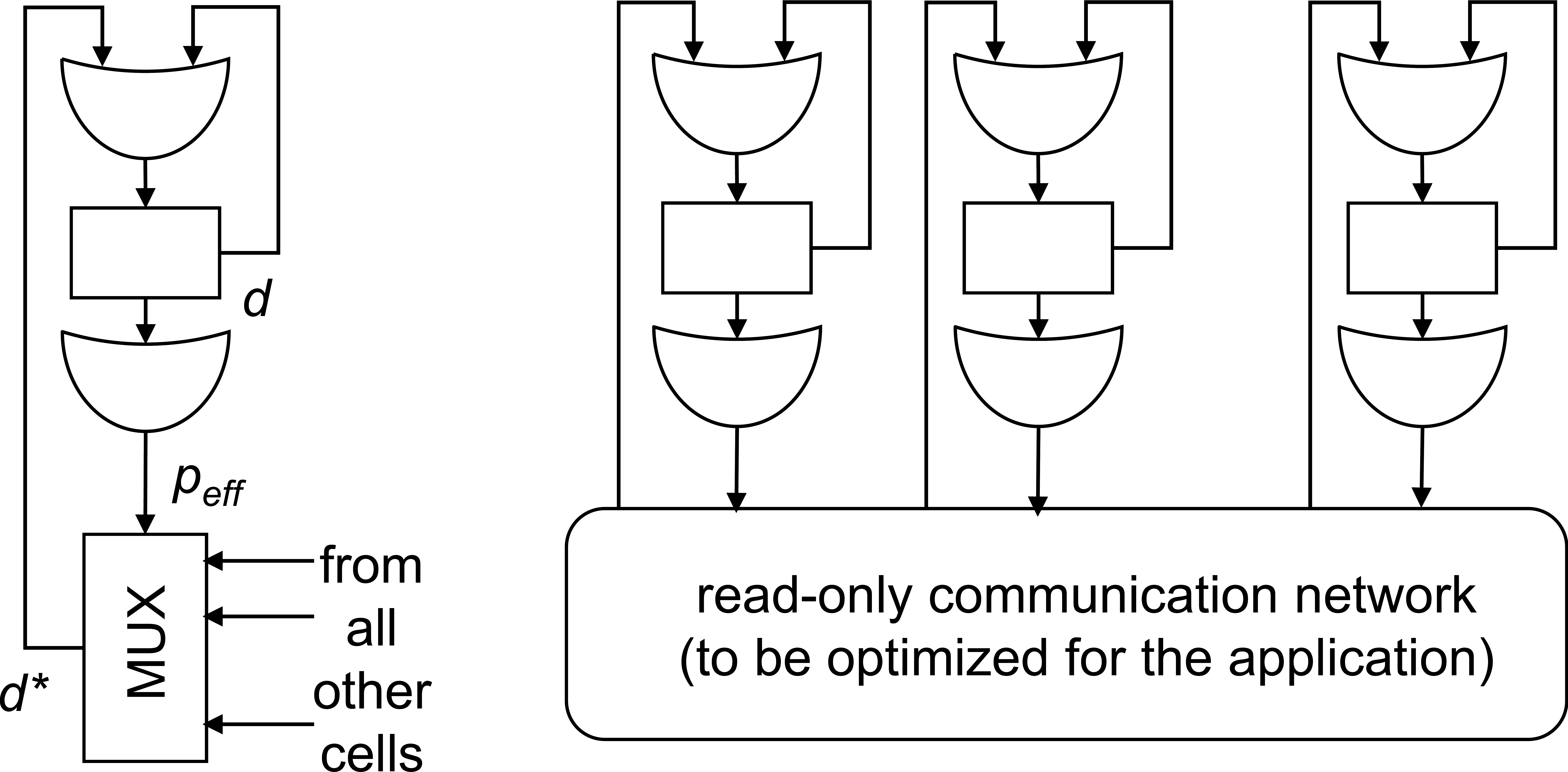}
			\caption{Fully parallel implementation. Communication implemented by a multiplexer in each cell (a). Communication implemented by a common network (b).}
	\label{fullpar}
\end{figure}

An important attribute is the \textit{degree of parallel processing}
(the number of processing/computation units) $p$
\footnote{In this Sect. \ref{GCA Hardware Architectures} 
about hardware architectures, $p$ stays for the \textit{degree of parallelism} and not for \textit{pointer}.}.
In other words, $p$ gives the number of results that can be computed and stored in parallel. 
A \textit{sequential} architecture is given by $p=1$,
a \textit{fully parallel} by $p=n$,
and a \textit{partial parallel} by $n>p>1$.

\textit{Fully parallel architecture} means that the whole GCA with $p=n$ is completely implemented in hardware (Fig. \ref{fullpar}) for a specific application.
The question is how many hardware resources are needed. The number of cells is $n$. 
Therefore the logic (computing the effective address and the next state) and the number of registers holding the cells' states are proportional to $n$. 
The local interconnections are proportional to $n$, too. 
As the GCA generally allows read-access from each cell to any other cell, 
the communication network needs $n\times (n-1)$ \textit{global} links, where a link consists of $V(n,m)$ bit-wires/channels.
$V(n,m)$ is the word length in bits of the cell's state.
The length of a global link is not a constant, it depends on the physical distance. 
In a ring layout, the average link length of $n/4 \times (\textit{space unit})$ has to be taken into account. 
See considerations about implementation complexity 
for the basic model in Sect. \ref{Basic Model with Stored Pointers} on page \pageref{page-complexity}.
Note that the longest distance also determines the maximal clock rate.

Many applications / GCA algorithms do not require a total interconnection fabric because only a subset of all communications (read accesses) are required for a specific application. Therefore the amount of wires and switches can be reduced significantly for one or a limited set of applications.  In addition, for each global link a switch is required. The switches can be implemented by a multiplexer in each cell, or by a common switching network (e.g. crossbar). Note that the number of switches of the network can also be reduced to the number of communication links used by the specific application. Another aspect is the multiple read (concurrent read) feature. In the worst case, one cell is accessed from all the other cells which may cause a fan-out problem in the hardware implementation.

\subsection{Sequential with Parallel Memory Access}
\label{Sequential with Parallel Memory Access}

\begin{figure}[hbp]
	\centering
		\includegraphics[width=8.2cm]{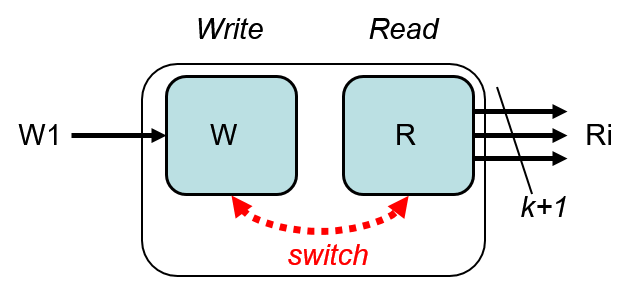} 
			\caption{Multiport memory. 
      When the computation of a new generation of cell states is completed, 
      the read and write access are switched.      
			}
	\label{MultiportAbb1}
\end{figure}

The goal in this and the next section is to design architectures with normal memories that work efficiently.
We assume that the GCA can access two global cells, $k=2$.
\footnote
{In this and the next section the number of pointers/links is denoted by ``\textit{k}'' and not by \textit{m}
as before.}
The cell state structure is
$(D, L1, L2)$ where $D$ is the data part and $L1, L2$ are the pointers. 
The array \textit{Cell} stores the whole set of cell states, 
and the array \textit{CellNew} is needed for buffering in synchronous mode. 
 
The computation of a new cell state at position $z$ needs the following four steps:

\begin{enumerate}
\item
(Fetch) The cell's state $a=\textit{Cell}[z]$ is fetched.

\item
(Get) The remote cell states $b=\textit{Cell}[L1]$, and $c=\textit{Cell}[L2]$ are fetched.

\item
(Execute)
The function $y=f(a,b,c)$ is computed.

\item
(Write)
The result (new state) is buffered $\textit{CellNew}[z]:=y$.

\end{enumerate}

Our first design assumes a virtual (or real) multiport memory (Fig. \ref{MultiportAbb1}) 
that can perform all necessary memory accesses in parallel.
The internal read memory $R$ is used to read the actual cell states $\textit{Cell}[i]$,
and the internal write memory $W$ is used to buffer the new cell states $\textit{CellNew}[i]$.
The read memory $R$ is a read multiport memory allowing $k+1$ parallel read accesses.
The read ports are $R1, R2, R3$.
$W$ is a write memory with one port $W1$. 

When the computation of a new cell generation  $\textit{CellNew}(t)$ is complete, 
it has to function as $\textit{Cell}(t+1)$ for the next time-step $t+1$.
One could alternate/interchange  the internal read memory with the internal write memory 
(\textit{switch}, using internal multiplexer hardware).
One could also use different pages and change read/write access for the ports.
In principle one could also copy the arrays $\textit{Cell} \leftarrow \textit{CellNew}$
to realize the required synchronous updating.

\begin{figure}[htbp]
	\centering
		\includegraphics[width=7cm]{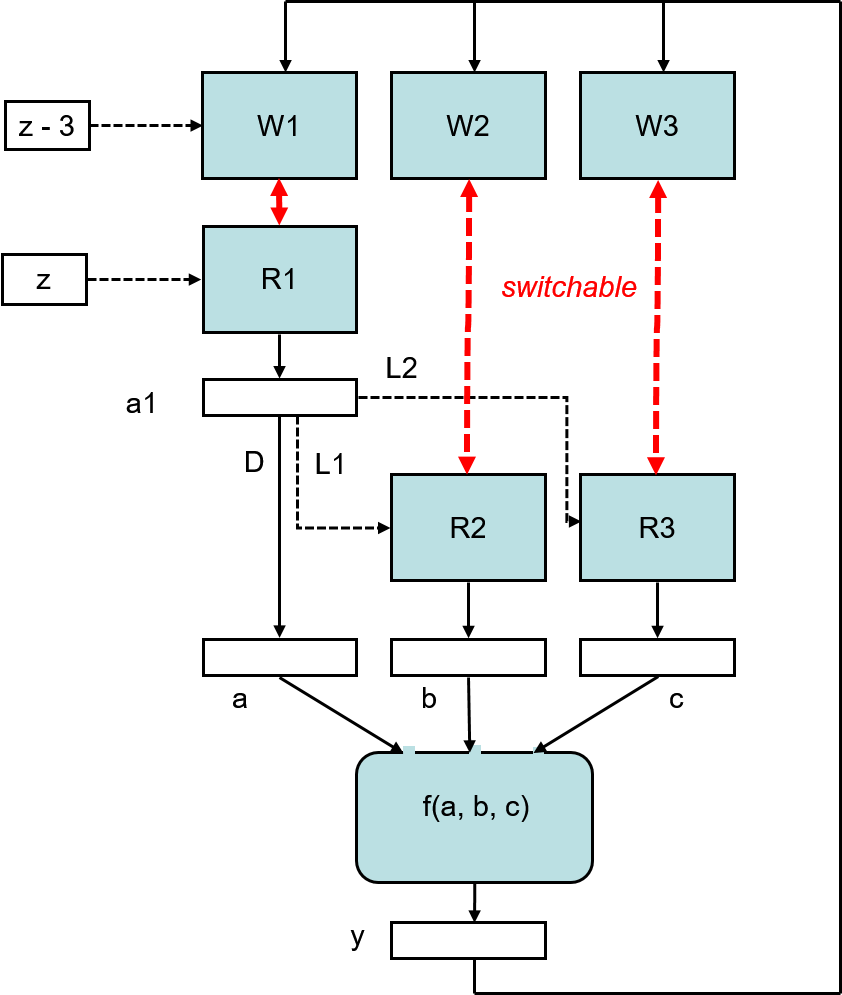} 
			\caption{Sequential architecture with pipelining. The multiport memory (parallel access)
      is emulated by the use of $2(k+1)$ normal memories. When a new generation is completed,
      the read and write memories are switched.
			}
	\label{SeqArchWithPipeliningAbb2}
\end{figure}
\begin{figure}[htbp]
	\centering
		\includegraphics[width=1\textwidth]{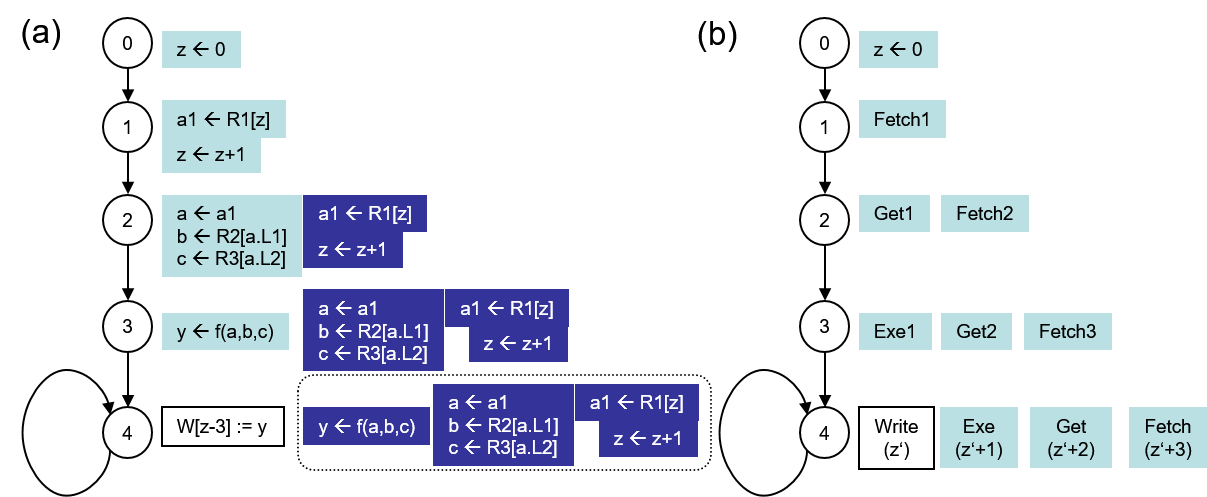} 
			\caption{
      Control algorithm controlling the execution of the pipeline shown in Fig. \ref{SeqArchWithPipeliningAbb2}.
      (a) Detailed with all register transfer operations. (b) Abstract representation, where $z'=z-3$.
			}
	\label{PipeControlAbb3}
\end{figure}

The multiport memory can be implemented using normal memories 
(Fig. \ref{SeqArchWithPipeliningAbb2}). 
Three read memories $R1, R2, R3$ and three write memories are used
(in general $2(k+1)$ memories).
Each new state is simultaneously written  into the write memories $W1, W2, W3$.
After switching the read and write memories, the new states are available in parallel from the read memories
for the next generation.

\textit{Control Algorithm} (Fig. \ref{PipeControlAbb3}).
The control algorithm for this pipelined architecture was developed by transformation of a purely sequential one. 

\small
\begin{itemize}
\item
\textit{\textbf{State 1}} 

  \begin{tabular}[h]{ll} 
  \textbf{Fetch1}: 
  &The cell's state at position $z$ is fetched and stored in $a1$. \\
  &The counter $z$ is incremented (synchronously).
  \end{tabular}
  
\item
\textbf{\textit{State 2}} 

  \begin{tabular}[h]{ll} 
  \textbf{Get1}: 
  &The global states $a1.L1$ and $a1.L2$ are fetched and $a1$ is shifted to $a$.\\

  \textbf{Fetch2}: 
  &The next cell's state is fetched.
  \end{tabular}
  
\item
\textbf{\textit{State 3}} 

  \begin{tabular}[h]{ll} 
  \textbf{Exe1}: 
  &The data values $a,b,c$ are available and the computation is performed. \\

  \textbf{Get2}:
  &For the next already fetched cell, the global cell states are accessed.\\

  \textbf{Fetch3}:
  &The next cell is fetched.\\
  \end{tabular} 
  
\item  \textbf{\textit{State 4}}: Four actions are performed in parallel when the pipeline is fully working.

  \begin{tabular}[h]{ll}
  \textbf{Write}: 
  &The result of cell $z-3$ is written. \\

  \textbf{Exe}: 
  &The result of cell $z-2$ is computed.\\

  \textbf{Get}:
  &The global cells' states, addressed by $z-1$, are read.\\

  \textbf{Fetch}: 
  &Cell $z$ is fetched.
  \end{tabular} 

\end{itemize}
\normalsize

\textit{Computation Time.}
If the number of cells is large enough, the latency (time to fill the pipeline in states 0--3)
can be disregarded.
Then a new result can be computed within one clock cycle, independently of the number $k$ of global cells:
$t(n,k)= nT$ , where $T$ is the duration of one clock cycle.\\

\textit{Implementation Complexity.}
The number of registers, functions (arithmetic and logic), and the local wiring according to the layout 
shown in Fig. \ref{SeqArchWithPipeliningAbb2} is relatively low and constant compared to the required memory capacity
(for a large number of cells).
The capacity $M_1$ (in bits) of one memory is

\blankline
$M_1(n,k) = n(bit(D) + k \cdot bit(L)) = n(bit(D) + k \cdot log_2 ~n)$ .

\blankline
\noindent The whole memory capacity for $2(k+1)$ memories is

\blankline
$M(n,k) = 2(k+1) M_1 = 2n(k+1) \cdot (bit(D) + k \cdot log_2 ~n)$ .
\blankline

The memory capacity is in $O(k^2 \cdot n \cdot log~n)$, therefore the number of pointers needs to be small,
usually $k=1$ or $k=2$ is sufficient for most applications.

\subsection{Partial Parallel Architectures}
\label{Partial Parallel Architectures}

\subsubsection{Data Parallel Architecture with Pipelining}
\label{Data Parallel Architecture with Pipelining}
\begin{figure}[htbp]
	\centering
		\includegraphics[width=0.45\textwidth]{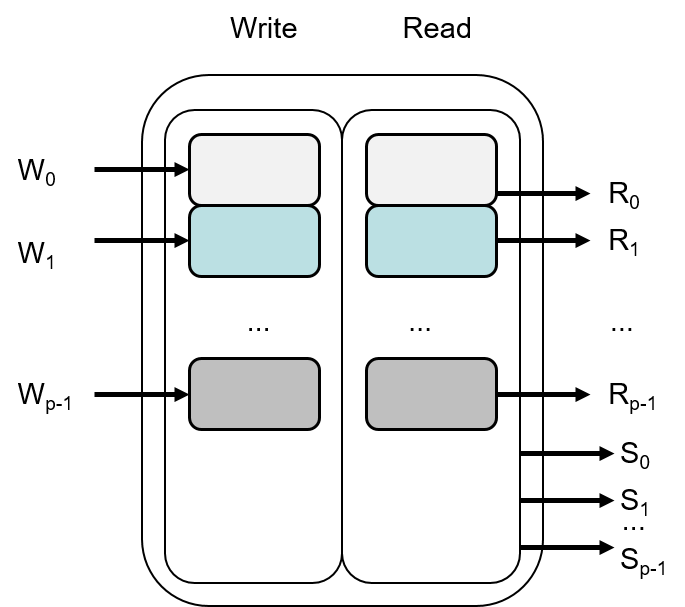} 
	\caption{Multiport memory that provides $p$ write and $p$  read ports to banks (address pages), and
  $p\cdot k$ read ports $S_j$ with the whole address range for accessing the neighbors.  Case $k=1$.} 
  \label{MultiportMemory2}  
\end{figure}

\begin{figure}[htbp]
	\centering
		\includegraphics[width=0.8\textwidth]{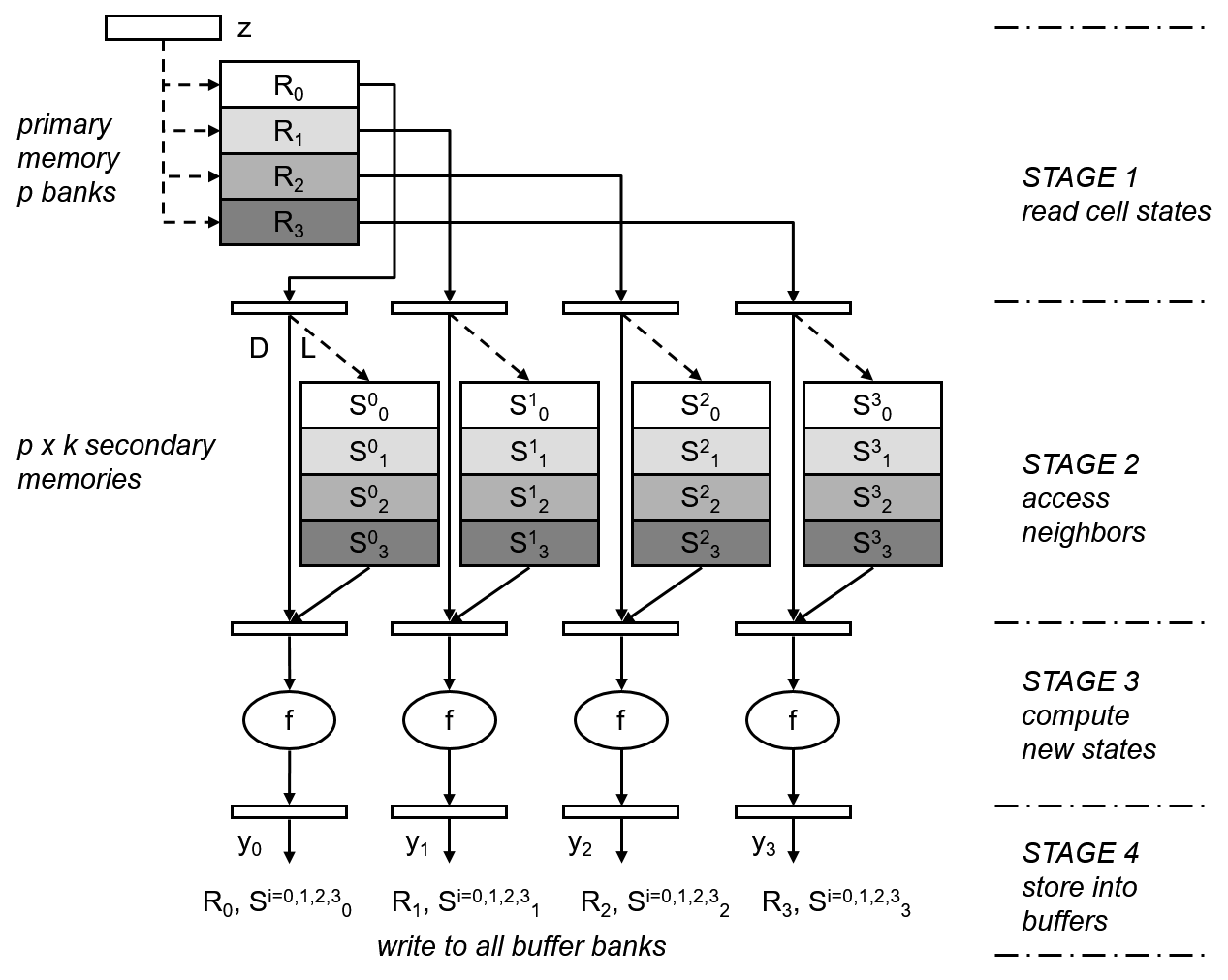} 
			\caption{Data parallel Architecture (DPA) with pipelining for $p=4$. 
      Stage 1: $p$ cells' states are read form the banks of the primary memory.
      Stage 2: the global neighbors are accessed. 
      Stage 3: $p$ new cell states  are computed. 
      Stage 4: The new states are written into all associated buffer banks (not shown).          
			}
	\label{ParallelPipeline4}
\end{figure}

We want to design a \textit{data parallel architecture} (DPA)  with pipelining for the parallel degree $p$,
 and with one pointer $k=1$.
We call the such an architecture ``\textit{data parallel}'', because $p$ data elements (cell states)
are computed in parallel. 
A special multiport memory (real or virtual) is needed (Fig. \ref{MultiportMemory2}).
It contains two sub memories that can be switched to allow alternating read/write access
in order  to emulate the synchronous updating scheme.
The sub memories are structured into $p$ banks/pages.
Each bank stores $n/p$ cells.
The banks can be accessed via $p$ write ports $W_0, W_1, \ldots W_{p-1}$ 
and  $p$ read ports $R_0, R_1, \ldots R_{p-1}$.
In addition, the read memory supplies $pk=p$  access ports $S^1, S^2, \ldots S^{p-1}$  with the whole address range,
dedicated to access the global neighboring cells.
The working principle for a new generation of cell states is:

\begin{enumerate}
\item
\textbf{for} $z:=0$ \textbf{to} $n/p - 1$ \textbf{do}

  \begin{enumerate}
  \item
  Read $p$ cell states from the $p$ banks in parallel from location $z$.
  \item
  Access $p \cdot k$ neighbors via the  whole range ports $S^{i=1..p}$.
 \item
  Compute $p$ results.
 \item
 Write the results to the $p$ banks of the write memory. 

  \end{enumerate}

\item
Interchange the read and write memory (switch) before starting a new generation.
\end{enumerate}

The write operations are without conflict, because each of the $p$ cells are assigned 
exclusively to a separate bank (like in the owner's write PRAM model).
The  memory capacity needed is just the space for the cells (doubled for buffering) 
and does not depend on $p$:

\blankline
$M_{multiport}(n,k) = 2n(bit(D) + k\cdot bit(Link)) = 2n(bit(D) + k\cdot log_2 ~n)$ ,
\blankline

\noindent
however we have to be aware that the hardware realization of such a multiport memory
is complex because it would need a special design with a lot of ports and wiring. 
Therefore we want to emulate it by using standard memories (Fig. \ref{ParallelPipeline4}).
For explanation we  assume the case $p=4$ and $k=1$.
We will use several bank memories of size $n/p$.

\begin{enumerate}
\item
In pipeline stage 1,  $p$ cells are fetched from the $p$ banks $R_{0,1,2,3}$ of the primary memory $R$
at position $z$ defined by a counter. 
\item
In stage 2, $pk$ (i.e. 4) global cells are accessed form the secondary memories $S^{i=0,1,2,3}$
with the whole address range.
Each $S^i$ memory is composed of $p$ banks $S^{i}_{j=0,1,2,3}$.

\item
In stage 3, $p=4$ results (new cell states) are computed.
\item
In stage 4, the results $y_j$ are transferred to each associated buffer bank (denoted by $^*$) at position $z-3$

$R^*_j[z-3], ~S^{*i}_j[z-3] \leftarrow y_j$ ~.

\end{enumerate}

After completion of one generation, the buffer memory banks and the used banks are interchanged:

\blankline
$R_j \leftrightarrow R^*_j$ ~~and ~~$S^{~i=0,1,2,3}_j \leftrightarrow S^{*~i=0,1,2,3}_j$ ~~for all banks $j$.
\blankline

After the start-up phase, $p$ new cell states are computed and stored for every time step. 
The number of bank memories needed is $(kp+1)p$, each holding $n/p$ cell bits.
The whole capacity needed is $(kp+1)p \cdot n/p = n(kp+1)$ cell bits, to be doubled because of
buffering. 

\blankline
$M = 2n(kp+1)\cdot(bits(D)+k~log_2 n)$ ~.
\blankline

\subsubsection{Generation of a Data Parallel Architecture}
\label{Generation of a Data Parallel Architecture}

The data parallel architecture (DPA)
(Sect. \ref{Data Parallel Architecture with Pipelining})
 uses $p$ pipelines in order to process $p$ cell rules in parallel.
It was implemented on FPGAs in different variants and for different applications up to $p=8$ 
(\cite{Heenes2005Neu,Jendrsczok2007Hirschberg,Jendrsczok2008HirschbergFPGA,JEH09,JHE09,JHL09,Jend2016}).
 
In \cite{JEH09,JHE09,JHL09} 
the whole address space is partitioned into (sub) arrays, also called ``\textit{cell objects}". 
In our implementation, a cell object represents either a cell vector or a cell matrix. 
A cell object is identified by its start address, 
and the cells within it are addressed relatively to the start address. 
The \textit{destination object} \textit{D} stores the cells to be updated, 
and the \textit{source object} \textit{S} stores the global cells to be read. 
Although for most applications \textit{D} and \textit{S} are disjunct, the may overlap or be the same. 

\begin{figure}[tbhp]
	\centering
		\includegraphics[width=1\textwidth]{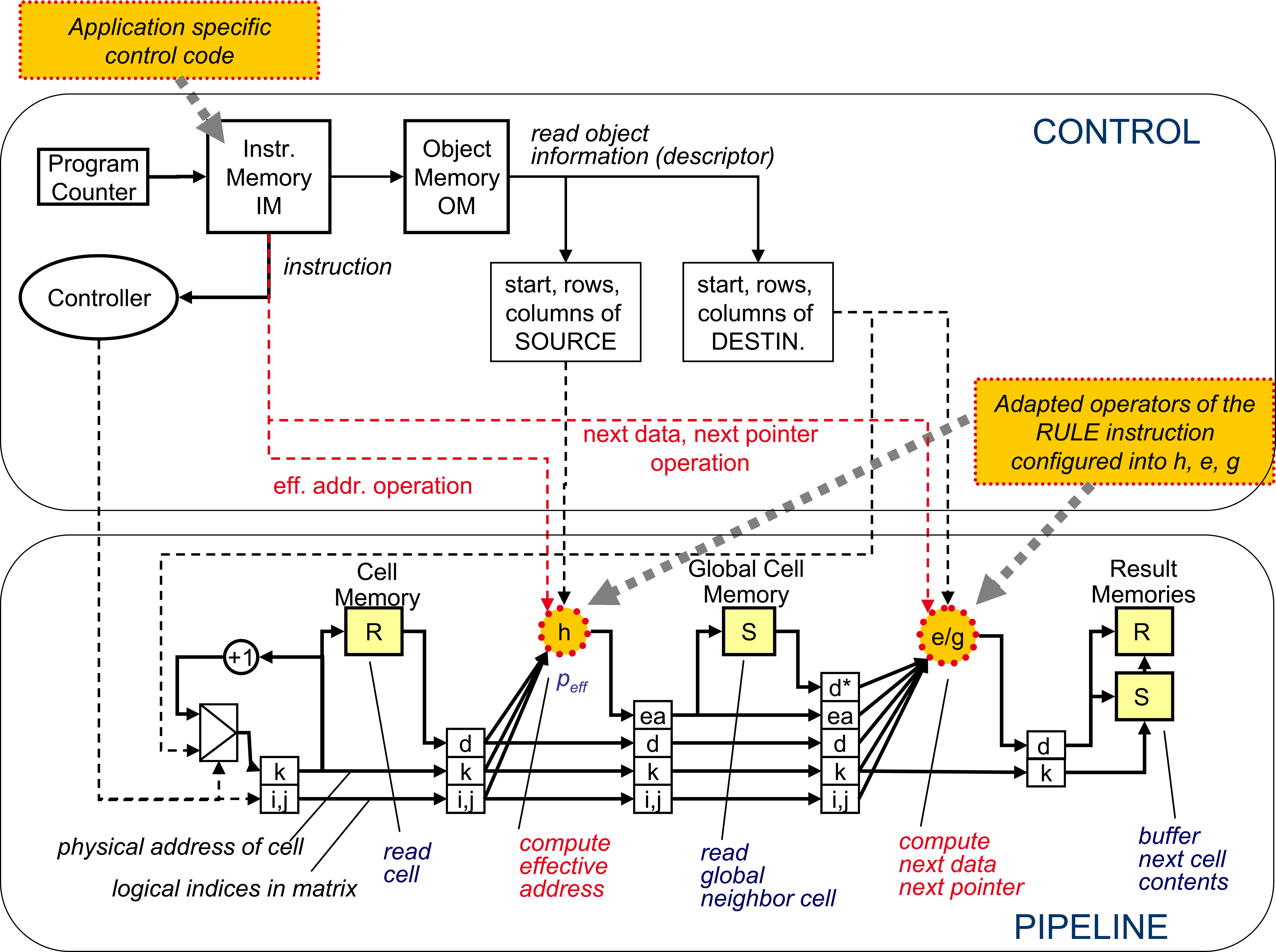} 
			\caption{Data parallel architecture (DPA) with one pipeline.}
	\label{pipe}
\end{figure}

The DPA consists of a control unit and $p$ pipelines, only one pipeline is shown in Fig. \ref{pipe}. 
In the case of one pipeline only, the cells of  \textit{S} are processed sequentially using a counter $k$. 
In the first pipeline stage the cell \textit{D[k]} is read from memory $R$. 
In the second stage the effective address \textit{ea} is computed by $h$. 
In the third stage the global cell \textit{S[ea]} is read. 
In the fourth stage the next cell state $d$ is computed. 
Then the next cell state is stored in the buffer memories $R'$ and $S'$ at location $k$. 
When all cells of the destination object are processed, the memories $(R,S)$ and $(R',S')$ are interchanged.   

\begin{figure}[tbhp]
	\centering
		\includegraphics[width=1\textwidth]{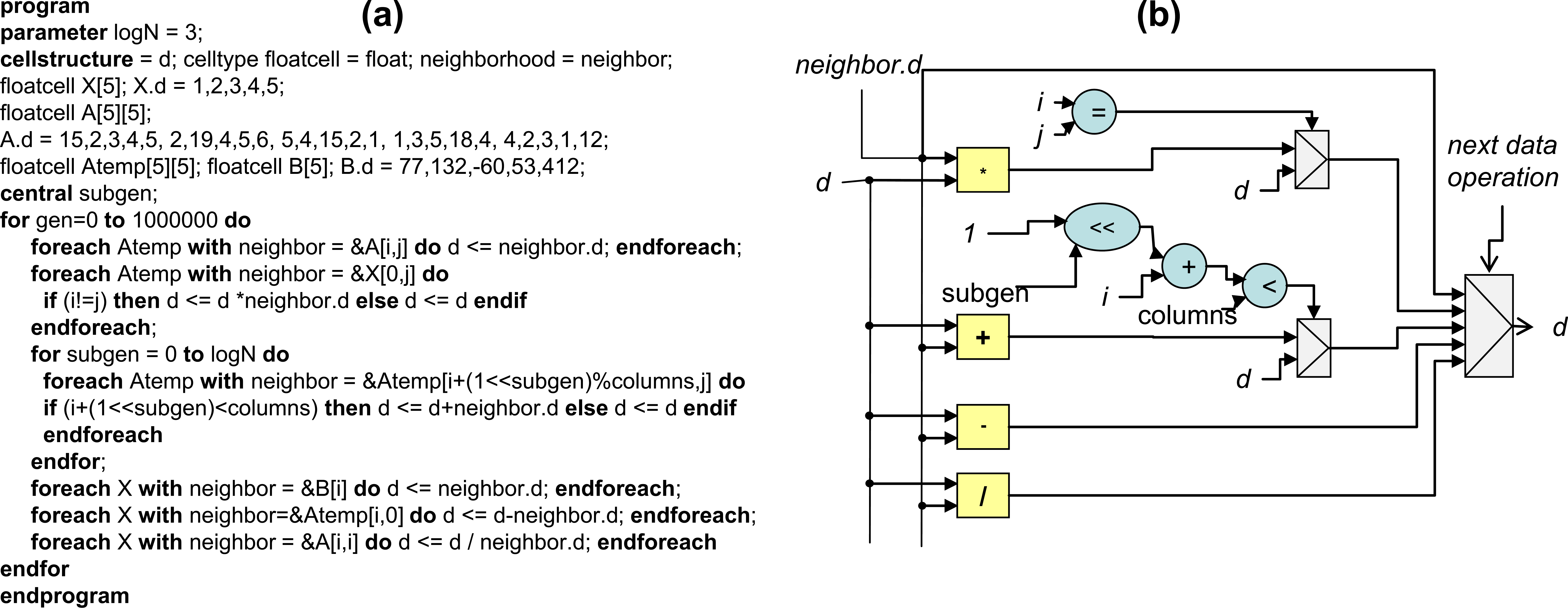} 
			\caption{(a) GCA-L program for the Jacobi iteration. (b) Next data operator $e$ automatically generated out of the progam. It
contains 4 floating point units and several integer units. 
}
	 \label{progandop}              
\end{figure}

An application specific DPA with $p$ pipelines can automatically be generated out of a high level description in the experimental language GCA-L \cite{JEH09}. The program (Fig. \ref{progandop}a) describes the Jacobi iteration \cite{JHE09} solving a set of linear equations.

The most important feature of GCA-L is the \textit{foreach D with neighbor =} \&\textit{S[..] do .. endforeach} construct. 
It describes the (parallel) iteration over all cells $D[i,j]$ using the global neighbors \&\textit{S[h(i,j)]}. 
Our tool generates Verilog code for the functions $h, e, g$ to be embedded in the pipeline(s). 
These functions are also pipelined. In addition control code for the control unit is generated. 
The most important control codes are the \textit{rule} instructions. 
A rule instruction triggers the processing of all cells in a destination object and applies the so called \textit{adapted operators} $h, e, g$ coded in the rule. 
All necessary application specific rule instructions are extracted from the source program. 

For the Jacobi iteration proigram \cite{JHE09}, 
Fig. \ref{progandop}b shows the generated next data operation used by a rule instruction. 
It contains 4 floating point units and several integer units. The floating point operations
are internally also pipelined (+(14 stages), -(14), *(11), /(33)).
Our tool generates Verilog code which  is then used further  for synthesis with Quartus II for  Altera FPGAs. 
For $p=8$ pipelines, normalized to the amount needed for one pipeline, 
the relative increments for the FPGA Altera Stratix II EP2S180  were: 
8.3 for the  ALUTs (logic elements), 7.5 for the registers, 
4.5 for the memory bits (note that the required memory bits are theoretically proportional to $(p+1)/2$ for the pipeline architecture). 
The speedup was 6.8 for 8 pipelines compared to one. Thus the scaling behavior was very good and almost linear for up to 8 pipelines.

\subsubsection{Multisoftcore}
\label{Multisoftcore}


\begin{figure}[htbp]
	\centering
		\includegraphics[width=8cm]{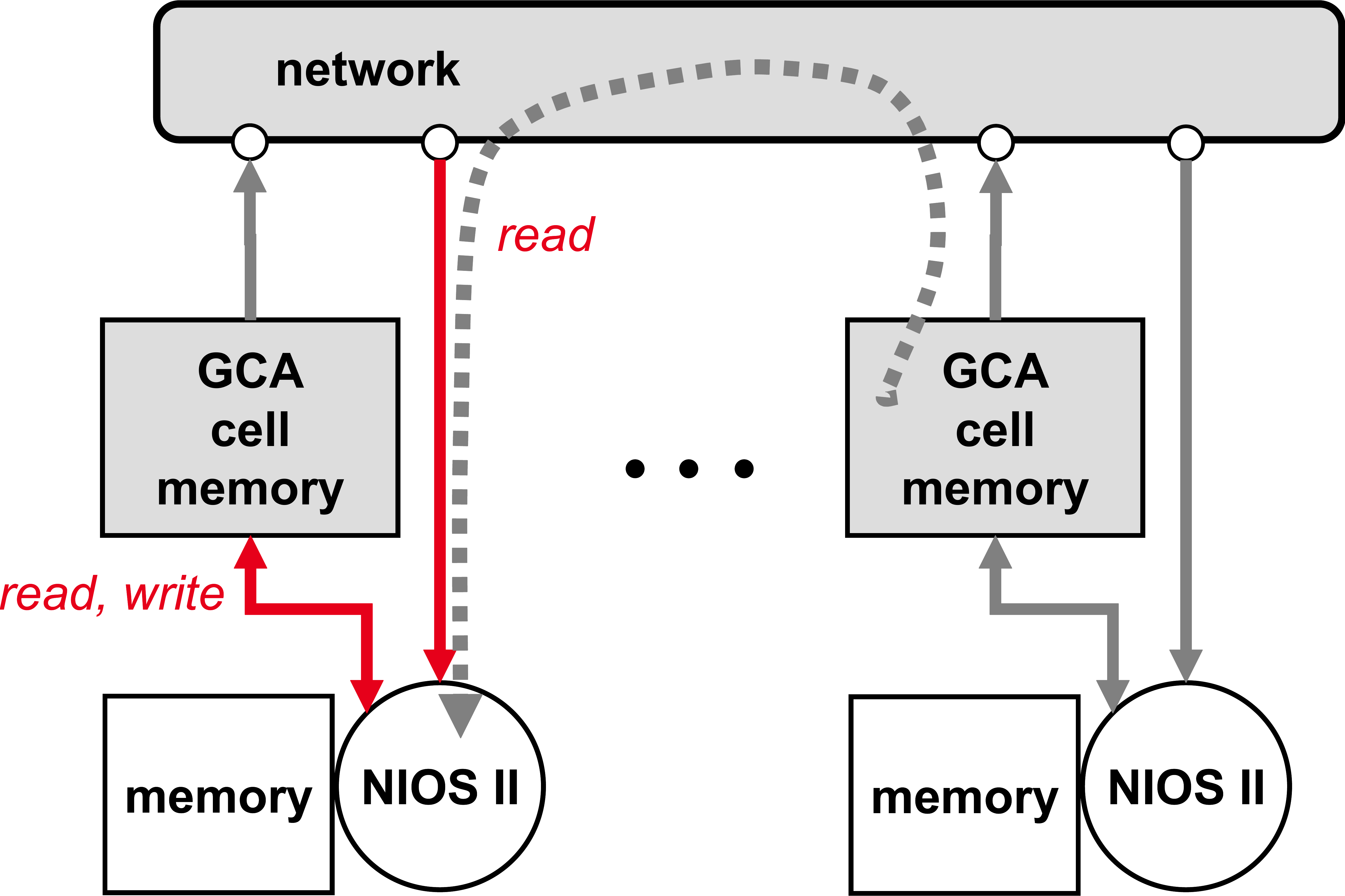} 
			\caption{
Multisoftcore system implemented on an FPGA. A local GCA cell memory is attached to each NIOS II softcore. Each core can read and write its own GCA cell memory and read from any other GCA cell memory via the network.  
}
	\label{nios}
\end{figure}
The basic idea is to use many standard softcores together with specific GCA support. Each core is responsible to handle a subset of all cells being processed in one generation. In our implementation, $p$ NIOS II softcores were used 
\cite{SHH09a}--\cite{Schak2011Corrected}. 
To each processor a GCA cell memory is attached (Fig. \ref{nios}).  A processor can read via the network the state of a global cell residing in another cell memory. Only the cells residing in the own cell memory need to be updated according to the GCA model. No write access  via the network is needed, thereby the network can be simplified. In case that only a specific application has to be implemented, the network can be minimized according to the communication links used by the application. The machine instruction set of the NIOS processors was extended (custom instructions), e.g. read a cell via the network, read/write local cell memory, floating point operations, synchronize and copy new cell states into the current cell states. 

A tool was developed that can automatically generate C code (extended by custom instructions) out of a GCA-L program for such a 
multisoftcore system. Then this C code is compiled and loaded into the cores of the system configured on an FPGA.

\newpage
\section{Conclusion}
\label{Conclusion}
\textit{Global Cellular Automata} (GCA) is a new data parallel programming model related to Cellular Automata (CA).
Applications are modeled as a set of cells which can dynamically connect to any other (global) cell. 
The global communication topology is dynamic but locally computed by the cells. 
In the \textit{basic model}, pointers are stored in the cell that point directly to global neighbors. 
They are updated by pointer rules taking the states of the cell and its neighbors into account. 
In the \textit{general model}, the pointers are modified before access. 
In the \textit{plain model}, the state of a cell is not structured into a data and a pointer part. 

The CROW PRAM model is related to the GCA model,
therefore CROW and CREW algorithms can be converted into GCA algorithms. 
The CROW model is \textit{processor based} ($n$ processors with instruction set, common memory),
whereas the GCA model is \textit{cell based} (state contains pointers, data and pointer rules, local memories).
Boolean Networks can be seen as a special GCA case where the state is binary and the individual links are fixed.

The range of GCA applications is very wide. Typical applications
besides CA applications are 
graph algorithms, 
hypercube algorithms,
matrix operations,
sorting,
PRAM algorithms,
particle and multi-agent simulation, 
logic simulation,
communication networks,
pointer structures, and dynamic topologies. 
Examples for GCA algorithms were given (maximum, reduction, prefix sum, bitonic merging, different XOR rules),
and the new application \textit{Synchronous Firing}.

GCA algorithms can easily be described in standard languages or in a special language like GCA-L,
and compiled to standard parallel platforms (like multicores, GPUs), or to special GCA target architectures.
GCA target architectures can relative easily be designed and generated for
FPGAs, like the \textit{fully parallel architecture}, 
the  \textit{data parallel architecture} with memory banks and pipelining, or
 a \textit{multisoftcore architecture}. 

The effort for the communication network between cells can be reduced by implementing only 
the required access pattern of the application,
or one could restrict the set of accessible global neighbors in advance by definition 
(e.g. hypercube or perfect shuffle connections) and then 
use for an algorithm the allowed connections only.

To summarize, the GCA model is a powerful and easy to use parallel programming model
based on cells with dynamic global neighbors, which
can efficiently be executed on standard and special parallel platforms. 
It fulfills to a large extent important requirements for a parallel programming model:
user-friendly, platform-independent, efficient, and system-design-friendly.

\newpage
\section{Appendix 0: Programs for the 1D Basic and General Model}
\label{Appendix 0}
\subsection{Basic Model}
\label{Appendix 0-Basic}

\footnotesize
The following program can be seen as a prototype for the 1D Basic GCA model.
The cell's state is $(c, p1,p2)$, where $c$ is the data state and $p1,p2$ are
the pointers.  
The pointer rules \textit{p1new\_PointerRule} and \textit{p1new\_PointerRule}
compute the new pointers (multiplying the current value by 2).
The data rule
\textit{DataRule\_with\_Data\_at\_Pointers} (XOR of left and right dynamic neighbor)
computes the new data state. 
The classical CA XOR rule can be emulated by setting the pointer constant to $p1=+1$ and $p2=-1$.

\tiny
\begin{verbatim}
{5.6.2022 RH. Simple 1D Basic GCA program, XOR with pointers doubled}
program prog_gca_basic_xor;
uses  SysUtils;
var   OUT_c, OUT_p1, OUT_p2, OUT_p1eff, OUT_p2eff: textfile;

const BlackSquare=#$E2#$96#$88#$E2#$96#$88;
const OutputZERO='  ';  OutputONE=' #'; // BlackSquare; can be used
const N=31; TMAX=5;       // number of cells, max number of generations

type  field = array [0..N-1] of integer;
type  cell = record c,cnew, p1,p1n2, p2,p2new, p1eff,p2eff : field end;  // cell's structure, not used here

var   c, cnew: field;      // data state, buffered sync operation
      p1, p1new: field;    // stored relative pointer, buffered sync operation
      p2, p2new: field;
var   t: integer;          // time-counter, generation
//=========================================================================== FUNCTIONS, PROCEDURES
function modN(a:integer):integer; begin modN:=(a+N)mod N; end;

function p1new_PointerRule(x:integer):integer;
const p1init= +1;
begin
       //_____________________ initial set pointer const at t=0
       if t=0 then p1new_PointerRule:=p1init;
       //_____________________ initial set pointer const at t=0
       //_____________________ for t=1,2, ...
       if t>0 then begin      
          p1new_PointerRule:=(p1[x]*2) mod N; // 1,2,4, ...
          if p1new_PointerRule=0 then p1new_PointerRule:= p1init; end;//don't use p=0, instead p1init       
       //_____________________ for t=1,2, ...
end;

function p2new_PointerRule(x:integer):integer;
const p2init= -1;
begin
       //_____________________ initial set pointer const at t=0
       if t=0 then p2new_PointerRule:=p2init;
       //_____________________ initial set pointer const at t=0
       //_____________________ for t=1,2, ...
       if t>0 then
       begin
          p2new_PointerRule:=(p2[x]*2) mod N; // -1,-2,-4, ...
          if p2new_PointerRule=0 then p2new_PointerRule:= p2init; //don't use p=0, instead p2init
       end;
       //_____________________ for t=1,2, ...
end;
//________________________________________ new Pointer for all cells
procedure p1new_p2new_Apply_PointerRule_at_t_for_tplus1;
var x: integer;   // cell's index/position
begin
   for x:=0 to N-1 do
      begin p1new[x]:=p1new_PointerRule(x);
            p2new[x]:=p2new_PointerRule(x); end;
end;
//________________________________________ new Pointer for all cells
//________________________________________ data rule at site x
function DataRule_with_Data_at_Pointers(x,p1,p2: integer):integer;
   function abs(p_relative:integer): integer;
   begin    
       abs:=modN(x+p_relative); 
   end;
  
begin
    // L exor R, abs(p1)=modN(x+p1), c[x] or c[modn(x+1) .. could also be used
    // may also depend on cell's state, fixed neighbors' states, time t, index x
    // new data cnew may depend on: t,x, (c, p1, p2), p1.(c,p1,p2), p2.(c,p1,p2)
    DataRule_with_Data_at_Pointers:=( c[abs(p1)]+c[abs(p2)] ) mod 2;
end;
//________________________________________ data rule at site x
//________________________________________ new cells' data states
procedure cnew_ApplyDataRule;
var x: integer;
begin for x:=0 to N-1 do cnew[x]:=DataRule_with_Data_at_Pointers(x, p1[x], p2[x]); end;
//________________________________________ new cells' data states
//________________________________________ init data state
procedure c_init(z:integer);
var x:integer;
begin for x:=0 to N-1 do c[x]:=z; end;
procedure c_init_Point_middle(background,color:integer);
begin c_init(background); c[N div 2]:=color; end;
//________________________________________ init data state
//________________________________________ print
procedure c_print;
var x, mid:integer;
begin
    mid:=N div 2;    // show pointers of cell at midddle
    for x:=0 to N-1 do   
      case  c[x]  of  0: write(OUT_c, OutputZERO); 1: write(OUT_c, OutputONE); otherwise write(OUT_c, ' ?'); end;      
    writeln(OUT_c,' t=',t:4, ' at[mid]: ', 'p1=', p1[mid]:4,' p2=', p2[mid]:4);
end;
procedure p_print(var ff:textfile; pointer:field);     // p1, p2
var x:integer; DIGITS:integer=3;
begin  if N<10 then DIGITS:=2 else if N<100 then DIGITS:=3 else if N<1000 then DIGITS:=4 else DIGITS:=5;       
       for x:=0 to N-1 do  write(ff, pointer[x]:DIGITS );  writeln(ff,' t=',t);
end;
//________________________________________ print
// ========================================================================== FUNCTIONS, PROCEDURES
// ========================================================================== MAIN
BEGIN
  assign(OUT_c,     'OUT_c.txt');     rewrite(OUT_c);
  assign(OUT_p1,    'OUT_p1.txt');    rewrite(OUT_p1); assign(OUT_p2,    'OUT_p2.txt');    rewrite(OUT_p2);  
  //______________________________________ init data at t=0
  c_init_Point_middle(0,1);  t:=0;
  //______________________________________ init data at t=0
  //______________________________________ init pointer at t=0
  p1new_p2new_Apply_PointerRule_at_t_for_tplus1; // init for t=0, see proc!
  p1:=p1new;  p2:=p2new;                         //syncupdate pointer t=0, init
  //______________________________________ init pointer at t=0
  //______________________________________ output initial at t=0
  c_print;
  p_print(OUT_p1,p1); p_print(OUT_p2,p2);
  //______________________________________ output initial at t=0
  for t:=1 to TMAX do
  begin
    //____________________________________ compute next generation
    //# state c and pointers p1,p2 are available (were computed at t-1)
    cnew_ApplyDataRule;                             // 1a. apply data rule
    p1new_p2new_Apply_PointerRule_at_t_for_tplus1;  // 1b. apply pointer rules
    c:=cnew;                                        // 2a. syncupdate data
    p1:=p1new;  p2:=p2new;                          // 2b. syncupdate pointer
    //____________________________________ compute next generation
    //____________________________________ output new generation at t after computation
    c_print;  p_print(OUT_p1,p1); p_print(OUT_p2,p2);
    //____________________________________ output new generation at t after computation
  end;
  close(OUT_c); close(OUT_p1); close(OUT_p2);
END.
// ========================================================================== MAIN END
output textfile OUT_c:
                               #                               t=   0 at[mid]: p1=   1 p2=  -1
                             #   #                             t=   1 at[mid]: p1=   2 p2=  -2
                         #   #   #   #                         t=   2 at[mid]: p1=   4 p2=  -4
                 #   #   #   #   #   #   #   #                 t=   3 at[mid]: p1=   8 p2=  -8
 #   #   #   #   #   #   #   #   #   #   #   #   #   #   #   # t=   4 at[mid]: p1=  16 p2= -16
 # # # # # # # # # # # # # # #   # # # # # # # # # # # # # # # t=   5 at[mid]: p1=   1 p2=  -1
\end{verbatim}
\normalsize

\newpage

\subsection{General Model with Address Modification}
\label{Appendix 0-General}

The following general GCA program computes the same result as the basic GCA program before. 
Two address bases $p1$ and $p2$ are used, that store the same value sequence 1, 2, 4, ...  .
(Therefore it would be sufficient to use one address base only.)
The effective addresses are $\textit{p1eff}=p1$ and $\textit{p2eff}=-p2$.

\tiny
\begin{verbatim}

{5.6.2022 RH. Simple 1D General GCA program, XOR, address modification}
program prog_gca_gneral_xor;
uses  SysUtils;
var   OUT_c, OUT_p1, OUT_p2, OUT_p1eff, OUT_p2eff: textfile;

const BlackSquare=#$E2#$96#$88#$E2#$96#$88;
const OutputZERO='  ';  OutputONE=' #'; // BlackSquare; can be used
const N=31; TMAX=5;       // number of cells, max number of generations

type  field = array [0..N-1] of integer;
type  cell = record c,cnew, p1,p1n2, p2,p2new, p1eff,p2eff : field end;  // cell's structure, not used here

var   c, cnew: field;      // data state, buffered sync operation
      p1, p1new: field;    // stored relative pointer, buffered sync operation
      p2, p2new: field;
      p1eff,p2eff: field;  // effective addresses, only temp variable
var   t: integer;          // time-counter, generation
//=========================================================================== FUNCTIONS, PROCEDURES
function modN(a:integer):integer; begin modN:=(a+N)mod N; end;

//________________________________________ effective address
procedure p1eff_p2eff_EffectiveAddress_at(x:integer);
begin
   p1eff[x]:=-1; p2eff[x]:=+1; // fixed nearest neighbors ok ECA, default
   // may also depend on cell's data state, fixed neighbors' states, time, index x
   begin p1eff[x]:= p1[x]; p2eff[x]:= -p2[x]; end;  // modified
end;
//________________________________________ effective address
//________________________________________ effective address for all cells
procedure p1eff_p2eff_Apply_EffectiveAddress_at_t_for_t;
var x: integer;    // index
begin for x:=0 to N-1 do p1eff_p2eff_EffectiveAddress_at(x); end;
//________________________________________ effective address for all cells
//________________________________________ new Pointer for all cells
procedure p1new_p2new_Apply_PointerRule_at_t_for_tplus1;
const p1init=1; p2init=1; var   x: integer;   // cell's index/position
begin
   for x:=0 to N-1 do
   begin
       // new pointer pnew may depend on: t,x, (c, p1, p2), p1.(c,p1,p2), p2.(c,p1,p2)
       //............................................................Pointer Rules
           //_____________________ initial set pointer const at t=0
           if t=0 then p1new[x]:=p1init; p2new[x]:=p2init;
           //_____________________ initial set pointer const at t=0
           //_____________________ for t=1,2, ...
           if t>0 then begin
             p1new[x]:=(p1[x]*2) mod N; p2new[x]:=(p2[x]*2) mod N;
             if p1new[x]=0 then p1new[x]:=p1init; //don't use p=0, instead pinit=1
             if p2new[x]=0 then p2new[x]:=p2init; end;             
           //_____________________ for t=1,2, ...
       //............................................................Pointer Rules
    end;  // for x
end;
//________________________________________ new Pointer for all cells
//________________________________________ data rule at site x
function DataRule_with_Data_at_Pointers(x,p1eff,p2eff: integer):integer;
   function abs(peff_relative:integer): integer;
   begin    abs:=modN(x+peff_relative); end;
begin
    // L exor R, abs(p1eff)=modN(x+p1eff), c[x] or c[modn(x+1) .. could also be used
    // may also depend on cell's state, fixed neighbors' states, time t, index x
    // new data cnew may depend on: t,x, (c, p1, p2), p1.(c,p1,p2), p2.(c,p1,p2)
    DataRule_with_Data_at_Pointers:=( c[abs(p1eff)]+c[abs(p2eff)] ) mod 2;
end;
//________________________________________ data rule at site x
//________________________________________ new cells' data states
procedure cnew_ApplyDataRule;
var x: integer;
begin for x:=0 to N-1 do cnew[x]:=DataRule_with_Data_at_Pointers(x, p1eff[x], p2eff[x]); end;
//________________________________________ new cells' data states
//________________________________________ init data state
procedure c_init(z:integer);
var   x:integer;
begin for x:=0 to N-1 do c[x]:=z; end;
procedure c_init_Point_middle(background,color:integer);
begin c_init(background); c[N div 2]:=color; end;
//________________________________________ init data state
//________________________________________ print
procedure c_print;
var x, mid:integer;
begin
    mid:=N div 2;    // show pointers of cell at mid
    for x:=0 to N-1 do
      case  c[x]  of  0: write(OUT_c, OutputZERO); 1: write(OUT_c, OutputONE);  otherwise write(OUT_c, ' ?'); end;
    writeln(OUT_c,' t=',t:4, ' at[mid]: ',
    'p1=', p1[mid]:4,' p2=', p2[mid]:4,' p1eff=',p1eff[mid]:4,' p2eff=',p2eff[mid]:4);
end;
procedure p_print(var ff:textfile; pointer:field);     //p1,p2,p1eff,p21eff
var x:integer;
var DIGITS:integer=3;
begin  if N<10 then DIGITS:=2 else if N<100 then DIGITS:=3
       else if N<1000 then DIGITS:=4 else DIGITS:=5;
       for x:=0 to N-1 do  write(ff, pointer[x]:DIGITS );  writeln(ff,' t=',t);
end;
//________________________________________ print
//=========================================================================== FUNCTIONS, PROCEDURES
// ========================================================================== MAIN
BEGIN
  assign(OUT_c,     'OUT_c.txt');     rewrite(OUT_c);
  assign(OUT_p1,    'OUT_p1.txt');    rewrite(OUT_p1);
  assign(OUT_p2,    'OUT_p2.txt');    rewrite(OUT_p2);
  assign(OUT_p1eff, 'OUT_p1eff.txt'); rewrite(OUT_p1eff);
  assign(OUT_p2eff, 'OUT_p2eff.txt'); rewrite(OUT_p2eff);
  //______________________________________ init data at t=0
  c_init_Point_middle(0,1);
  //______________________________________ init data at t=0
  //______________________________________ init pointer at t=0
  t:=0;
  p1new_p2new_Apply_PointerRule_at_t_for_tplus1; // init for t=0, see proc!
  p1:=p1new;  p2:=p2new;      //syncupdate pointer t=0, init
  // peff depends on p init, to be printed at t=0
  p1eff_p2eff_Apply_EffectiveAddress_at_t_for_t;
  //______________________________________ init pointer at t=0
  //______________________________________ output initial at t=0
  c_print;
  p_print(OUT_p1,p1); p_print(OUT_p2,p2);
  p_print(OUT_p1eff,p1eff); p_print(OUT_p2eff,p2eff);
  //______________________________________ output initial at t=0
  for t:=1 to TMAX do
  begin
    //____________________________________ compute next generation
    //# state c and pointer p are computed
    p1eff_p2eff_Apply_EffectiveAddress_at_t_for_t;  // 1.  compute peff
    cnew_ApplyDataRule;                             // 2a. apply data rule
    p1new_p2new_Apply_PointerRule_at_t_for_tplus1;  // 2b. apply pointer rule
    c:=cnew;                                        // 3a. syncupdate data
    p1:=p1new;  p2:=p2new;                          // 3b. syncupdate pointer
    //____________________________________ compute next generation
    //____________________________________ output new generation at t after computation
    c_print;  p_print(OUT_p1,p1); p_print(OUT_p2,p2);
    p_print(OUT_p1eff,p1eff);     p_print(OUT_p2eff,p2eff);
    //____________________________________ output new generation at t after computation
  end;
  close(OUT_c); close(OUT_p1); close(OUT_p2); close(OUT_p1eff); close(OUT_p2eff)
END.
// ========================================================================== MAIN END
output text file OUT_c:
                               #                               t=   0 at[mid]: p1=   1 p2=   1 p1eff=   1 p2eff=  -1
                             #   #                             t=   1 at[mid]: p1=   2 p2=   2 p1eff=   1 p2eff=  -1
                         #   #   #   #                         t=   2 at[mid]: p1=   4 p2=   4 p1eff=   2 p2eff=  -2
                 #   #   #   #   #   #   #   #                 t=   3 at[mid]: p1=   8 p2=   8 p1eff=   4 p2eff=  -4
 #   #   #   #   #   #   #   #   #   #   #   #   #   #   #   # t=   4 at[mid]: p1=  16 p2=  16 p1eff=   8 p2eff=  -8
 # # # # # # # # # # # # # # #   # # # # # # # # # # # # # # # t=   5 at[mid]: p1=   1 p2=   1 p1eff=  16 p2eff= -16
\end{verbatim}
\normalsize
\newpage
\section{Appendix 1: Program for Synchronous Firing within Two Rings}
\label{Appendix 1}
\begin{figure}[hbt]
	\centering
		\includegraphics[width=12cm]{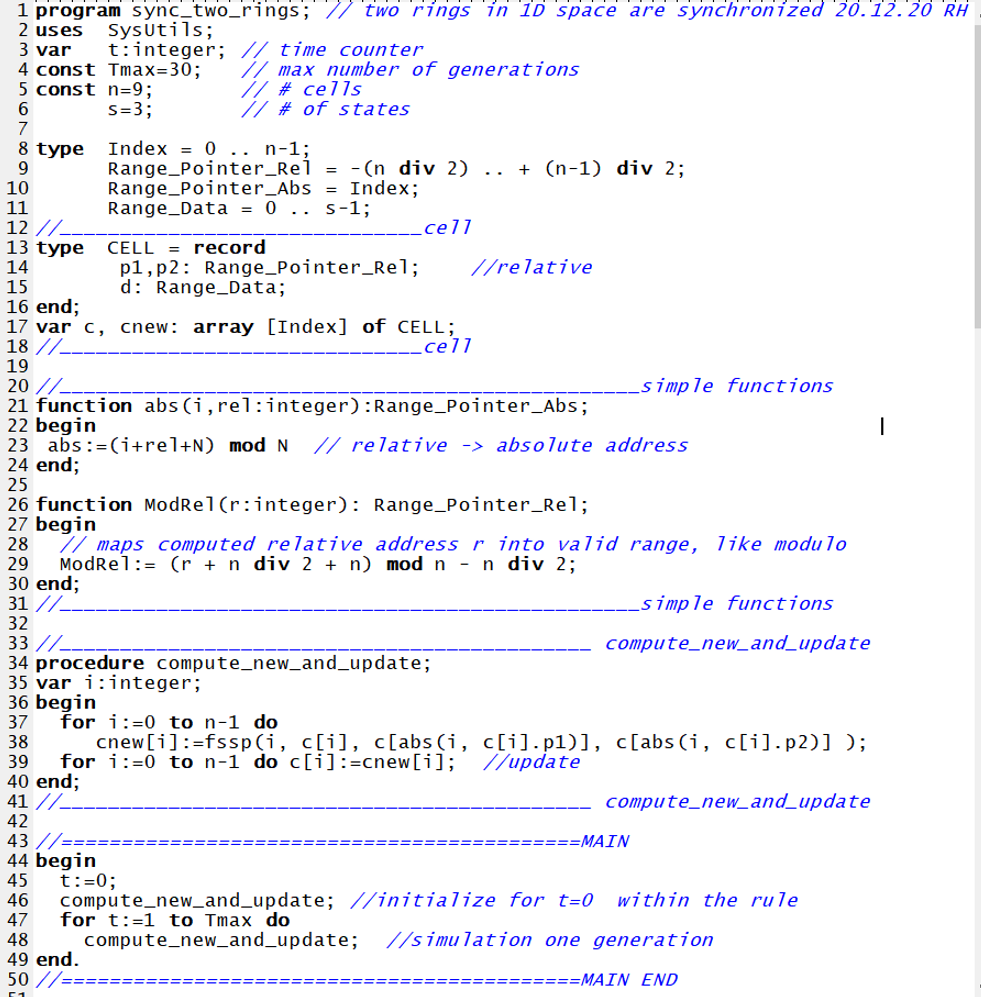}
			\caption{Pascal program part 1: Main. Synchronous Firing within two rings in 1D using waves as described in 
			Sect. \ref{Synchronous Firing with Spaces}.
			}
	\label{program-rule-fssp-main}
\end{figure}

\begin{figure}[hbt]
	\centering
		\includegraphics[width=12cm]{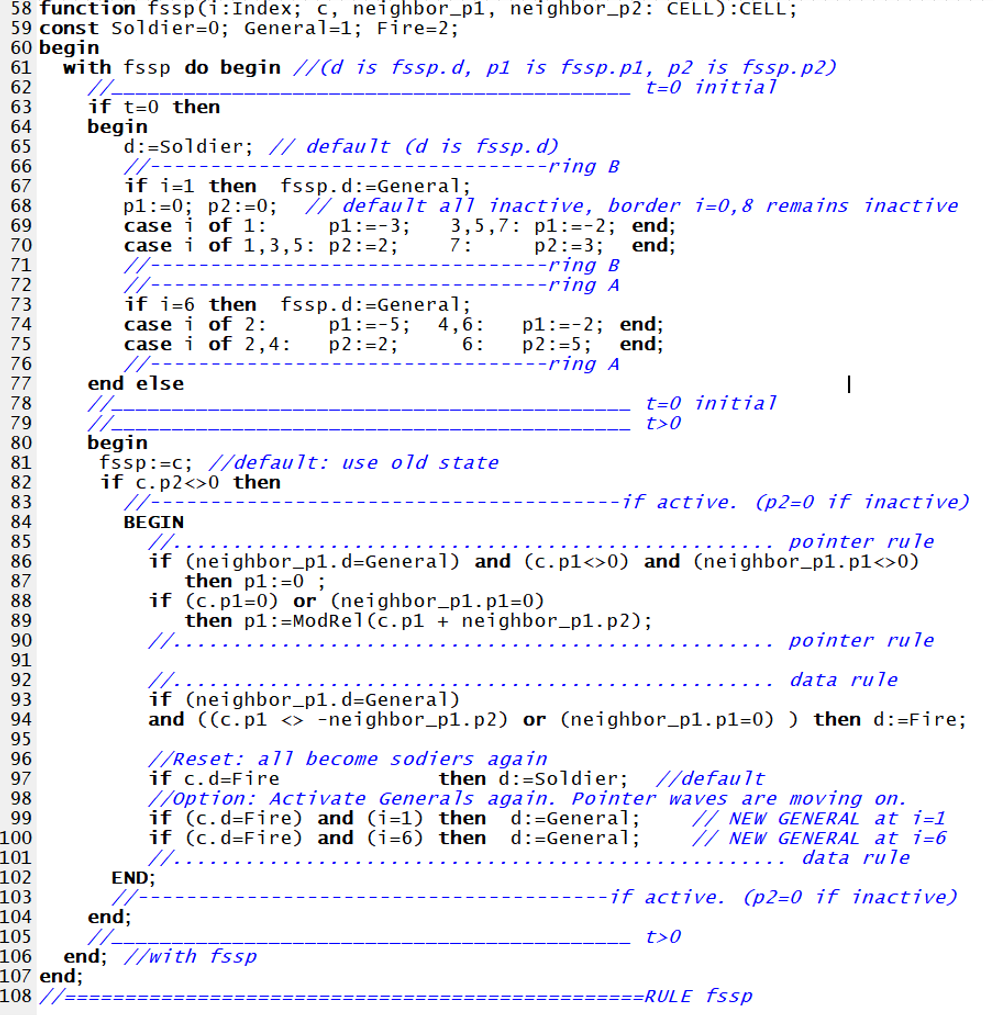}
			\caption{Pascal program part 2: The Rule fssp. Synchronous Firing within two rings in 1D using waves as described in
			Sect. \ref{Synchronous Firing with Spaces}.
			}
	\label{program-rule-fssp}
\end{figure}

\clearpage
\section{Appendix 2: First Paper \cite{HVW00} Introducing the GCA Model}
\label{Appendix 2}

\begin{center}
\large{Global Cellular Automata GCA:\\
An Universal Extension of the CA Model}
\normalsize

\emph{Rolf Hoffmann,
Klaus--Peter V\"olkmann,
Stefan Waldschmidt}\\
\footnotesize
Darmstadt University of Technology, Germany\\
{\tt (hoffmann,voelk,waldsch)@informatik.tu-darmstadt.de}
\end{center}


\begin{abstract}

A model called global cellular automata (GCA) will be introduced. The
new model preserves the good features of the cellular automata but overcomes
its restrictions. In the GCA the cell state consists of a data field and
additional pointers. Via these pointers, each cell has read access to any
other cell in the cell field, and the pointers may be changed from
generation to generation. Compared to the cellular automata the
neighbourhood is dynamic and differs from cell to cell. For many
applications parallel algorithms can be found straight forward and can
directly be mapped on this model. As the model is also massive parallel
in a simple way, it can efficiently be supported by hardware.
\footnote{The section numbering has changed here because the old paper was integrated into this comprising publication.}

\end{abstract}

\subsection{Motivation}

The classical cellular automata model (CA) can be characterized by the
following features

\begin{itemize}

\item The CA consists of a {\it n}--dimensional field of cells. Each cell can be
identified by its coordinates.

\item The neighbours are fixed and are defined by relative coordinates.

\item Each cell has local read access to the states of its neighbours. Each
cell contains a local rule. The local rule defines the next state depending
on the cell state and the states of the neighbours.

\item The cells are updated synchronously, the new generation of cells (new
cell states) depend on the old generation (old cell states).

\item The model is massive parallel, because all next states can be computed
and updated in parallel.

\item Space or time dependent rules can be implemented by the use of special
space or time information coded in the state.

\end{itemize}

The CA is very well suited to problems and algorithms, which need only
access to their fixed local neighbours \cite{TOFFOLI}. Algorithms with
global (long distance) communication can only indirectly be implemented by
CA. In this case the information must be transported step by step along the
line from the source cell to the destination cell, which needs a lot of
time. Therefore the CA is not an efficient model for global algorithms.

We have searched for a new model, which preserves the good features of the
CA but overcomes the local communication restriction. The new model shall be
still massive parallel, but at the same time suited to any kind of global
algorithm. Thus we will be able to describe a more general class of
algorithms in a more efficient and direct way. We also claim that this model
can efficiently be implemented in hardware.

\subsection{The GCA model}

The model is called {\it global automata model} (GCA). The GCA can be
characterized by the following features

\begin{itemize}

\item A GCA consists of a {\it n}--dimensional field of cells. Each cell can
be identified by its coordinates.

\item Each cell has {\it n} individual neighbours which are variable and may
change from generation to generation. The neighbours are defined by relative
coordinates (addresses, pointers).

\item The state of a cell contains a data field and {\it n} address fields.
\begin{itemize}
\item[] {\it State = (Data, Address1, Address2, ...)}
\end{itemize}

\item Each cell has global read access to the states of its neighbours by
the use of the address fields.

\item Each cell contains a {\it local rule}. The local rule defines the next
state depending on the cell state and the states of the neighbours. By
changing the state, the addresses may also be changed, meaning that in the
next generation different neighbours will be accessed.

\item The cells are updated synchronously, the new generation of cells
depends on the old generation.

\item The model is massive parallel, because all next states can be computed
and updated in parallel.

\item Space or time dependent rules can be implemented by the use of special
space or time information coded in the state.

\end{itemize}

A one--dimensional GCA with two address fields will be defined in a formal way, using a
PASCAL like notation:

\begin{enumerate}

\item The cell field
\begin{verbatim}
Cell = array [0..n-1] of State
\end{verbatim}

\item The State of each cell
\begin{verbatim}
State = record
  Data: Datatype
  Address1: 0..n-1
  Address2: 0..n-1
endrecord
\end{verbatim}

\item The definition of the local rule
\begin{verbatim}
function Rule(Self:State, Neighbour1:State,Neighbour2:State)
\end{verbatim}

\item The computation of the next generation
\begin{verbatim}
for i=0..n-1 do in parallel
    Cell[i]:= Rule(Cell[i], Cell[Address1], Cell[Address2])
endfor
\end{verbatim}

\end{enumerate}

\begin{figure}[ht]
  \begin{center}
    \includegraphics[width=0.8\textwidth]{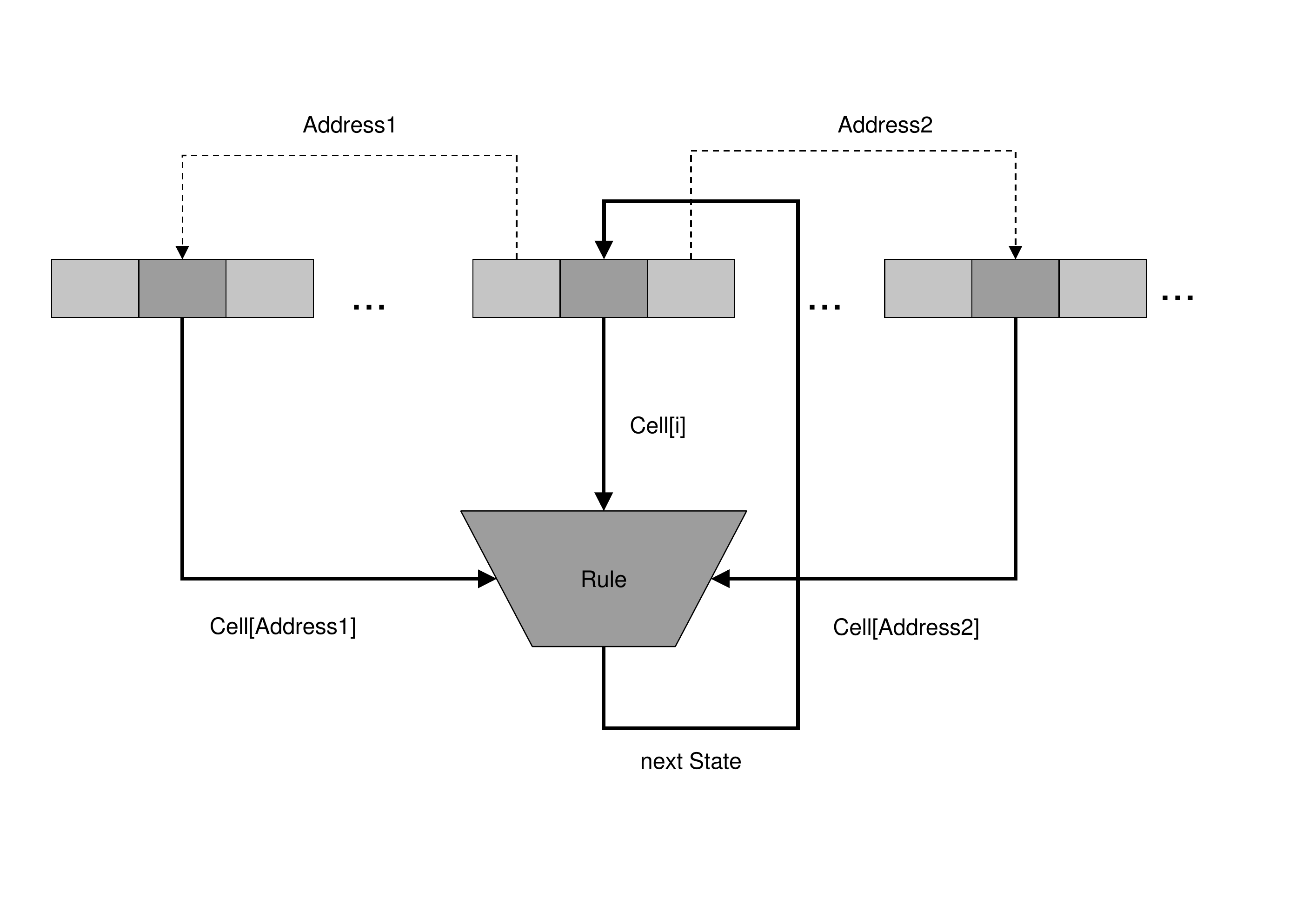}
  \end{center}
  \vspace{-5em}
  \caption{The GCA model.}
  \label{fig:one}
\end{figure}

Fig.~\ref{fig:one} shows the principle of the GCA model. Cell[i] reads two
other cell states and computes its next state, using its own state and the
states of the two other states in access. In the next state, Cell[i] may
point to two different cells.

The above model can be defined in a more general way with respect to the
following features

\begin{itemize}

\item The number {\it k} of addresses can be 1, 2, 3... If {\it k=1} we call it a
{\it one-handed} GCA, if {\it k=2} we call it a {\it two-handed} GCA and so forth.

\item The number {\it k} may vary in time and from cell to cell, in this
case it will be a {\it variable-handed} GCA.

\item Names could be used for the identification of the cells, instead of
ordered addresses. In this case the cells can be considered as an unordered
set of cells.

\item A special {\it passive} state may be used to indicate that the cell state
shall not be changed any more. It can be used to indicate the end of the
computation or the deletion of a cell. A cell which is not in the {\it
passive} state is called {\it active}. An active cell may turn a passive cell to
active.

\end{itemize}

Similar models have been proposed before\cite{KU63}. Usually they are
theoretically oriented and lack any aspects of applications and
implementations\cite{Sch-real-tsm}.

\subsection{Mapping problems on the GCA model}

The GCA has a very simple and direct programming model. The programming model
is the way how the programmer has to think in order to map an algorithm to a
certain model, which is interpreted by a machine. In our case, the
programmer has to keep in mind, that a machine exists which interpretes and
executes the GCA model.

Many problems can easily and efficiently be mapped to the GCA model, e.g.

\begin{itemize}

\item sorting of numbers

\item reducing a vector, like sum of vector elements

\item matrix multiplication

\item permutation of vector elements

\item graph algorithms

\end{itemize}

The following examples are written in the cellular programming language
CDL\cite{PACT95}. CDL was designed to facilitate the description of cellular
rules based on a rectangular n-dimensional grid with a local neighbourhood.
The locality of the neighbourhood radius was asserted and controlled by the
declaration of {\tt distance}={\it radius}. For the GCA the new keyword {\tt
infinity} was introduced for the declaration of the {\it radius}.

In CDL the unary operator \verb+*+ is used (like in C) to dereference the
relative address of a cell in order to obtain the state of the referenced
cell. The following
examples are {\it one--handed} GCAs, showing how useful unlimited
read-access to any other cell is.

\subsubsection{Example 1: Firing Squad Problem}

This is an implementation of the firing squad algorithm on a one dimensional
array. The set of possible states for every cell is described by the type
{\tt celltype} in lines \listing(7) to \listing(11).

The sequence of soldier cells ({\tt kind=soldier}) has to be enclosed by
edge--cells ({\tt kind=edge}) to mark the border of the squad. The {\tt
wave} component of all cells should be initialised with {\tt [-1]} which is
the relative address, pointing to the neighbour on the left.

\footnotesize
\begin{myverbatim}
cellular automaton firing_1;

const  dimension = 1;
       distance  = infinity;     // allows unlimited access
       init=[-1];

type  celltype=record
        kind : (soldier,edge);  // have an edge on each side
        wave : celladdress;     // init with [-1]
        fire : boolean;         // init with false
      end;

var   n:celladdress;   

rule begin
  n:=(*[0]).wave;  // address of the cell our wave points to

  if (n=init) then            // we are in init state
    if (*n.kind=edge) or (*n.wave!=init) then 
                              // the wave is just coming
      *[0].wave:=[0]
  else                        // the wave passed already
    if *n.kind=edge then
      *[0].fire:=true         // the wave reached the edge
    else 
      *[0].wave:=[n.
end;
\end{myverbatim}
\normalsize

At the beginning ({\tt n=[-1]}) every soldier is looking to his direct
neighbour on the left. If his neighbour is a edge cell or a soldier which is
not in the init state anymore (line \listing(19)) the soldier himself will
leave the init state (line \listing(21)) and defines the front of the wave.

\begin{figure}[ht]
  \begin{center}
    \includegraphics[width=0.7\textwidth]{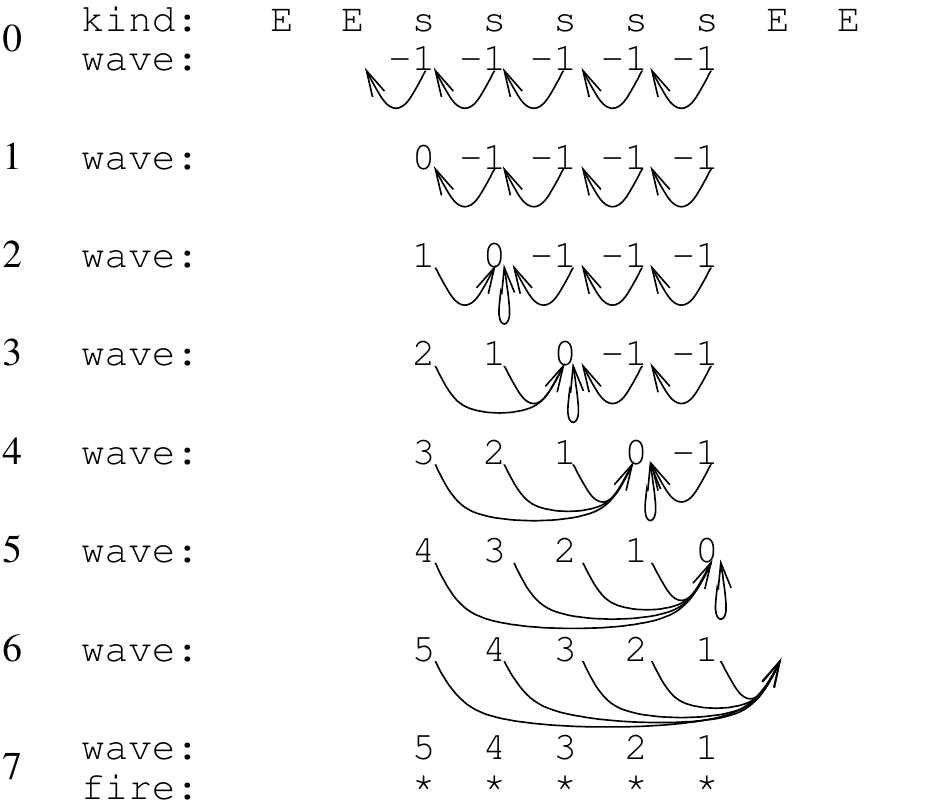}
  \end{center}
  \vspace{-1em}
  \caption{The firing squad.}
  \label{fig:fs_1}
\end{figure}

If the wave already passed the soldier (lines \listing(23) to \listing(26))
the variable {\tt n} points to the wave front. All soldiers fire when the
wave reaches the right edge, otherwise the wave rolls on one more step.

\subsubsection{Example 2: Fast Fourier Transformation}

The fast Fourier transformation (FFT) is another, more complex example. We
do not want to explain the algorithm in this paper, it is described in
details in \cite{book:fft}. The example is used to demonstrate that a complex
algorithm can

\begin{itemize}
\item easily be mapped onto the GCA model
\item concisely be described
\item efficiently be executed in parallel
\end{itemize}

\begin{figure}[ht]
  \begin{center}
    \includegraphics[width=0.7\textwidth]{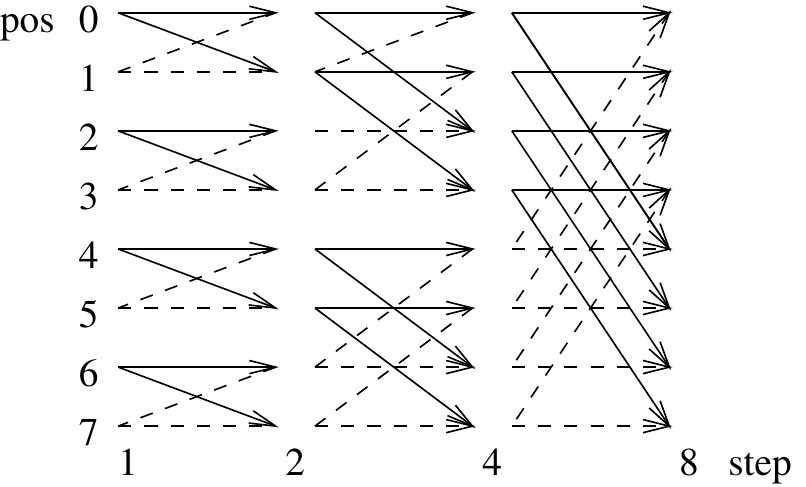}
  \end{center}
  \vspace{-1.0em}
  \caption{The FFT access pattern.}
  \label{fig:fft}
\end{figure}

Each cell contains a complex number ({\tt r,i}) which is calculated in every
time step from its own number and the number contained in another cell.
The address of the other cell depends on its own absolute address ({\tt
position}) and the time step in the way shown in fig.~\ref{fig:fft}.

For example, the cell at position 2 reads the cell at position 3 in the
first step, the cell at position 0 in the next step, and the cell at
position 6 in the last time step. Obviously this access pattern can not be
implemented efficiently on a classical cellular automaton using strict
locality.

\footnotesize
\begin{myverbatim}
cellular automaton FFT;

const  dimension=1;
       distance=infinity;  // global access, radius of neighborhood

type  celltype=record
        r,i       : float;   // the complex value
        step      : integer; // initialised with 1
        position  : integer; // init with 0..(2^k)-1
      end;

#define this  *[0] // the cell's state, contents(*) of rel. address 0

var  other:celladdress;
     a,wr,wi:float;

rule begin
  // calculate relative address of other cell
  other := [ (this.position exor this.step)-this.position ];

  // calculate new values for local r and i
  a:= -pi / this.step * (this.position and (this.step-1));
  wr:=cos(a);
  wi:=sin(a);
  if ( other > 0 )
  {                      // other cell has higher number
    this.r := this.r + wr* *other.r - wi* *other.i;
    this.i := this.i + wr* *other.i + wi* *other.r;
  }
  else
  {                      // other cell has lower number
    this.r := *other.r - ( wr* this.r - wi* this.i );
    this.i := *other.i - ( wr* this.i + wi* this.r );
  }

  this.step    := 2 * this.step;    // step=1,2,4,8...
  this.position:= this.position;    // carry own position
end
\end{myverbatim}
\normalsize

The algorithm is concise and efficient because the address of the neighbour
is calculated (line \listing(19)) and thereby an individual neighbour is
accessed (lines \listing(27) and \listing(28)). The listing of the FFT
without using this feature would at least be twice as long and the
calculation would take significantly more time.

\subsection{Conclusion}

We have introduced a powerful model, called {\it global cellular
automata} (GCA). The cell state is composed of a data field and {\it n} pointers
which point to {\it n} arbitrary other cells. The new cell state is computed
by a local rule, which takes into account its own state and the states of
the other cells which are in access via the pointers. In the next generation
the pointers may point to different cells. Each cell changes its state
independently from the other cells, there are no write conflicts. Therefore
the GCA model is massive parallel meaning that it has a great potential to
be efficiently supported by hardware. We plan do implement the GCA model on
the CEPRA-S processor \cite{Hoffmann:2000:SPA}.

Parallel algorithms can easily be described and mapped onto the GCA.
Compared to the CA model it is much more flexible although it is only a
little more complex.

\subsection{References of First Paper (Appendix 2)}

\newpage
\section{References of Sections 1 -- 6}

\end{document}